\pdfoutput=1

\documentclass[11pt]{article}
\usepackage{amssymb}
\usepackage{amsmath}
\usepackage{verbatim}
\usepackage{url}
\usepackage{tikz}
\usepackage{graphicx}
\usepackage{amsthm}
\usepackage{etoolbox}
\usepackage{bbm}
\usepackage[margin=1.25in]{geometry} 
\usepackage{abstract}
\usepackage{standalone}
\usepackage{url}
\usepackage{booktabs}
\usepackage{multicol}
\usepackage{array}
\usepackage{longtable}
\usepackage{supertabular}
\usepackage{makeidx}

\makeindex

\usepackage[nottoc,notlot,notlof]{tocbibind}

\usepackage{listings,xcolor}
\usepackage{asymptote}
\usepackage[font=small,labelfont=bf]{caption}
\usepackage{chngcntr}
\counterwithin{figure}{section}
\usepackage{subcaption}
\usepackage{setspace}
\def\Xint#1{\mathchoice
{\XXint\displaystyle\textstyle{#1}}%
{\XXint\textstyle\scriptstyle{#1}}%
{\XXint\scriptstyle\scriptscriptstyle{#1}}%
{\XXint\scriptscriptstyle\scriptscriptstyle{#1}}%
\!\int}
\def\XXint#1#2#3{{\setbox0=\hbox{$#1{#2#3}{\int}$ }
\vcenter{\hbox{$#2#3$ }}\kern-.6\wd0}}

\def\dashint{\Xint-}

\DeclareCaptionFont{black}{\color{black}}
\DeclareCaptionFormat{listing}{{\hfill \parbox{\textwidth}{\hfill #1 $|$ #2#3}}}

\usepackage{hyperref}
\usepackage{color}
\hypersetup{
    colorlinks,
    citecolor=black,
    filecolor=black,
    linkcolor=black,
    urlcolor=black
}

\newcommand{\bb}{\\\\\vspace{-15pt}\\}

\newcommand{\SL}{\text{SL}}

\renewcommand{\div}{\text{div}}
\newcommand{\grad}{\text{grad}}

\renewcommand{\epsilon}{\varepsilon}
\newcommand{\eps}{\varepsilon}
\newcommand\getcurrentref[1]{%
 \ifnumequal{\value{#1}}{0}
  {??}
  {\the\value{#1}}%
}    
\newcounter{itemcounter}
\newcommand{\cc}{\getcurrentref{section}.\getcurrentref{subsection}.\theitemcounter.\stepcounter{itemcounter}}
\begin{document}
\nocite{*}
\pagenumbering{gobble}
\begingroup 
\hbox{ \hspace*{0.05\textwidth} 
\rule{1pt}{\textheight} 
\hspace*{0.05\textwidth} \parbox[b]{0.8\textwidth}{ 
{\noindent\Large\bfseries 
Quantum Ergodicity and  the Analysis \vspace{2pt}\\
 of Semiclassical Pseudodifferential Operators 
}\\[2\baselineskip] 
Felix J. Wong$^*$
\\\\
A thesis presented 
to the Department of Mathematics, Harvard University, Cambridge, Massachusetts, in partial fulfillment of the honors requirements for the A.B. degree in Mathematics.\\\\
\emph{Advisor: Clifford H. Taubes}
\\
\emph{April 2014}
\\\\\\\\\\\\\\\\\\\\\\\\\\\\\\\\\\

\begin{spacing}{0.8}
\begin{longtable}{@{}p{0.04in}@{}p{5in}}
{\footnotesize $^*$} & {\footnotesize \emph{Present address: Department of Applied Physics, Harvard School of Engineering and Applied Sciences, Cambridge, MA 02138, USA}. Email: {felixjwong@seas.harvard.edu}.}
\\
\end{longtable}
\end{spacing}
}}
\endgroup

\section*{Acknowledgements} 
I have had the fortune to meet some of my closest friends and wisest advisors during my time as an undergraduate at Harvard. 

I humbly thank my advisor, Clifford Taubes, without whose advice and patience this thesis would not have been possible. Aside from suggesting useful sources in the literature, discussing the role of PDEs in geometry, promptly responding to my questions, correcting my many misunderstandings in spectral and geometric analysis, and reviewing numerous drafts of my thesis, Professor Taubes has taught me a great many things, not the least of which is how to become a better mathematician. 

I thank my concentration advisor, Wilfried Schmid, for his keen insight and wise direction throughout my time as a mathematics concentrator. Professor Schmid has always made himself available for me at the most ungodly hours, and his advice for both my studies and life in general has never fallen short of amazing.

I thank several great mathematicians from whom I've had the delightful opportunity to learn, including my research mentor Jeremy Gunawardena and professors Sukhada Fadnavis, Benedict Gross, Joe Harris, Peter Kronheimer, Siu Cheong Lau, Martin Nowak, and Horng-Tzer Yau. With their guidance, never once have I felt lost in the vast sea of mathematics.

I thank professors Ronald Walsworth, Amirhamed Majedi, and Joe Blitzstein for teaching and guiding me in some the most rewarding courses I have ever taken. 

Finally, I thank my family and friends for their support over the years. 
\pagenumbering{arabic}

\pagebreak
\begin{abstract}
This thesis is concerned with developing the tools of differential geometry and semiclassical analysis needed to understand the the quantum ergodicity theorem of Schnirelman (1974), Zelditch (1987), and Colin de Verdi\`{e}re (1985) and the {quantum unique ergodicity conjecture} of Rudnick and Sarnak (1994). The former states that, on any Riemannian manifold with negative curvature or ergodic geodesic flow, the eigenfunctions of the Laplace-Beltrami operator equidistribute in phase space with density 1. 
Under the same assumptions, the latter states that the eigenfunctions induce a sequence of Wigner probability measures on fibers of the Hamiltonian in phase space, and these measures converge in the weak-$*$ topology to the uniform Liouville measure. If true, the conjecture implies that such eigenfunctions equidistribute in the high-eigenvalue limit with no exceptional ``scarring" patterns. This physically means that the finest details of chaotic Hamiltonian systems can never reflect their quantum-mechanical behaviors, even in the semiclassical limit. 

The main contribution of this thesis is to contextualize the question of quantum ergodicity and quantum unique ergodicity in an elementary analytic and geometric framework. 
In addition to presenting and summarizing 
numerous important proofs, 
such as Colin de Verdi\`{e}re's proof of the quantum ergodicity theorem, 
we perform graphical simulations of certain billiard flows and expositorily discuss several themes in the study of quantum chaos. 
\end{abstract}

\pagebreak
\tableofcontents

\pagebreak

\section{Introduction}

Much of the research done in mathematical physics over the past few decades has concerned itself with describing the bridge between the classical world and the quantum regime. How does the transition from classical dynamics to quantum mechanics occur, and when is chaotic behavior in the classical world generated by quantum effects? Investigations into these questions have resulted in a proliferation of mathematical techniques and insights that have deep implications not only in quantum chaos and geometric analysis, but also ergodic theory and number theory. For instance, resolving the problem of \emph{quantum ergodicity} has aided analytic geometers in understanding the equidistribution of {Laplacian eigenfunctions}. Developing a procedure for operator quantization has helped in a number of applications, including spectral statistics and \emph{semiclassical analysis}.

The field that deals with the relationship between quantum mechanics and classical chaos has naturally been termed \emph{quantum chaos}. \index{quantum chaos} The mathematical theory behind quantum chaos---which has to some extent been guided by physics intuition---is surprisingly rich. In quantum mechanics, for example, the rigorous formulation of Weyl's functional calculus has led to a theory of pseudodifferential operators, operators which simplify a wide range of partial differential equations. By allowing the manipulation of operators as if they were scalars, this functional calculus has also rigorously justified the {correspondence principle} \index{Egorov's theorem!correspondence principle} in physics through a more mathematical formulation known as \emph{Egorov's theorem}. \index{Egorov's theorem}

Common to both quantum chaos and recent trends in geometric analysis is the Laplace operator $\Delta$, which on the Euclidean space $\mathbb{R}^n$ with coordinates $(x_1,...,x_n)$, is defined as $\Delta=\sum_{i=1}^n\frac{\partial^2}{\partial x_i^2}$. There is a natural analogue of $\Delta$ on any Riemannian manifold (c.f. \S1.1), and it is well-understood that the eigenvalue spectrum of $\Delta$ provides geometric information about the source manifold. One of the most widely known investigations into the geometric properties of the Laplacian dates back to Weyl, and was presented in Kac's seminal 1966 paper in \emph{American Mathematical Monthly} \cite{Kac1}. Given the \emph{Helmholtz equation}\index{Weyl's law}
$$\Delta \psi_n+k^2_n\psi_n=0$$
where $\psi_n$ denotes an eigenfunction of $\Delta$ with eigenvalue $k^2_n$ (and frequency $k_n$), Kac asked if one could determine the geometric shape of a Euclidean domain or manifold knowing the spectrum of $\Delta$. Since the Helmholtz equation \index{Helmholtz equation} is a special case of the wave equation $\Delta\psi-c^{-2}\partial_t^2 \psi=0$ and the eigenfunctions of $\Delta$ correspond to sound waves, this question has been popularly rephased as ``can one hear the shape of a drum?" \index{eigenvalue} \index{eigenfunction}

In 1964, Milnor showed that the eigenvalue sequence for $\Delta$ does not, in general, characterize a manifold completely by exhibiting two 16-dimensional tori that are distinct as Riemannian manifolds but share an identical sequence of eigenvalues \cite{Mil1}. A similar result was shown for the two-dimensional case in 1992, for which Gordon, Webb, and Wolpert constructed two different regions in $\mathbb{R}^2$ sharing the same set of eigenvalues \cite{Gor1}. Nevertheless, it is known from a proof of Weyl's that one can still ``hear" the area of a domain $D\subset \mathbb{R}^2$; i.e. 
$$N(\lambda)\sim \frac{\text{Area}(D)}{4\pi}\lambda$$ 
in the limit $\lambda \to \infty$, where $N(\lambda)$ is the number of eigenvalues of $\Delta$ less than $\lambda$ \cite{Wey1}. This observation suggests that the geometry of the underlying manifold is somehow connected with the spectrum of $\Delta$, and we will devote the upcoming chapters to exploring this relation. 

If we return to the Helmholtz equation, we can see an immediate connection to quantum mechanics by taking $k_n^{-1}=\hbar_n$, where $\hbar_n$ is an ``effective Planck's constant," so that we have
$$-\frac{\hbar_n^2}{2}\Delta \psi_n=\frac{1}{2}\psi_n,$$
or the time-independent Schr\"{o}dinger equation \index{Schr\"{o}dinger equation} for a non-relativistic particle of unit mass and total energy $1/2$. Thus the eigenvalue $k_n^2$ can be interpreted as the energy of a particle. We can then ask another question that relates $\Delta$ to the behavior of quantum systems: how are the eigenfunctions $\psi_n$ of $\Delta$ distributed in the high-energy limit? That is, if we arrange the spectrum $\text{Spec}(\Delta)$ in ascending order to get a sequence of nonnegative eigenvalues $\{k_n^2\}$, does the corresponding sequence of eigenfunctions $\{\psi_n\}$ ``fill up" our underlying manifold $M$ uniformly as $n$ tends towards infinity ($\hbar_n \to 0$)? Or do the eigenfunctions localize on some subset of $M$ and exhibit periodic, ``scarring" behavior? As an aside, we note that the condition $\hbar_n \to 0$ reflects the semiclassical limit of quantum mechanics because our effective Planck's constant $\hbar_n$ reflects the degree of energy quantization in a physical system. 

The question of eigenfunction distribution is what \emph{quantum ergodicity} (QE) \index{quantum ergodicity} is concerned with. If we maintain that our eigenfunctions $\{\psi_n\}$ are $L^2$-normalized so that they have a natural interpretation as wavefunctions, then equidistribution in the limit $\hbar_n \to 0$ would suggest that $|\psi_n|^2d\nu$ as a probability measure converges to the uniform measure. This is fundamentally what quantum ergodicity states. On the other hand, \emph{quantum unique ergodicity} \index{quantum ergodicity!quantum unique ergodicity} (QUE) asserts that the induced measures converge \emph{uniquely} to the uniform measure. If in the semiclassical limit $\hbar_n \to 0$ a system exhibits quantum ergodicity, then there is only a small proportion of \emph{exceptional} wavefunctions---eigenfunctions that are scarring or periodic---so that almost all the eigenfunctions and their linear combinations are equidistributed. If a system exhibits quantum unique ergodicity, then there cannot exist any sequence of exceptional eigenfunctions that do not converge to the uniform measure in the semiclassical limit. 

It turns out that if a classical system is ergodic, then the corresponding quantum system is quantum ergodic. This result, known as the quantum ergodicity theorem, \index{quantum ergodicity!quantum ergodicity theorem} was proven by Schnirelman (1974), Zelditch (1987), and Colin de Verdi\`{e}re (1985) for manifolds without boundary and in subsequent works for manifolds with boundary (in particular, G\'{e}rard-Leichtman in 1993 and Zelditch-Zworski in 1996) \cite{Shn1,Zel3,Col1,Ger1,Zel2}. The analogous statement for QUE, however, is demonstrably not true. For example, Hassell proved in 2010 that QUE does not hold for almost all \emph{Bunimovich stadiums}, \index{billiards!Bunimovich stadium} two-dimensional domains composed of rectangles of arbitrary lengths capped by two semicircles \cite{Has1}. Although Bunimovich had demonstrated QE for his stadiums, the failure of QUE had previously been suggested in an earlier study by Heller (1984), where he phenomenologically observed that certain eigenfunctions localize along unstable geodesics in some Bunimovich stadiums (a phenomenon called ``strong scarring") \cite{Bun1,Hel1}. 

There could plausibly be certain cases in which QUE is true. This is rather unintuitive, as in the classical case unique ergodicity is a very strong condition: one periodic classical orbit is enough to make a system fail to be classically uniquely ergodic. Due to the linear superposition of eigenfunctions, however, quantum mechanics is not quite as sensitive to individual orbits, and it is only if an orbit remains stable that a quantum system concentrates around it. In the case that the underlying manifold has negative curvature and exhibits certain arithmetic symmetries, we can actually infer more about the localization of eigenfunctions. We may, for instance, consider \emph{arithmetic surfaces}, which are quotients of a hyperbolic space by a congruence subgroup. For these manifolds, it turns out that there exists an algebra of \emph{Hecke operators} \index{Hecke operator} which commute with the Laplacian, so that examining the orthonormal eigenfunctions of Hecke operators tells us information about the eigenfunctions of $\Delta$. In 1994, Rudnick and Sarnak showed that there can be no strong scarring on certain arithmetic congruence surfaces \cite{Rud1}. Along with numerical computations that confirmed the plausibility of QUE on negatively curved Riemannian manifolds, Rudnick and Sarnak proposed the now-famous \emph{quantum unique ergodicity conjecture}, which roughly states: 
\bb
\textbf{Conjecture.} \emph{If $(M,g)$ is a Riemannian manifold \index{Riemannian manifold} with negative curvature or ergodic geodesic flow, \index{ergodicity} 
then the only quantum limit measure for any orthonormal basis of eigenfunctions of $\Delta$ is the uniform Lebesgue measure.} \index{Liouville measure} \index{quantum ergodicity!quantum unique ergodicity conjecture}
\bb
This conjecture, in addition to implying the absence of strong scars, claims that there is only one measure to which the eigenfunction-induced Wigner measures converge. A positive resolution of this conjecture would show that in the semiclassical limit, the quantum mechanics of strongly chaotic systems does not reflect the finest small-scale classical behavior. 

Although the conjecture has been outstanding for almost twenty years, several advances have recently been made. Aside from the aforementioned result for Bunimovich stadiums, there have been several contributions not only in showing that certain measures can never be quantum limits, but also in proving the QUE conjecture outright in the arithmetic case \cite{Ana1,Lin1}. Lindenstrauss was notably awarded the Fields Medal in 2010 for his work leading to a proof of QUE for arithmetic manifolds, a proof which was completed in 2009 for the modular surface $\SL_2\mathbb{Z}/\mathbb{H}$ by Soundararajan \cite{Sou1}. 

Looking forward, there is much to be done in regard to proving the full conjecture. Because of the assortment of techniques that QUE research involves, the subject is relevant to many areas, including number theory, geometry, and analysis. 
\bb
\textbf{Our focus.} This thesis begins by rigorously introducing the Laplace-Beltrami operator, a second-order linear differential operator that acts on a dense subset of $L^2$ functions. 
We then introduce many fundamentals of spectral and semiclassical analysis, 
including the Fourier transform, symbol quantization, pseudodifferential operators, and Weyl's law. 
With a background in semiclassical analysis in hand, we rigorously formulate the foregoing ideas from quantum chaos and prove the quantum ergodicity theorem. 
We conclude the exposition with a survey of the work done in quantum unique ergodicity, and in particular we note  
Hassell's disproof of quantum unique ergodicity on Bunimovich stadiums.

\emph{We emphasize geometric intuition over straightforward proofs.} This will be illustrated by certain key themes that recur throughout our exposition: for example, \emph{reparameterizing} with a small constant allows us to modify familiar definitions and obtain their semiclassical counterparts, and relating symbols to their pseudodifferential operators gives us the ability to alternate between classical and quantum mechanics. 
Although these themes are introduced in Section 1.3 and developed early on in our text, we will constantly illustrate how they create a unified framework for thinking about problems related to quantum ergodicity.

This thesis is accessible to any student with a first course in differential geometry who aims to understand the generalized Laplace-Beltrami operator and its associated questions, especially as they pertain to semiclassical analysis and quantum ergodicity. 
\bb
\textbf{Structure.} The current chapter formalizes the Laplace-Beltrami operator and the motivation behind quantum ergodicity. If the reader is not familiar with the field of quantum chaos, this exposition should be a sufficient introduction to its guiding principles. 

Chapter 2 introduces the theory of semiclassical analysis, which includes Weyl quantization and the symbol calculus. We define pseudodifferential operators with the objective of proving 
Weyl's law for the asymptotic behavior of eigenvalues of the Laplacian and Egorov's theorem for the correspondence principle. This chapter provides the basic notions needed to address quantum ergodicity and quantum unique ergodicity. 

Chapter 3 realizes our objective to rigorously introduce QE and QUE. 
With the previously developed formalism, we state the 
QE theorem and QUE conjecture and exhibit Schnirelman, Zelditch, and Colin de Verdi\`{e}re's proof of the former. 

Chapter 4 concludes with a survey of recent results in quantum unique ergodicity: first, we note Hassell's proof that QUE fails on Bunimovich stadiums, and second, we briefly discuss current research areas in QUE ranging from 
Barnett's numerical computation of billiard eigenfunctions to spectral statistics. 
The concepts and tools developed in Chapters 2 and 3 are indispensable for these latter accounts.


\subsection{The Laplace-Beltrami Operator}

Recall the notion of a \emph{spectrum}:\index{eigenvalue!spectrum} if $T:V\to V$ is a symmetric, nonnegative linear transformation of inner product spaces (taken to be compact if $V$ is infinite-dimensional), then there exists an orthonormal basis $\{e_1,...,e_n\}$ of eigenvectors of $V$ with eigenvalues $0\leq \lambda_1\leq ...\leq \lambda_n$; the set of eigenvalues $\{\lambda_i\}$ is called the \emph{spectrum of\hspace{2pt} $T$}, and is denoted by $\text{Spec}(T)$. From linear algebra, we remind ourselves that $\lambda \notin \text{Spec}(T) \Longleftrightarrow (T-\lambda I)^{-1}$ exists $\Longleftrightarrow \ker(T-\lambda I)=0$, and that each eigenvalue has finite multiplicity and accumulate only at $0$ if $T$ is compact. 
This definition can be extended to the case of a symmetric, unbounded operator $D$ acting on an infinite-dimensional Hilbert space $H$ (c.f. Appendix I). For the purposes of this thesis, we say that $\text{Spec}(D)$ is the set of all $\lambda$ such that $D-\lambda I$ has a kernel. 

The spectrum of the unbounded Laplace-Beltrami operator (Laplacian) $\Delta$ lies at the center of our exposition. It will, however, be useful to remind ourselves of several facts before defining the Laplacian. We begin with the ideal space of functions that the Laplacian acts on: \index{Laplace-Beltrami operator}
\bb
\textbf{Definition \cc} (\emph{$L^2$ space of functions}) \index{$L^2$ space} A function $f$ defined on a measure space (or Riemannian manifold) $(M,\mu)$ is of class $L^2$ if $f:M\to \mathbb{C}$ is square-integrable, i.e.
$$\int_M |f|^2d\mu<\infty.$$
We then write $f\in L^2(M)$, but note that elements of $L^2(M)$ are actually equivalence classes of functions that differ on a set of measure zero. 
$L^2(M)$ is a Hilbert space with inner product $\langle f,g\rangle=\int_M f\overline{g}d\mu$ 
and norm $||f||=\langle f,f\rangle^{1/2}=(\int_M |f|^2 d\mu)^{1/2}$.
\bb 
Introducing some geometry now allows us to define $\Delta$ for any Riemannian manifold. \index{Riemannian manifold} We recall that a smooth manifold $M$ is said to be \emph{Riemannian} if it is endowed with a {metric} $g$, a family of smoothly varying, positive-definite inner products $g_x$ on $T_xM$ for all points $x\in M$. Two Riemannian manifolds $(M,g)$ and $(N,h)$ are \emph{isometric} if there exists a smooth diffeomorphism $f:M\to N$ that respects the metric; in particular, $g_x(X,Y)=h_{f(x)}(d_xf(X),d_xf(Y))$ for all $X,Y\in T_xM$ and $x\in M$. The metric $g_x$ is a bilinear form on $T_xM$, so it is an element of $T^*_xM\otimes T_x^*M$. Since $g_x$ must be smoothly varying, $g$ is a smooth section of the tensor bundle $T^*M\otimes T^*M$ and a positive-definite symmetric $(0,2)$-tensor. A local construction and partition of unity argument shows that any manifold is metrizable. 
 Any Riemannian manifold is naturally a measure space, with distance function
 \vspace{-5pt}
$$d(x,y)=\inf \left\{ \int_a^b||\dot{\gamma}(t)||_{g(\gamma(t))}dt \text{ s.t. }\gamma:[a,b]\to M\text{ is a }C^1\text{ curve joining }x,y\in M\right\}.$$

It will also be helpful to introduce \emph{local coordinates} so that we can compute with our metric $g$. First we take a point $x\in M$. If $(x^1,...,x^n)$ is a local coordinate chart in the neighborhood of $x$ for all $v,w\in T_xM$, then there exist scalars $\alpha_i,\beta_i$ such that \index{Riemannian manifold!local coordinates}
$$v=\alpha_i \frac{\partial}{\partial x^i}\ \ \ \ \ \text{ and }\ \ \ \ \ w=\beta_i\frac{\partial}{\partial x^i}$$ 
in Einstein notation, so that $g_x(v,w)=\alpha_i\beta_j g_x(\partial_{i},\partial_{j})$ for $\partial_i=\partial_{x^i}:=\frac{\partial}{\partial x^i}$. The metric $g_x$ is then determined by the symmetric positive-definite matrix $(g_{ij}(x)):=(g_x(\partial_i,\partial_j))$. Moreover, by unrestricting ourselves from $x$, we can express $g$ in terms of the dual basis $\{dx^i\}$ of the cotangent bundle as
$g=\sum_{i,j}g_{ij}dx^i\otimes dx^j.$
The construction above is stable under coordinate transformation: if $(h_{ij}(y))=(g(\partial_{y^i},\partial_{y^j}))$ is the matrix of $g$ in another coordinate chart with coordinates $(y^1,...,y^n)$, then on the overlap of the coordinate charts we have $g_{ij}=\sum_{k,l} \frac{\partial y^k}{\partial x^i}\frac{\partial y^l}{\partial x^j}h_{kl}$. 
\bb
\textbf{Example \cc} (\emph{Riemannian metrics}) Recall that, for $M=\mathbb{R}^n$ and the natural identification $T_x\mathbb{R}^n=\mathbb{R}^n$, we can identify the standard normal basis $\{e_i\}$ with $\{\partial_i\}$ so that  $g_x(\alpha_i\partial_i,\beta_j\partial_j)=\sum_i \alpha_i\beta_i$. This defines the \emph{Euclidean metric} with metric tensor given by the identity, e.g. $(g_{ij})=(\langle e_i,e_j\rangle)=(\delta_{ij})$. There are also natural metrics for submanifolds, products, and coverings of Riemannian manifolds: for example, if $i:N\to M$ is an immersion of a submanifold $N\subset M$, then the induced metric on $N$ is $g_N=i^*g_M$, where $g_M$ is the metric on $M$. If $(M,g_1)$ and $(N,g_2)$ are Riemannian manifolds, then their product $M\times N$ exhibits the metric $g:=g_1\oplus g_2$. If $(M,g)$ is a Riemannian manifold and $\pi:\widetilde{M}\to M$ is a covering map, then $\widetilde{g}:=\pi^*g$ is a metric on $\widetilde{M}$ that is preserved by covering transformations.
\bb
Now let $(M,g)$ be a Riemannian manifold, with covariant metric tensor $(g_{ij})$ and contravariant metric tensor $(g^{ij}):=(g_{ij})^{-1}$. For $M=\mathbb{R}^n$, it is clear that the Laplacian can be defined as $\Delta =\div\ \grad=\partial_{1}^2+...+\partial_{n}^2$ where $\grad:C^\infty(M)\to TM$ and $\div:TM\to C^\infty(M)$. $\Delta$ is then a second-order differential operator that takes functions to functions. We can extend this construction to any Riemannian manifold using the \emph{musical isomorphisms}, which for coordinates $\{\partial_i\}$ of $TM$, $\{dx^i\}$ of $T^*M$, and metric tensor $g=g_{ij}dx^i\otimes dx^j$, are defined by \index{Riemannian manifold!metric tensor} \index{musical isomorphism}
$$\flat:\Gamma^\infty(TM)\to \Gamma^\infty(T^*M),\ \ \  \flat(X)=g_{ij}X^idx^j=X_jdx^j$$ for $X=X^i\partial_i$ and 
$$\sharp:=\flat^{-1}:\Gamma^\infty(T^*M)\to \Gamma^\infty(
TM
)
,\ \ \  \sharp(\omega)=g^{ij}\omega_i\partial_j$$ 
for $\omega=\omega_idx^i$. For $X=X^i\partial_i$ and $Y=\partial_j$, $\flat$ satisfies the relation $\flat(X)(Y)=g(X,Y)$, since $\flat(X)(\partial_j)=g(X,\partial_j)=g_{ij}X^i\Longleftrightarrow \flat(X^i\partial_i)=g_{ij}X^idx^j.$ Likewise, for a covector field $\omega=\omega_idx^i$, $\sharp$ satisfies the relation $g(\sharp(\omega),Y)=\omega(Y)$. 

If $f$ is a smooth function on $(M,g)$, then the gradient of $f$, $\nabla f=\grad f$, is the vector field $\sharp(df)$ in which the $\sharp$ operator raises an index from the one-form $df$; this is because $\nabla$ is defined by the relation $g(\nabla f,X)=df(X)\Longleftrightarrow \nabla f=\sharp(df)$ for all $X\in TM$. In other words, $\nabla f$ is the vector field associated to the one-form $df$ via the $\sharp$ operator. $\nabla$ can therefore be written in local coordinates on any Riemannian manifold as $\nabla f=g^{ij}\partial_i f\partial_j$. 

We can define the divergence similarly. If $\dim M=n$ and the volume form (or element if $M$ is not oriented) on $M$ is given by $\mu=\sqrt{|g|}dx^1\wedge...\wedge dx^n$ where $|g|:=\det(g_{ij})$, then the divergence operator satisfies the relation $\div(X)\mu=d(\iota_X\mu)$, where $\iota_X\mu$ is the $n-1$ form coming from the contraction of $X$ with $\mu$. If $X=X^i\partial_i$ locally, then 
$$\div(X)\mu=d(\iota_{X^i\partial_i}\mu)=d((-1)^{i-1}X^i\sqrt{|g|}dx^1\wedge...\wedge \widehat{dx^i}\wedge...\wedge dx^n)=\partial_i(X^i\sqrt{|g|})dx^1\wedge...\wedge dx^n,$$
and as a result $\div$ can be written in local coordinates as 
$\div(X)=(\sqrt{|g|})^{-1}\partial_i(X^i\sqrt{|g|}).$\index{Laplace-Beltrami operator!in local coordinates}
The foregoing details allow us to define the Laplacian for functions on any Riemannian manifold:
\bb
\textbf{Definition \cc} (\emph{Laplace-Beltrami operator}) The positive \emph{Laplacian} is the second-order linear, elliptic differential operator defined on any Riemannian manifold $(M,g)$ as
$$\Delta:L^2(M,g)\to L^2(M,g), \ \ \ \ \ \Delta: f\mapsto \text{div }\text{grad }f=\frac{1}{\sqrt{|g|}}\partial_i(\sqrt{|g|}g^{ij}\partial_j f).$$
From this expression in local coordinates, we see that $\Delta$ and $g$ determine each other on any  manifold. This means that every Riemannian manifold has a Laplacian, and $\Delta$ determines the contravariant metric tensor $(g^{ij})$ when evaluated on a function that is locally $x^ix^j$. 
\bb
\emph{Remark.} Since $\Delta$ is an unbounded, closed graph operator, we must actually define it on a dense subset of $L^2(M)$. We relegate these functional-analytic technicalities to Appendix I.
\bb
Before proceeding, we verify that $\Delta$ is well-defined. To see this, it suffices to show that $\nabla$ and $\div$ are well-defined under change of coordinates. In the case of $\div$, we let $(y^1,...,y^n)$ be another set of coordinates on some neighborhood of $M$, so that $X=X^i\partial_{x^i}=Y^j\partial_{y^j}$. Then
$$\div\ X=\frac{1}{\sqrt{|g|}}\partial_{x^i}(X^i\sqrt{|g|})=\frac{1}{\sqrt{|g|}}\partial_{y^j}(Y^j\sqrt{|g|}),$$
and the result for $\nabla$ can be checked analogously. We observe that in the case $M=\mathbb{R}^n$, $(g_{ij})=(\delta_{ij})\Longrightarrow |g|=1$ for the standard orthonormal coordinates $x^i$, so that
$\Delta f=(\sqrt{|g|})^{-1}\partial_i\sqrt{|g|}\partial_if=\partial_i^2f$
and we recover the original Euclidean Laplacian.  
\bb
There are several important properties of $\Delta$ that make it a nice example of a second-order elliptic differential operator. For example, Green's second identity $\int_\Omega(f\Delta g-g\Delta f)dx=0$ for $\Omega\subset M$ and $\partial \Omega=\emptyset$ 
implies that $\Delta$ is essentially self-adjoint on $L^2(M)$:
\bb
\textbf{Proposition \cc} (\emph{Green's second identity}) If $f$ and $g$ are smooth functions on $M$, $\partial M=\emptyset$, and both are compactly supported, then
$\int_M f\Delta g\mu=-\int_M \langle \nabla f,\nabla g\rangle\mu=\int_Mg\Delta f\mu.$

\emph{Proof.} 
$\div(fX)=f\div(X)+\langle \nabla f,X\rangle$, so $\div(f\nabla g)=f\Delta g+\langle\nabla f,\nabla g\rangle.$
The result follows from the divergence theorem since $f\nabla g$ is compactly supported. $\hfill \blacksquare$
\bb
Thus, we have $\int_M \Delta f\mu=0$ for $f\in C^\infty_c(M)$, the space of smooth, compactly supported functions on $M$, and $\langle \Delta f,g\rangle=\langle f,\Delta g\rangle$. 
Similarly, Green's first identity $\langle \Delta f,f\rangle=-||\nabla f||^2$, along with Friedrich's inequality $||f||^2\leq c||\nabla f||^2$ for $c>0$, shows that $\Delta$ is negative-definite. 

These properties of $\Delta$ relate to its spectrum, which consists of all eigenvalues $\lambda \in \mathbb{R}$ for which there is a corresponding nonzero function $f\in L^2(M)$ with $-\Delta f=\lambda f$.
(Note that we have included a negative sign in the eigenvalue equation  $-\Delta f=\lambda f$ to make $\Delta$ a positive-definite operator, so that all nontrivial eigenvalues $\lambda$ are also positive.) 
If $M$ is compact, then the spectrum $\text{Spec}(\Delta)$ is unbounded with finite multiplicity and no accumulation points;
 the eigenvalues of $\Delta$
 can therefore be arranged into a discrete sequence $0=\lambda_0\leq \lambda_1\leq ...\to \infty$. 
The following theorem summarizes these results, in addition to telling us when the eigenfunctions of $\Delta$ form an orthonormal basis of $L^2(M)$ (c.f. Appendix I and \cite{Jos1}):
\bb
\textbf{Theorem \cc} (\emph{spectral theorem and eigenfunction basis of $\Delta$}) \index{spectral theorem}\index{eigenvalue}\index{eigenfunction}
Let $(M,g)$ be a compact Riemannian manifold. Then the eigenvalue problem \index{eigenvalue!eigenvalue equation}
$$\Delta f+\lambda f=0, \ \ \ \ \ f\in L^2(M)$$
has countably many eigenvalue-eigenfunction pair solutions $(\lambda,f)=(\lambda_n,f_n)$ for which $\langle f_n,f_m\rangle=\delta_n^m$ and $\lambda_n\delta_n^m=\langle \Delta f_n,f_m\rangle=-\langle \nabla f_n,\nabla f_m\rangle=\lambda_m\delta_n^m$. If $\partial M=\emptyset$, then the eigenvalue $\lambda_0=0$ is attained only when its eigenfunction is a constant; otherwise all eigenvalues are positive, and $\lim_{n\to \infty}\lambda_n=\infty$. We also have
$$g=\sum_{i=0}^\infty \langle g,f_i\rangle f_i\ \ \ \text{ and } \ \ \ \langle \Delta h,h\rangle=-\langle \nabla h,\nabla h\rangle=\sum_{i=1}^\infty \lambda_i\langle h,f_i\rangle^2$$
for all $g\in L^2(M)$ and $h\in C^\infty_c(M)$. 
\bb
If $M$ is a manifold with boundary, we further assume either \emph{Dirichlet} ($f=0$ on $\partial M$) or \emph{Neumann boundary conditions} ($\partial_{\vec v}f=0$, where $\vec v$ is normal to $\partial M$). The restriction of $\Delta$ to the space of functions $\{f\in L^2(M):f|_{\partial M}=0\}$ will be called the \emph{Dirichlet Laplacian} $\Delta_D$, and the restriction of $\Delta$ to $\{f\in L^2(M):\partial_{\vec{v}}f=0\text{ for }\vec{v}\text{ normal to }\partial M\}$ the \emph{Neumann Laplacian} $\Delta_N$. 
These Laplacians can be defined on $\mathbb{R}^n$ by the \emph{Friedrichs extension procedure}, but we will not deal with such functional-analytic considerations in this thesis.  \index{Laplace-Beltrami operator!Dirichlet Laplacian}\index{Laplace-Beltrami operator!Neumann Laplacian}
\bb
\textbf{Example \cc} (\emph{Laplacian on $S^1$}) We note that the simplest differential operator on the circle is $d/d\theta$, but its spectrum is empty since it identifies with the exterior derivative. Instead, we consider $\Delta:={d^2}/{d\theta^2}$. Since $-\Delta e^{in\theta}=n^2e^{in\theta}$ and $n^2$ for $n\in \mathbb{Z}$ are the eigenvalues of $-\Delta$, an orthonormal basis of eigenfunctions for $L^2(S^1)$ is $\{e^{in\theta}\}$ for $n\in \mathbb{Z}$. Clearly, $0$ occurs with multiplicity $1$ and all other eigenvalues occur with multiplicity 2. The eigenfun-
\begin{figure}[!htb]
  \begin{subfigure}[b]{0.49\textwidth}\centering
    \includegraphics[height=3.5cm]{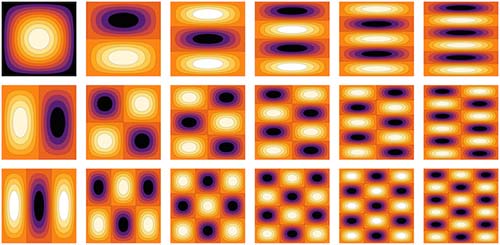}
    \captionsetup{font=footnotesize}
          \caption{$f_{j,k}(x,y)=\sin\left(\frac{j\pi}{a}x\right)\sin\left( \frac{k\pi}{b}y\right)$}
          \label{subfig-1:dummy}
      \end{subfigure}
      \begin{subfigure}[b]{0.49\textwidth}\centering
    \includegraphics[height=3.5cm]{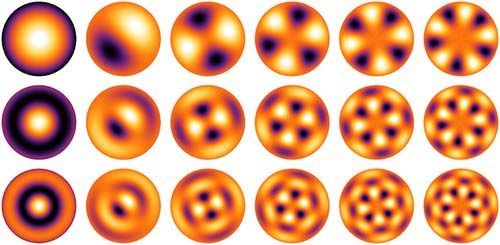}
    \captionsetup{font=footnotesize}
        \caption{$f^\lambda_{k}(r,\theta)=(\cos(k\theta)+\sin(k\theta))J_k(\sqrt{\lambda}r)$}
          \label{subfig-2:dummy}
      \end{subfigure}
      \vspace{-15pt}
      \quad\renewcommand{\thefigure}{\arabic{section}.\arabic{figure}\ $|$\ }
    \captionsetup{format=hang,position=bottom,labelsep=none,font=footnotesize}
      \caption{Contour plots of Dirichlet Laplacian eigenfunctions on a square $[0,1]^2\subset \mathbb{R}^2$ for $j\in \{1,2,3\}$, $k\in\{1,...,6\}$ and the unit disk $\{x\in \mathbb{R}^2:|x|\leq 1\}$ for $k\in \{0,...,5\}$ and $\sqrt{\lambda}$ being the first, second, and third zeros of $J_k$. Lighter colors denote higher values.}
\end{figure}
$\null$\vspace{-10pt}\\
\vspace{-5pt}\\
ction decomposition of $f\in L^2(S^1)$ is given by the usual Fourier decomposition:
$$f=\sum_n a_ne^{in\theta}=\sum_n \langle f,e^{in\theta}\rangle e^{in\theta}, \text{ where }\langle f,g\rangle := \frac{1}{2\pi}\int_{S^1}f(\theta)\overline{g(\theta)}d\theta.$$ \vspace{-5pt}\\
\textbf{Example \cc} (\emph{Laplacian on a rectangular domain}) 
On a rectangle $R=[0,a]\times [0,b]\subset \mathbb{R}^2$, a straightforward calculation using separation of variables shows us that, with Dirichlet boundary conditions, the eigenfunctions of $-\Delta$ are given by
$$f_{j,k}(x,y)=\sin\left(\frac{j\pi}{a}x\right)\sin\left( \frac{k\pi}{b}y\right)$$
for $j,k\geq 1$, with eigenvalues $\lambda_{j,k}=(j\pi/a)^2+(k\pi/b)^2$. With Neumann boundary conditions, the eigenfunctions are
$$f_{j,k}(x,y)=\cos\left(\frac{j\pi}{a}x\right)\cos\left(\frac{k\pi}{b}y\right)$$
for $j,k\geq 0$, with eigenvalues $\lambda_{j,k}=(j\pi/a)^2+(k\pi/b)^2$. 
\bb
\textbf{Example \cc} (\emph{Laplacian on $\mathbb{D}$})
On the unit disc $\mathbb{D}=\{x\in \mathbb{R}^2:|x|\leq 1\}$, we have $-\Delta=\partial_r^2+\frac{1}{r}\partial_r+\frac{1}{r^2}\partial_\theta^2$ in polar coordinates. Separating $f(r,\theta)=g(r)\phi(\theta)$, we see that 
$$-\Delta f=\lambda f\Longrightarrow g''(r)\phi(\theta)+\frac{1}{r}g'(r)\phi(\theta)+\frac{1}{r^2}g(r)\phi''(\theta)=\lambda g(r)\phi(\theta).$$
This can ultimately be written in the form of Bessel's equation
$x^2J''(x)+xJ'(x)+(x^2-k^2)J(x)=0$, for which the solutions are given by the Bessel functions
$$J_k(x)=\sum_{i=0}^\infty \frac{(-1)^i}{(i!)\Gamma(k+i+1)}\left(\frac{x}{2}\right)^{k+2i}.$$
One can verify that the general eigenfunctions of $-\Delta$ are of the form 
$f^\lambda_{k}(r,\theta)=\phi_k(\theta)J_k(\sqrt{\lambda}r),$
where $\phi_k(\theta)=a_k\cos(k\theta)+b_k\sin(k\theta)$ and the eigenvalue corresponding to $f^\lambda_{k}$ is $\lambda$. With Dirichlet boundary conditions, $\sqrt{\lambda}$ must be a zero of $J_k$; with Neumann boundary conditions, $\sqrt{\lambda}$ must be a zero of $J_k'$. Any $f\in L^2(R)$ can be decomposed into these eigenfunctions.
\bb
It is apparent that the eigenfunctions of $\Delta$ or $-\Delta$ are difficult to calculate for not-so-vanilla manifolds,  as we are limited only to separation of variables and Wentzel-Kramers-Brillouin (WKB) approximation methods from quantum mechanics. 
Oftentimes, however, the inverse problem of describing what geometric information we can get from the eigenpairs of $\Delta$ is more important. For instance, the semiclassical analysis we develop in Chapter 2 will be useful for proving \emph{Weyl's law} for the asymptotic distribution of
eigenvalues in the limit $\lambda\to \infty$. \index{Weyl's law}
We will revisit the eigenvalue problem in greater depth after we introduce Weyl quantization  and the associated symbol calculus.


\subsection{On Classical and Quantum Chaos}
\setcounter{itemcounter}{1}

Quantum chaos deals with the quantum mechanics of classically chaotic systems. This term is, however, a misnomer: quantum systems are usually much less sensitive to initial conditions than classically chaotic systems, so instead what we refer to as quantum chaos (or 
``quantum chaology") focuses on the semiclassical limit of systems whose classical counterparts are chaotic \cite{Ber2}. Qualitatively speaking, a classically chaotic Hamiltonian system is one whose orbits exhibit extreme sensitivity to perturbations in initial conditions. 

Current research in quantum chaos concentrates on roughly two areas: \emph{spectral statistics}, which compares the statistical properties of Laplacian eigenvalue (energy) distribution to the classical behavior of the Hamiltonian, and \emph{semiclassical analysis}, which relates the classical motion of a dynamical system to its quantum mechanics. We focus on semiclassical analysis, and  give only a brief survey of spectral statistics in Chapter 4. \index{quantum chaos}

\subsubsection{Billiard Flows and Hamiltonian Systems}

We first provide some intuition by introducing a model system known as the \emph{billiard flow}. A \emph{billiard} is a bounded, planar domain $\Omega\subset \mathbb{R}^2$ with a smooth (possibly piecewise) boundary. The billiard flow refers to the classical, frictionless motion of a particle in $\Omega$ in which its angle of incidence to the boundary $\partial \Omega$ is the same as its angle of reflection off $\partial \Omega$ (Figure 1.3). Clearly, the total kinetic energy is conserved by this motion. The classical trajectory of the particle depends largely on the shape of $\partial \Omega$, and can itself be very complicated.\index{billiards}
\bb
\textbf{Example \cc} (\emph{circular and Sinai billiard flows}) Consider a particle starting somewhere on the boundary of a circular billiard, with angular coordinate $\varphi_0$ and initial velocity vector at an angle $\alpha_0$ to the circle's tangent line (Figure 1.2a). Then the $n$th incident boundary point is $\varphi_n=(\varphi_0+2n\alpha_0)\text{ mod }2\pi$, and all the tangent angles are $\alpha_0$. Perturbations affect the circular billiard linearly: if $\alpha_0':=\alpha_0+\epsilon$, then the new incident points are $\varphi_n'=(\varphi_n+2\epsilon n)\text{ mod }2\pi$, and indeed the distance $|\varphi_n'-\varphi_n|$ only grows linearly with $n$. 

The flow on the \emph{Sinai billiard} (Figure 1.2b), however, is both unstable and generally difficult to calculate. With the same notation as above, the orbit described by the conditions $\varphi_n=\varphi_0=0,\alpha_n=\alpha_0=\frac{\pi}{2}$ is trivially periodic. If we perturb $\alpha_0$ by $\epsilon_0$, then for $\varphi_0'>\epsilon_0$ we have \index{billiards!Sinai stadium}
$\alpha_0'=\frac{\pi}{2}-(\epsilon_0+\varphi_0')$. This gives $\epsilon_1=\epsilon_0+2\varphi_0'>3\epsilon_0$, assuming that $\epsilon_0$ is small enough so that the particle collides with the circular boundary. Repeating this procedure yields $\epsilon_n>3\epsilon_{n-1}>...>3^n\epsilon_0$, so $\varphi_0'>3^n\epsilon_0$. 
Thus $\varphi_n'\geq 3^n\epsilon_0$ for all $n$ less than any $N\in \mathbb{N}$. 
The periodic orbit we started with ($\varphi_n=\varphi_0=0,\alpha_n=\alpha_0=\frac{\pi}{2}$) is  {exponentially unstable}, and indeed it turns out that all orbits have this property \cite{DeB1}.\pagebreak
\begin{figure}
  \begin{minipage}[b]{0.13\textwidth}
        \captionsetup{position=top,labelsep=none,font=footnotesize}
    \renewcommand{\thefigure}{\arabic{section}.\arabic{figure}\ $|$\vspace{2pt}\newline }
          \caption{Diagrams for calculating trajectories on the circular and Sinai billiards.}
          \label{figure}
      \vspace{68pt}
      
    \end{minipage}\hspace{7pt}
      \begin{subfigure}[b]{0.33\textwidth}\centering
    \includegraphics[height=3.5cm]{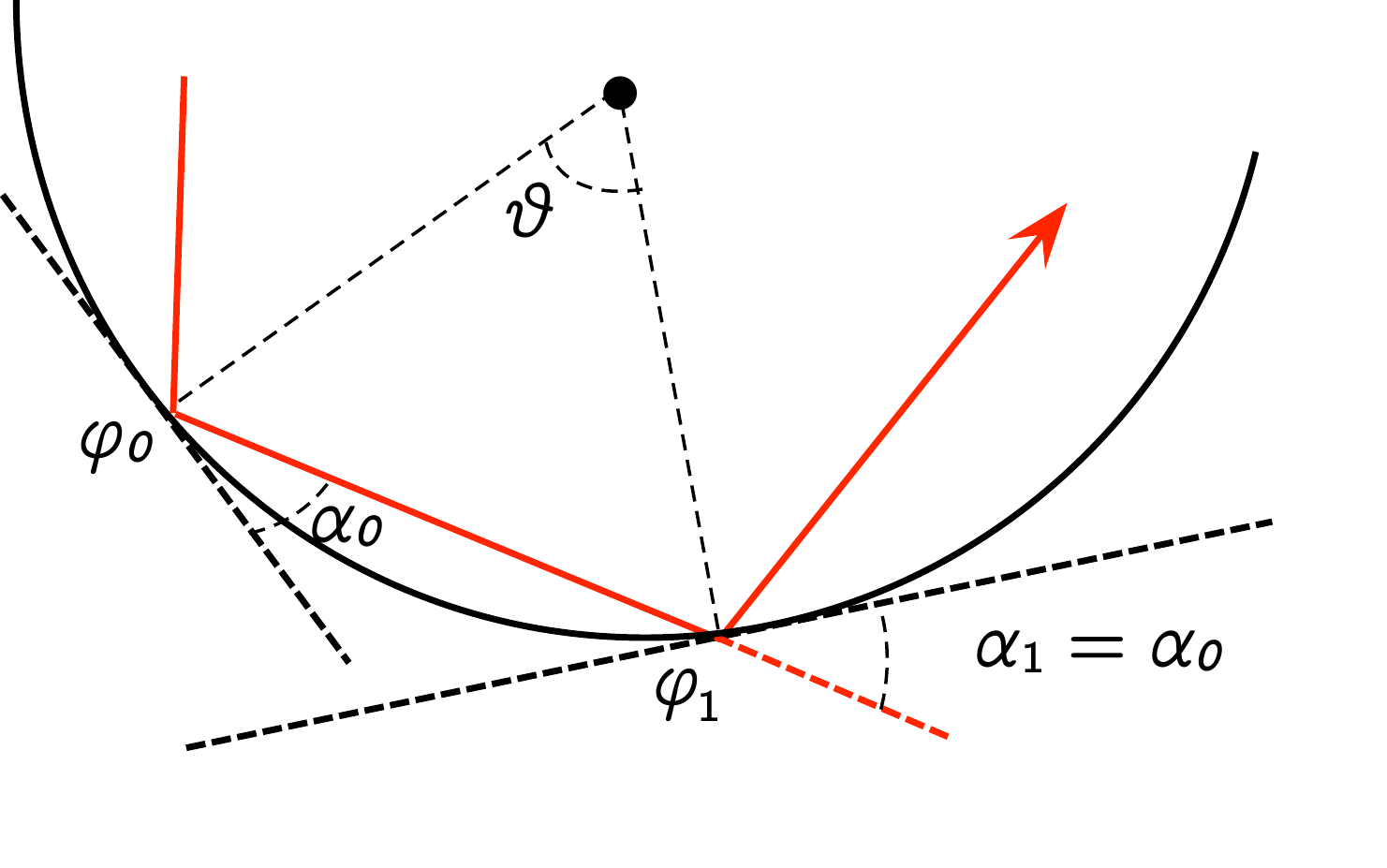}
    \captionsetup{font=footnotesize}\vspace{-20pt}
          \caption{$\varphi_1=\varphi_0+\vartheta$, where $\varphi_n$ is the angular coordinate after $n$ incidents and $\alpha_0$ is the initial incident angle. By plane geometry, $\vartheta=2\alpha_0$ and $\alpha_1=\alpha_0$.}
          \label{subfig-1}
      \end{subfigure}\hspace{7pt}
      \begin{subfigure}[b]{0.48\textwidth}\centering
    \includegraphics[height=3.7cm]{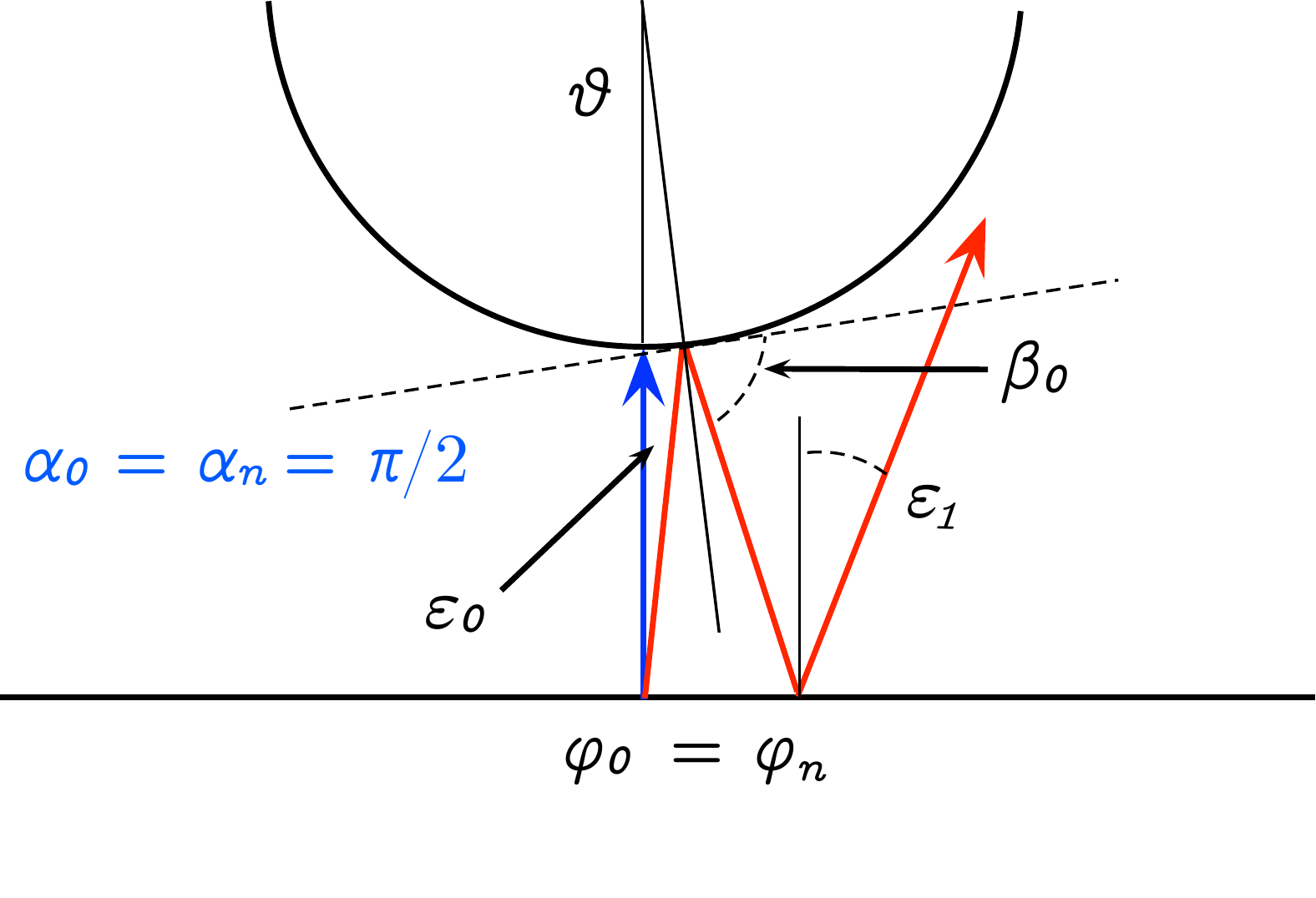}
    \captionsetup{font=footnotesize}
    \vspace{-10pt}
        \caption{The blue trajectory is trivially periodic, but exponentially sensitive to perturbations. By perturbing this trajectory by $\varepsilon_0$ and identifying the first incident point $\varphi_0'$ with the angular coordinate $\vartheta$, we have $(\pi/2)-\beta_0=\varepsilon_0+\vartheta$, for $\beta_0:=\alpha_0'$.}
          \label{subfig-2}
      \end{subfigure}
\vspace{15pt}

  \begin{subfigure}[b]{0.31\textwidth}\centering
    \includegraphics[height=3cm]{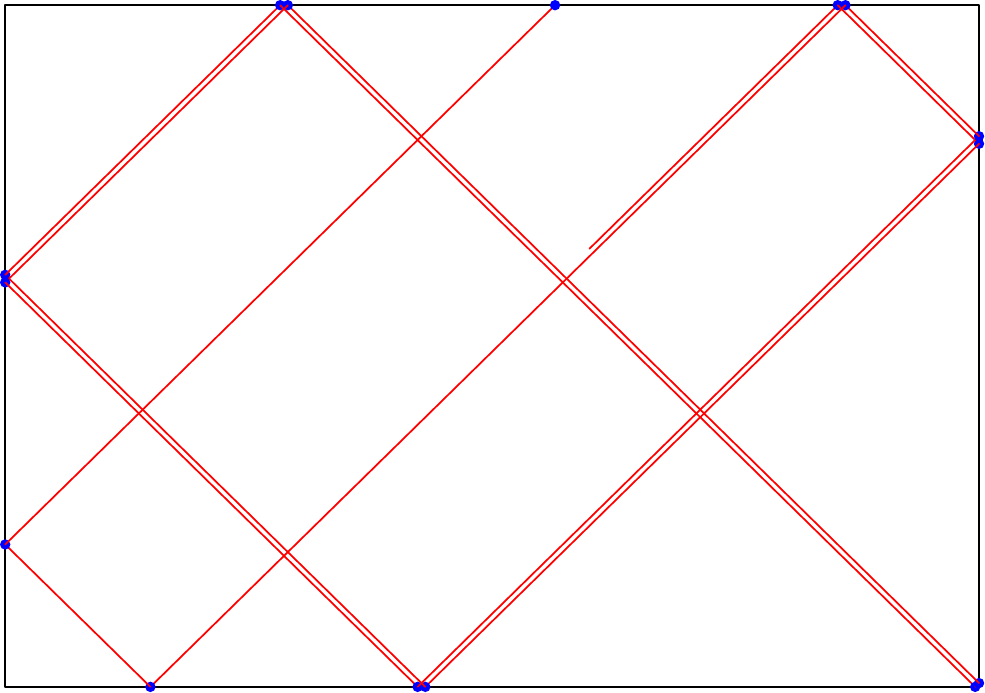}
    \setcounter{subfigure}{0}
    \captionsetup{font=footnotesize}
          \caption{Rectangular stadium}
          \label{subfig-1}
      \end{subfigure}\hspace{-5pt}
      \begin{subfigure}[b]{0.3\textwidth}\centering
    \includegraphics[height=3cm]{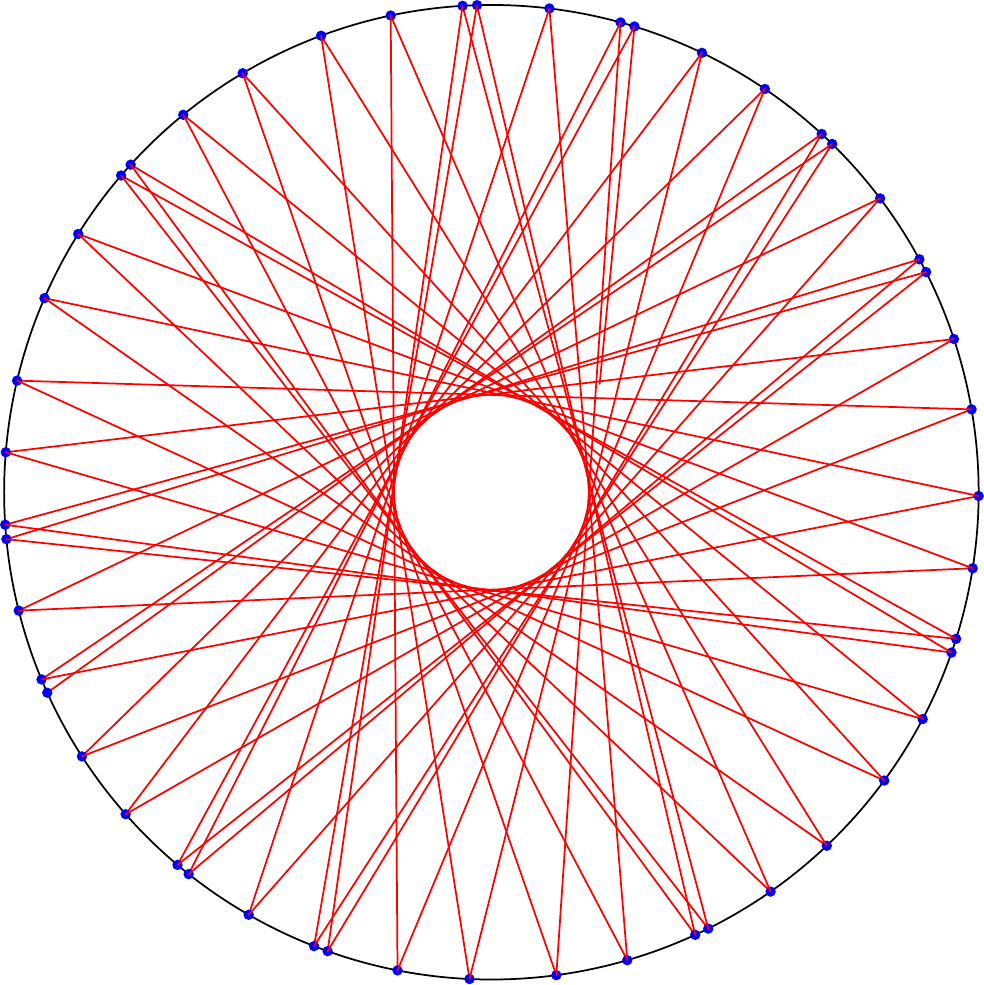}
    \captionsetup{font=footnotesize}
        \caption{Circular stadium}
          \label{subfig-2}
      \end{subfigure}\hspace{-5pt}
  \begin{subfigure}[b]{0.39\textwidth}\centering
    \includegraphics[height=3cm]{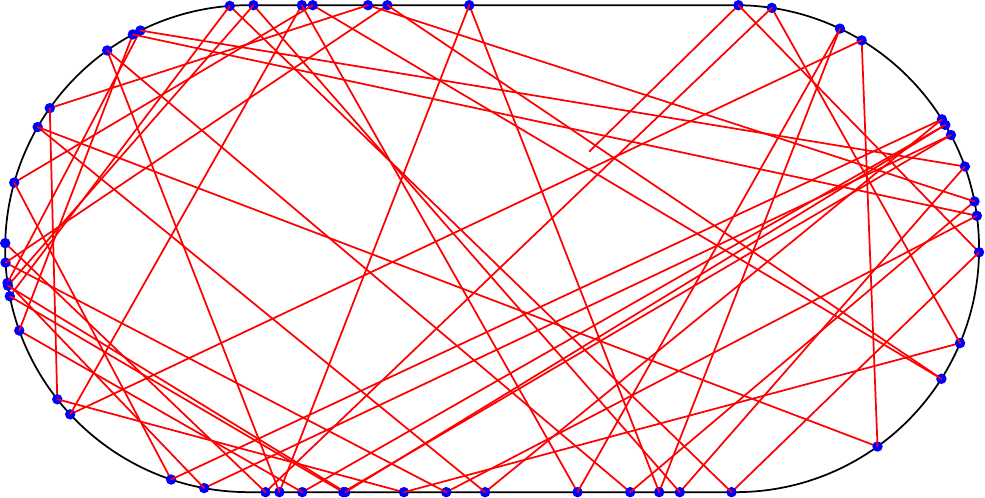}
    \captionsetup{font=footnotesize}
        \caption{Bunimovich stadium}
          \label{subfig-2}
      \end{subfigure}
  \vspace{15pt}

  \begin{minipage}[b]{0.23\textwidth}
        \captionsetup{position=top,labelsep=none,font=footnotesize}
    \renewcommand{\thefigure}{\arabic{section}.\arabic{figure}\ $|$\vspace{2pt}\newline }
          \caption{The billiard flow on different domains. Observe that certain stadiums exhibit stable, periodic trajectories, while the other stadiums are evenly filled with unstable, ``chaotic" trajectories.}
          \label{fig:dummy}\vspace{-5pt}
    \end{minipage}\hspace{5pt}
      \begin{subfigure}[b]{0.25\textwidth}\centering
    \includegraphics[height=3cm]{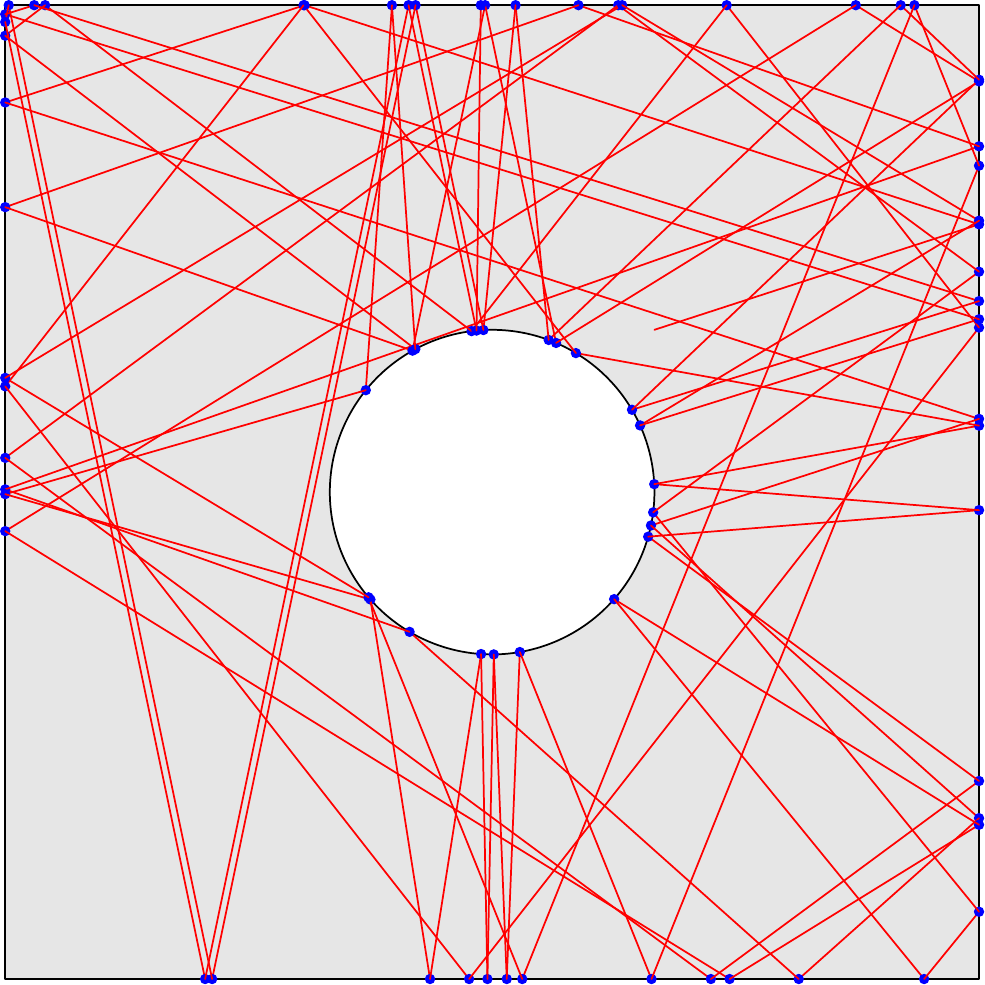}
    \setcounter{subfigure}{3}
    \captionsetup{font=footnotesize}
          \caption{Sinai stadium}
          \label{subfig-4}
      \end{subfigure}\hspace{-10pt}
      \begin{subfigure}[b]{0.25\textwidth}\centering
    \includegraphics[height=3cm]{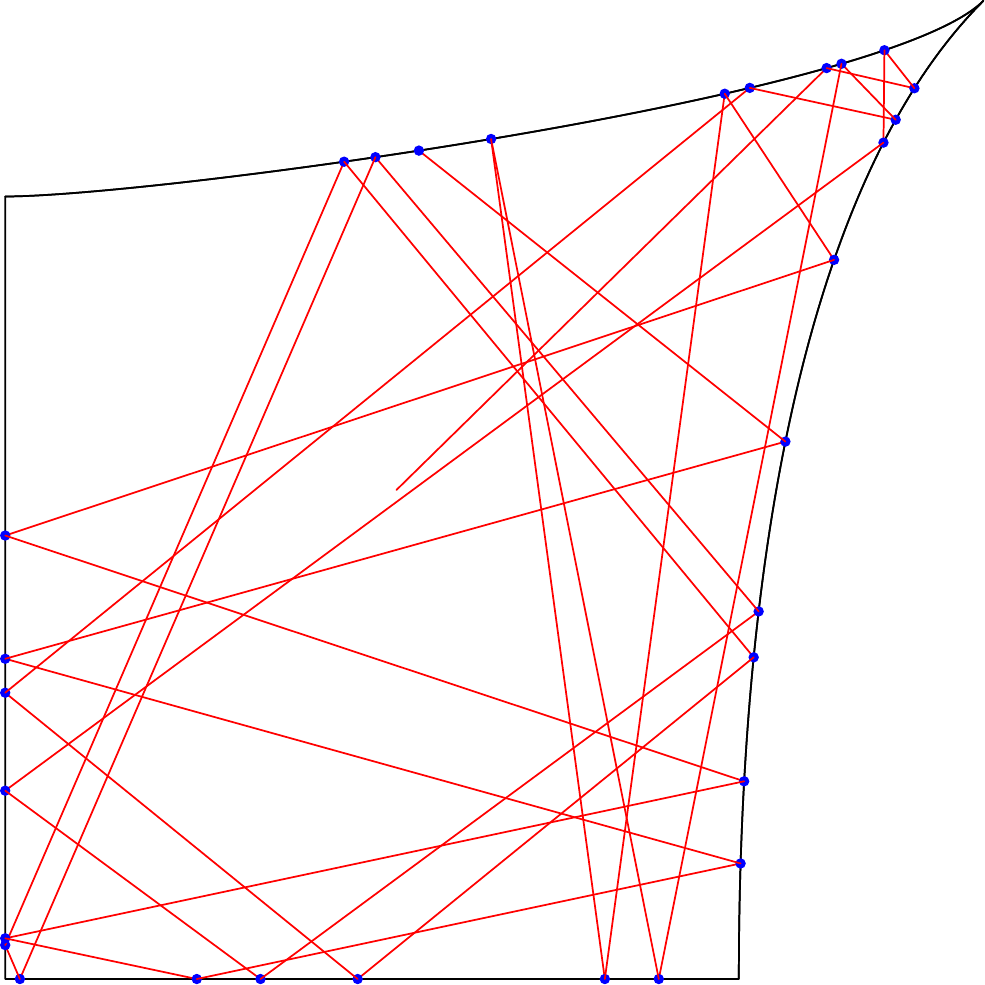}
    \captionsetup{font=footnotesize}
        \caption{Barnett stadium}\index{billiards!Barnett stadium}
          \label{subfig-5}
      \end{subfigure}\hspace{-0pt}
  \begin{subfigure}[b]{0.25\textwidth}\centering
    \includegraphics[height=3cm]{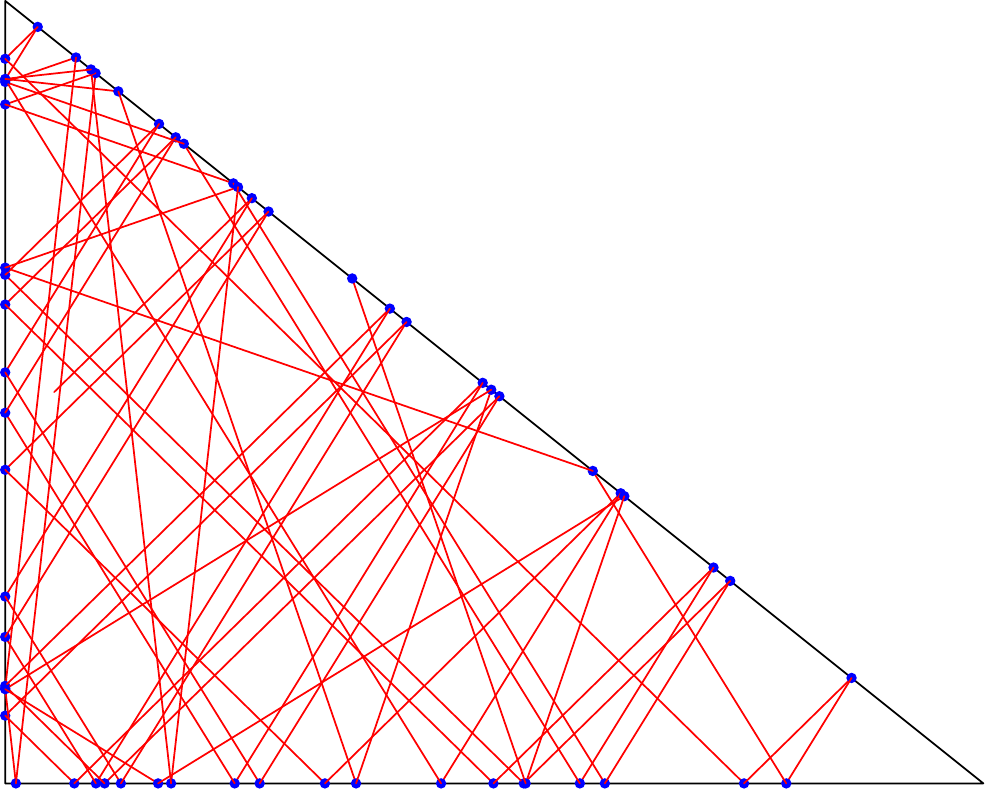}
    \captionsetup{font=footnotesize}
        \caption{Triangular stadium}
          \label{subfig-6}
      \end{subfigure}
\end{figure}
$\null$
\bb
In the quantum regime, billiards are described by wavefunctions $\psi_n$ whose time-evolution is given by the Schr\"{o}dinger equation
$$i\hbar \frac{\partial}{\partial t} \psi_n(x,t)=-\frac{\hbar^2}{2m}\Delta \psi_n(x,t),$$
where $\Delta=\Delta_D$ is the Dirichlet Laplacian, $\psi_n \in L^2(\Omega)$, $\hbar$ is Planck's constant, and $m$ is the mass of the particle. We know from quantum mechanics that the time-dependent solutions are of the form $\psi_n(x,t)=\exp(-{it E_n}/{\hbar})\phi_n(x)$, where 
the $E_n$ are quantum energy levels and $\phi_n$ satisfies the eigenvalue equation 
$-\frac{h^2}{2m}\Delta \phi_n=E_n\phi_n.$
By taking $\lambda_n:=2mE_n/\hbar^2$, we see that {the quantum-mechanical billiard flow corresponds to the eigenvalue problem $-\Delta \psi_n=\lambda_n\psi_n$.} Thus the probability distributions of a particle's position on the rectangular or circular billiards are reflected in the contour plots of Figure 1.1; furthermore, they vary depending on the eigenvalue---or energy---under consideration. The convergence between the classical billiard flow and the eigenvalue problem is an example of what semiclassical analysis deals with. 

Before discussing quantum ergodicity, we backtrack to formulate some of the mathematical notions behind classical and quantum mechanics more rigorously. We discuss classical mechanics and symplectic geometry in this section, and briefly recall certain aspects of quantum mechanics and its mathematical formulation as we progress in \S2. \index{symplectic geometry}

Recall that a \emph{symplectic manifold} is a smooth, even-dimensional manifold $M$ equipped with a closed, nondegenerate two-form $\omega$.  \index{Hamiltonian formulation of mechanics}
A \emph{Hamiltonian} is a smooth function $H:M\to \mathbb{R}$ or $\mathbb{C}$. If $(M,\omega)$ is a symplectic manifold, then there is a fiberwise isomorphism $\omega:TM\to T^*M$ by the nondegeneracy of $\omega$. This identifies vector fields on $M$ with one-forms on $M$, 
so every Hamiltonian $H$ determines \index{Hamiltonian formulation of mechanics!Hamiltonian}
a unique \emph{Hamiltonian vector field} $X_H$ on $M$ where the contraction $\iota_{X_H}(\omega)$ with $\omega$ is the one-form $dH$, 
i.e. $\omega(X_H,Y)= dH(Y)\Longleftrightarrow\iota_{X_H}\omega=dH\Longleftrightarrow\iota_X\omega$ is exact. Moreover, integrating the Hamiltonian vector field $X_H$ generates a one-parameter family of \emph{integral curves}, which represent solutions to the equations of motion.
These integral curves are diffeomorphisms $\Phi^t:M\to M$ which preserve the symplectic form in the sense that $(\Phi^t)^*\omega=\omega$ for all $t$, and 
$$\begin{cases}\Phi^0=\text{Id}_M,\\
\partial_t\Phi^t\circ (\Phi^t)^{-1}=X_H.
\end{cases}$$
We also remind ourselves that any Hamiltonian vector field $X_H$ preserves the Hamiltonian in the sense that $\mathcal{L}_{X_H}H=\iota_{X_H}dH=\iota_{X_H}\iota_{X_H}\omega=0$ ,where $\mathcal{L}_{X_H}$ denotes the Lie derivative along $X_H$ (i.e. the commutator $[X_H,\cdot]$). By analogy, we call a vector field preserving the symplectic form $\omega$ (e.g., $\mathcal{L}_X\omega=0 \Longleftrightarrow \iota_X\omega$ is closed) \emph{symplectic}. The first de Rham cohomology group $H^1(M)$ measures when $\iota_{X_H}$ is closed and exact, or when a symplectic vector field is Hamiltonian. If $H^1(M)=0$, then these two notions coincide globally; else they only coincide locally on contractible open sets. \index{Hamiltonian formulation of mechanics!Hamiltonian flow}

This Hamiltonian formalism allows us to describe classical systems in terms of Hamiltonian flows. In particular, we take a Riemannian manifold $M$ as the configuration space of a classical system and the cotangent bundle $T^*M$ as its phase space.\index{Hamiltonian formulation of mechanics!phase space} \index{Hamiltonian formulation of mechanics!configuration space}
By convention, we set $T^*M=\mathbb{R}^{2n}$ as our symplectic manifold with the canonical symplectic structure $\omega=\sum_{j}dx_j\wedge dp_j$, where $x=(x_j)$ and $p=(p_j)$ respectively denote position and momentum coordinates. 
$\Phi^t=(x(t),p(t))$ is then an integral curve of the Hamiltonian vector field $X_H$ if \emph{Hamilton's equations} hold:
$$\begin{cases}
\partial_t x_j=\partial_{p_j}H, \\
\partial_t p_j = -\partial_{x_j}H.
\end{cases}$$
This is because if 
$$X_H=\sum_{j=1}^n\left(\frac{\partial H}{\partial {p_j}}\frac{\partial }{\partial{x_j}}-\frac{\partial H}{\partial{x_j}}\frac{\partial}{\partial{p_j}}\right),$$ 
then applying $\iota_{X_H}$ to $\omega=\sum_j dx_j\wedge dp_j$ gives
$$
\iota_{X_H}\omega  =
\sum_{j=1}^n\iota_{X_H}(dx_j\wedge dp_j)=\sum_{j=1}^n ((\iota_{X_H}dx_j)\wedge dp_j-dx_j\wedge (\iota_{X_H}dp_j)) $$$$
 =\sum_{j=1}^n\left(\frac{\partial H}{\partial{p_j}}dp_j+\frac{\partial H}{\partial{x_j}}dx_j\right)=dH.
$$
\textbf{Example \cc} (\emph{Newton's second law}) 
If $n=3$, then Newton's second law states that
$m\frac{d^2x}{dt^2}=-\nabla V(x)$
for $x=(x_1,x_2,x_3)\in \mathbb{R}^3$ and $m$ the mass of a particle moving along a curve $x(t)$ under a potential $V(x)$. If the momentum variables are defined as $p_i=m\frac{dx_i}{dt}$ and the Hamiltonian is $H(x,p)=\frac{1}{2m}|p|^2+V(x)$, then in the phase space $T^*\mathbb{R}^3=\mathbb{R}^6$ with coordinates $(x_1,x_2,x_3,p_1,p_2,p_3)$, we have
$$
\frac{dx_i}{dt}=\frac{p_i}{m}=\frac{\partial H}{\partial p_i} \ \ \ \text{ and }\ \ \ 
\frac{dp_i}{dt}=m\frac{d^2x_i}{dt^2}=-\frac{\partial V}{\partial x_i}=-\frac{\partial H}{\partial x_i}.
$$
So Hamilton's equations are equivalent to Newton's second law, and it is clear that 
$H$ is conserved by the motion.
\bb
What is the algebraic structure of Hamiltonian vector fields? First we recall that vector fields are differential operators on functions in the sense that $Xf=df(X)=\mathcal{L}_Xf$.
We also recall that, if the Lie bracket of two vector fields $X$ and $Y$ is denoted by $[X,Y]=XY-YX$ and $X$ and $Y$ are symplectic, then $[X,Y]$ is itself a Hamiltonian vector field with Hamiltonian function $\omega(X,Y)$. The Lie bracket $[\cdot,\cdot]$ then endows a bilinear form on vector fields on $(M,\omega)$, showing that they are in fact Lie algebras with the following inclusions:\index{Lie bracket}\index{Poisson bracket}
$$\begin{array}{c}
(\text{Hamiltonian vector fields},[\cdot,\cdot])\subset 
(\text{symplectic vector fields},[\cdot,\cdot])\subset (\text{vector fields},[\cdot,\cdot]). \end{array}$$
Similarly, we recall that the \emph{Poisson bracket} of $f,g\in C^\infty(M,\mathbb{R})$ is given by 
$$\{f,g\}=\omega(X_f,X_g)=\sum_{i=1}^n \left(\frac{\partial f}{\partial {x_i}}\frac{\partial g}{\partial {p_i}}-\frac{\partial f}{\partial{p_i}}\frac{\partial g}{\partial{x_i}}\right),$$
where $(x,p)$ are canonical (Darboux) coordinates. 
This has the property that $X_{\{f,g\}}=[X_g,X_f]\Longrightarrow X_{\{f,g\}}=-[X_f,X_g]$. 
Like the Lie bracket, the Poisson bracket is antisymmetric and satisfies the Jacobi identity $\{f,\{g,h\}\}+\{g,\{h,f\}\}+\{h,\{f,g\}\}=0$. A \emph{Poisson algebra} $(P,\{\cdot,\cdot\})$ is then a commutative associative algebra with a Poisson bracket that satisfies the Leibniz rule $\{f,gh\}=\{f,g\}h+g\{f,h\}$. If $(M,\omega)$ is a symplectic manifold, then $(C^\infty(M),\{\cdot,\cdot\})$ is a Poisson algebra and $M$ is said to be a \emph{Poisson manifold}. 

Finally, we remind ourselves that a \emph{Hamiltonian system} is specified by the data $(M,\omega,H)$, where $(M,\omega)$ is a symplectic (Poisson) manifold and $H$ is a Hamiltonian function. Any other function $f$ is said to be \emph{in involution} with $H$ if $\{H,f\}=0$; this means that $f$, called a \emph{constant of motion}, remains constant (or ``is conserved") along the integral curves of the Hamiltonian vector field $X_H$. By the Jacobi identity, we see that if $f$ and $g$ are constants of motion, then $\{f,g\}$ is a constant of motion as well. A Hamiltonian system is \emph{integrable} if it admits a maximal set of independent constants of motion $f_1,f_2,...$ that are in involution with each other. Here the functions $f_1,f_2,...$ are said to be \emph{independent} if their differentials $df_1,df_2,...$ are linearly independent on a dense subset of $M$. 
If our symplectic manifold $M$ is of dimension $2n$, then by symplectic linear algebra we know that the ``maximal set" contains $n$ elements. \index{Hamiltonian formulation of mechanics!Hamiltonian system}
\bb
\textbf{Definition \cc} (\emph{integrability}) A $2n$-dimensional Hamiltonian system $(M,\omega,H)$ is \emph{(completely, classically, or Liouville) integrable} if it admits $n$ independent constants of motion $H=f_1,f_2,...,f_n$ such that $\{f_i,f_j\}=0$ for all $i$ and $j$. \index{Hamiltonian formulation of mechanics!integrability}
\\
Though there are interesting cases where $\dim M=\infty$ and an infinite number of constants of motion exist (see, for example, the KdV hierarchy \cite{Gal1}), for the purposes of quantum ergodicity our prototypical examples will involve Hamiltonian systems in lower finite dimensions. 
The \emph{Liouville-Arnold theorem} tells us that if a system is integrable, then there is a canonical change of variables to action-angle coordinates on $M$ in which the Hamiltonian flow behaves like quasiperiodic flows on tori:\index{Hamiltonian formulation of mechanics!Liouville-Arnold theorem}
\bb
\textbf{Theorem \cc} (\emph{Liouville-Arnold}, \cite{Can1}) Let $(\mathcal{M},\omega,H)$ be an integrable system of dimension $2n$ with integrals of motion $f_1=H,f_2,...,f_n$. Let $c\in \mathbb{R}^n$ be a regular value of $F:=(f_1,...,f_n)$. The corresponding level set (or \emph{energy shell}) $F^{-1}(c)$ is a Lagrangian submanifold of $\mathcal{M}$ (i.e. a submanifold of dimension $n$ where $\omega$ restricts to zero), and\vspace{-5pt}
\begin{enumerate}
\item If the flows of $X_{f_1},...,X_{f_n}$ starting at $p\in F^{-1}(c)$ are complete, then the connected component of $F^{-1}(c)$ containing $p$ is of the form $\mathbb{R}^{n-k}\times \mathbb{T}^k$ for some $k$ where $0\leq k\leq n$. With respect to this affine structure, this component has angle coordinates $(\alpha_1,...,\alpha_n)$ in which the flows of $X_{f_1},...,X_{f_n}$ are linear. \vspace{-5pt}
\item There are action coordinates $(\beta_1,...,\beta_n)$ where the $\beta_i$'s are integrals of motion and $(\alpha_1,...,\alpha_n,\beta_1,...\beta_n)$ form a Darboux chart.
\end{enumerate}
$\null$
\vspace{-25pt}
\bb
In other words, the phase space of an integrable system is foliated by invariant tori, and the Hamiltonian flow reduces to translations on these tori. If a system is ``stable," then two similar initial conditions would correspond to points on nearby tori, and the orbits of the Hamiltonian flow coming from these tori would correspond to translations in slightly different directions. The trajectories will therefore separate slowly (or linearly). It should not be surprising that integrability is a rather strong condition: the probability that a randomly chosen system with more than one degree of freedom is integrable is zero \cite{San1}. 
\bb
\textbf{Example \cc} (\emph{integrability of two-dimensional systems}) Any two-dimensional Hamiltonian system is trivially integrable because the Hamiltonian is conserved: examples of this include the simple pendulum and harmonic oscillator. 
If $\mathcal{M}=\mathbb{R}^{2n}$ with the canonical symplectic form, then any system in which $H$ varies only with momentum coordinates $p_i$ is integrable, as the $p_i$ themselves are independent constants of motion in pairwise involution. 
\bb
\textbf{Example \cc} (\emph{Hamiltonian nature of billiards}) \emph{Billiards are Hamiltonian systems, and certain ones are integrable.} The angular momentum $\alpha_0$ is a constant of motion for the circular billiard, since it remains unchanged throughout the motion. Similarly, the elliptical and rectangular billiards are integrable, as their angular momentum and linear momentum are respectively preserved. As we may expect, the triangular, Sinai, Barnett, and Bunimovich billiards are ``chaotic" and not integrable; see \cite{DeB1} for a more detailed discussion and proof. 
\bb
With the set-up as above, we note that the level set $\Sigma_c:=H^{-1}(c)$ carries a natural flow-invariant measure. This is called the \emph{Liouville measure}, and is constructed as follows. If $\dim \mathcal{M}=2n$, then $\frac{1}{n!}\omega^n$, where $\omega$ is the symplectic structure, is a volume form on $\mathcal{M}$. Since $dH$ is a nonzero one-form in a neighborhood of $\Sigma_c$, we can locally write\index{Liouville measure}\index{level set}
$\frac{1}{n!}\omega^n=\eta \wedge dH$
for some $2n-1$ form $\eta$. The pullback of $\frac{1}{n!}\omega^n$ to $H^{-1}(c)$ is clearly independent of the choice of $\eta$, and is therefore a well-defined volume form. Furthermore, this measure is preserved by the Hamiltonian flow because $\omega$, $dH$, and $\Sigma_c$ are.
\bb
\textbf{Definition \cc} (\emph{Liouville measure}) The \emph{Liouville measure} $\mu_L^c$ is the flow-invariant volume form on any energy shell $\Sigma_c$ of a Hamiltonian system. In particular, for each $c\in [a,b]$, $\mu_L^c$ is characterized by the formula
$$\iint_{H^{-1}[a,b]}f dxdp=\int_a^b\int_{\Sigma_c}f d\mu_L^cdc$$
for all $a<b$ and $f:T^*M\to \mathbb{R}$ (or $\mathbb{C}$), where $M$ is a Riemannian manifold.
\bb
If $\mathcal{M}=\mathbb{R}^{2n}$ with the canonical symplectic form, then the Liouville measure on $H^{-1}(c)$ is given by 
$d\mu_L^c={d\sigma}/{||\nabla H||},$
where $d\sigma$ is the hypersurface area element and $\nabla$ is the metric-induced gradient. 
To see this, we consider the volume element on $H^{-1}[c,c+\delta c]$ for small $\delta$, and note that the  ``thickness" of this shell is proportional to $1/||\nabla H||$ (c.f. \cite{Pet1}).

We need one more geometric ingredient. Recall that the \emph{geodesic flow} $g^t$ on a Riemannian manifold $(M,g)$ is a local $\mathbb{R}$-action on $TM$ defined by\index{geodesic flow}
$g^t(v)=\dot{\gamma}_v(t),$
where $\dot{\gamma}_v(t)$ is the unit tangent vector to the geodesic $\gamma_v(t)$ for which $\dot{\gamma}_v(0)=v$. It is well-understood that any geodesic flow is a Hamiltonian flow given a suitable Hamiltonian:
\bb
\textbf{Theorem \cc} (\emph{geodesic and cogeodesic flow}, \cite{Mil2}) Let $(M,g)$ be a Riemannian manifold and endow the tangent bundle $T^*M$ with the canonical symplectic structure $\omega=\sum_{i}dx_i\wedge dp_i$, where $(x_1,...,x_n,p_1,...,p_n)$ are local coordinates in $T^*M$. If $H:T^*M\to \mathbb{R}$ is given by 
$$H(x,p)=\frac{1}{2}|p|^2=\frac{1}{2} g^{ij}p_ip_j,$$
then the Hamiltonian flow of $H$ is called the \emph{cogeodesic} flow. The trajectories of the cogeodesic flow are geodesics when projected to $M$, and the cogeodesic flow identifies with the geodesic flow of $H$ on $(TM,\omega)$ via the metric-induced isomorphism $\flat:TM\cong T^*M$. 
Thus the integrability of a geodesic flow is a well-defined notion.
\bb
\textbf{Example \cc} 
(\emph{geodesic integrability of surfaces of revolution}, \cite{Kiy1}) Define a \emph{surface of revolution} as a two-dimensional Riemannian manifold that admits a faithful $S^1$-action by isometries. Let $M$ be a surface of revolution, $X$ the corresponding Killing field, and $\pi:TM\to M$ the canonical projection. From the classical Clairaut's theorem, the function $f:TM\to \mathbb{R}$ defined by $f(x,\xi)=\langle \xi,X(\pi(x,\xi))\rangle_{g(x)}$ is invariant under the geodesic flow $g^t$ generated by the vector field of $H=\frac{1}{2}g^{ij}p_ip_j$. (Explicitly, if $\gamma(t)=(r(t),\vartheta(t))$ is a geodesic on $M$ and $\alpha(t)$ is the angle between $\gamma'(t)$ and $\frac{\partial}{\partial \vartheta}|_{\gamma(t)}$, then the quantity $F=r(t)\sin \alpha(t)$ remains constant along $\gamma$.) Thus $g^t$ is integrable, with constants of motion $F$ and $H$. 
\bb
Other examples of surfaces with integrable geodesic flows include circular billiards, 
the hyperbolic half-plane (or any Hadamard manifold, \cite{Bal1}), and the two-dimensional sphere $S^2$. Although we will not explain these examples, detailed discussions of many integrable geodesic flows are readily available in the literature \cite{Arn1,Mil2}.

\subsubsection{Fundamentals of Ergodic Theory}
We briefly discuss the meaning of \emph{chaotic}. If integrable systems exhibit stable trajectories, then systems that only conserve the Hamiltonian must be the opposite: they must exhibit evenly distributed trajectories and be ``chaotic." Indeed, this idea is equivalent to saying that the flow as a measure-preserving transformation is \emph{ergodic}.\index{ergodicity}
\bb
\textbf{Definition \cc} (\emph{ergodicity and mixing}) Let $(M,\mathfrak{A},\mu)$ be a measure space, $\mathfrak{A}$ a $\sigma$-algebra, and $T:M\to M$ a measurable, measure-preserving map. Then $T$ is \emph{ergodic} if:\vspace{-5pt}
\begin{enumerate}
\item the only $T$-invariant measurable sets are $\emptyset$ and $M$;
\vspace{-7pt}
\item every $T$-invariant function ($f\circ T=f$) is constant except on a set of measure zero;
\vspace{-7pt}
\item or almost every orbit is equidistributed, i.e. for almost all $x\in M$, $$\lim_{N\to \infty} \frac{\# \{n\in \mathbb{N}:0\leq n\leq N-1, T^n(x)\in A\}}{N}=\frac{\mu(A)}{\mu(M)}$$
for every measurable subset $A\in \mathfrak{A}$. 
\end{enumerate}\vspace{-5pt}
It is straightforward to show that these conditions are equivalent \cite{Ste1}. $T$ is said to be \emph{(strong) mixing} if 
$\lim_{n\to \infty}\mu(A\cap T^{-n}(B))=\mu(A)\mu(B)$
for all measurable subsets $A,B\in \mathfrak{A}$. Mixing implies ergodicity: if $A$ is a $T$-invariant measurable set, then setting $A=B$ yields $\mu(A)=\mu(A)^2\Longrightarrow \mu(A)=0$ or $1$. 
\bb
Thus we are able to discuss the ergodicity of the geodesic flow $g^t$ by taking $M$ as the (compact) energy shell $\Sigma_c$, $\mu$ as the Liouville measure $\mu_L^c$, and $T$ as $g^t$. $g^t$ is then an \emph{ergodic flow} if the ergodicity conditions hold for all $t\in \mathbb{R}$, and $g^t$ is \emph{ergodic on $H^{-1}[a,b]$} if they hold for all $c\in [a,b]$. As mentioned before, an immediate example of an ergodic geodesic flow is the billiard flow on the Sinai stadium.

We take this opportunity to state a fundamental result from ergodic theory. 
If $z=(x_1,...,x_n,p_1,...,p_n)\in \Sigma_c$ and $f:T^*M\to \mathbb{R}$ or $\mathbb{C}$, then for $T>0$ we define the \emph{time average} as
$$\langle f\rangle_T:=\frac{1}{T}\int_0^T f(g^t(z))dt=\dashint_0^T f(g^t(z))dt,
$$
where the slash through the second integral denotes averaging. Note that $\langle f\rangle_T$ depends on $z\in T^*M$. Our first theorem, a weaker version of Birkhoff's ergodic theorem, relates the time-average to the space-average:
\bb
\textbf{Theorem \cc} (\emph{weak ergodic theorem}) If $g^t$ is ergodic on $(\Sigma_c,\mathfrak{A},\mu_L^c)$, then \index{ergodicity!weak ergodic theorem}
$$\lim_{T\to \infty}\int_{\Sigma_c}\left( \langle f\rangle_T-\dashint_{\Sigma_c}fd\mu_L^c\right)^2d\mu_L^c=0,$$
for all $f\in L^2(\Sigma_c)$. The space-average is denoted here by $\dashint_{\Sigma_c} fd\mu_L^c=\int_{\Sigma_c} f d\mu_L^c/\mu_L^c(\Sigma_c)$. 

\emph{Proof.} Following \cite{Zwo1}, we normalize $\mu_L^c$ so that $\mu_L^c(\Sigma_c)=1$. Let $X_H$ denote the Hamiltonian vector field that identifies with the geodesic flow $g^t$, and let 
$$A=\{f\in L^2(\Sigma_c):(g^t)^*f=f\ \forall t\},\ \  B_0=\{X_H\phi:\phi\in C^\infty(\Sigma_c)\},\ \ \text{and }\   B=\overline{B}_0\subset L^2(\Sigma_c),$$ 
where wlog all functions are $\mathbb{C}$-valued. 
We claim that $B_0^\perp=A$, where the orthogonal complement is in $L^2(\Sigma_c)$: if $h\in A$ and $f=X_H\phi\in B$, then by the flow-invariance of $\mu_L^c$,
\begin{eqnarray*}
\langle h,f\rangle&=&\int_{\Sigma_c}h\overline{f}d\mu_L^c=\int_{\Sigma_c} h\overline{X_H\phi}d\mu_L^c=\frac{\partial}{\partial t}\int_{\Sigma_c}h\overline{(g^t)^*\phi}d\mu_L^c|_{t=0} \\
&=&\frac{\partial}{\partial t}\int_{\Sigma_c}(g^{-t})^*h\overline{\phi}d\mu_L^c|_{t=0}=\frac{\partial}{\partial t}\int_{\Sigma_c}h\overline{\phi}d\mu_L^c|_{t=0}=0,
\end{eqnarray*}
so $h\in B_0^\perp$. Likewise, for $h\in B_0^\perp$, we have
$$0=\int_{\Sigma_c}h\overline{X_H(g^{-t})^*\phi}d\mu_L^c=-\frac{\partial}{\partial t}\int_{\Sigma_c}h\overline{(g^{-t})^*\phi}d\mu_L^c=-\frac{\partial}{\partial t}\int_{\Sigma_c}(g^t)^*h\overline{\phi}d\mu_L^c
$$
for any $\phi \in C^\infty(\Sigma_c)$. So for all $t\in \mathbb{R}$ and $\phi \in C^\infty(\Sigma_c)$,
$$\int_{\Sigma_c}(g^t)^*h\overline{\phi}d\mu_L^c=\int_{\Sigma_c}h\overline{\phi}d\mu_L^c,$$
and we take $(g^t)^*h=h\in A$. Then $B_0^\perp=A\Longrightarrow L^2(\Sigma_c)=A\oplus B$. Decomposing $f=f_A+f_B$ for $f_A\in A, f_B\in B$, we see that $\langle f_A\rangle_T=f_A$. 
Now to show that $\langle f\rangle_T=\langle f_A\rangle_T+\langle f_B\rangle_T\to f_A$ in $L^2(\Sigma_c)$, it suffices to show that $\langle f_B\rangle_T\to 0$ as $T\to \infty$ where $f_B\in B$. This follows easily:
\begin{eqnarray*}
\int_{\Sigma_c}|\langle X_H \phi \rangle_T|^2d\mu_L^c &= &\frac{1}{T^2}\int_{\Sigma_c}\left| \int_0^T \frac{d}{dt} (g^t)^*\phi dt\right|^2d\mu_L^c=\frac{1}{T^2}\int_{\Sigma_c}|(g^T)^*\phi-\phi|^2d\mu_L^c\\
&\leq& \frac{4}{T^2}\int_{\Sigma_c}|\phi|^2d\mu_L^c\to 0
\end{eqnarray*}
as $T\to \infty$. So indeed we have $\langle f\rangle_T\to f_A$ in $L^2(\Sigma_c)$. 

Finally, we note that the ergodicity of $g^t$ implies that $A$ is precisely the set of constant functions. This is because for all $f_A\in A$, the set $f_A^{-1}[c,\infty)$ is invariant under $g^t$ and has either full or zero measure. Since functions are unique in $L^2(\Sigma_c)$ up to a set of measure zero, $f_A$ is identically constant. Observing that the projection $f\mapsto f_A$ is identical to space-averaging w.r.t. $\mu_L^c$, we have $\langle f\rangle_T=f_A=\dashint_{\Sigma_c}fd\mu_L^c$ as $T\to \infty$. $\hfill \blacksquare$
\bb
Birkhoff's ergodic theorem tells us that in fact
$\langle f\rangle_T\to \dashint_{\Sigma_c}fd\mu_L^c$
as $T\to \infty$, but
we will only use the weak ergodic theorem in Chapter 3 for proving the quantum ergodicity theorem. Surprisingly, there are few other prerequisites we need. 

One final thing that is important for understanding the quantum chaos literature is the statement that the geodesic flow on any negatively curved Riemannian manifold is ergodic. Though the full result requires the machinery of smooth ergodic theory and the introduction of such notions as the \emph{Anosov property} and \emph{hyperbolicity}, for the sake of brevity we will only cite the theorem as follows: \index{ergodicity!Anosov property}
\bb
\textbf{Theorem \cc} (\emph{ergodicity of geodesic flow on negatively curved manifold}, \cite{Bal1}) Let $(M,g)$ be a compact Riemannian manifold with a $C^3$ metric and negative sectional curvature. Then the geodesic flow $g^t:TM\to TM$ is ergodic. \index{ergodicity!Hopf theorem}

\emph{Sketch of proof.} This exact result is proved in the source cited above. The proof relies on a \emph{Hopf argument}, which uses the Birkhoff ergodic theorem, the density of continuous functions among integrable functions, and the foliation of the tangent space into stable and unstable manifolds to show that $g^t$ is Anosov (a chaotic property stronger than ergodicity). It can be shown from this that all $g^t$-invariant functions on $\Sigma_c$ are constant w.r.t. $\mu_L^c$, except on a set of measure zero. $\hfill \Box$
\bb
Since these conditions are equivalent, QE and QUE theorems in the literature assume either the ergodicity of a manifold's geodesic flow or the negativity of its sectional curvature.







\subsection{Key Themes in Semiclassical Analysis and Quantum Ergodicity}

We conclude our introductory chapter with a broad overview of the problem at hand. 
{Semiclassical analysis} examines how a chaotic system's classical description is reflected in its quantum behavior in the semiclassical $\hbar \to 0$ limit: more precisely, \emph{how does the ergodicity of the geodesic flow on a Riemannian manifold determine the distribution of high-eigenvalue Laplacian eigenfunctions?}

As we have seen, the time evolution of any classical system on a Riemannian manifold $(M,g)$ is given by the Hamiltonian flow $\Phi^t$ on the phase space $T^*M$, and the flow on the energy shell $\Sigma_c$ simply identifies with the geodesic flow $g^t$ on $M$. If $(M,g)$ is compact with negative curvature, then we also know that $g^t$ is ergodic with respect to the Liouville measure $\mu_L^c$ on $\Sigma_c$. The corresponding quantum dynamics is the unitary flow generated by the Laplace-Beltrami operator on $L^2(M)$, as the quantum-mechanical time evolution is determined by solutions to the eigenvalue problem $-\Delta \psi=\lambda_n\psi$ \cite{Lan1}.  We may expect that the ergodicity of $g^t$ influences the spectral theory of the Laplacian by making its eigenfunctions equidistributed: if the eigenpair sequence $\{(\psi_n,\lambda_n)\}$ is ordered by increasing eigenvalues, then as $n\to \infty$ the sequence of probability measures given by\index{Wigner measure}
$$\mu_n(B):=\int_B |\psi_n(y)|^2dy$$
for $B\subset M$ may converge to the uniform measure over $M$. 
This is essentially what the \emph{quantum ergodicity theorem} states, and 
we will formulate this result more rigorously in Chapter 3.
\bb
Having introduced these requisite notions, we pause to reflect on certain key themes that appear throughout the rest of our thesis. We will continue to see that these themes create a coherent framework for thinking about problems in semiclassical analysis. Moreover, the following comparisons are useful for readers not already familiar with the mathematical formulation of quantum mechanics.

\begin{enumerate}

\item \textbf{The classical-quantum correspondence.}\vspace{-5pt}
\begin{itemize}
\item Classical states are points of a symplectic manifold $(\mathcal{M},\omega)$, where $\mathcal{M}$ is the cotangent bundle of a Riemannian manifold $(M,g)$, i.e. $\mathcal{M}=T^*M$. Quantum states are elements in $\mathbb{P}\mathcal{H}$ (the projectivization of a Hilbert space $\mathcal{H}$) or $\mathbb{CP}^n$. This is because both $\psi$ and $c\psi$ for $c>0$ represent the same physical state. Since $\mathcal{M}=T^*M$ in the classical case, here we take $\mathcal{H}=L^2(M)$. 
\item Classical observables are functions $f:\mathcal{M}\to \mathbb{R}$ (or $\mathbb{C}$). As we know from quantum mechanics, quantum observables are self-adjoint operators on $\mathcal{H}$. An example of a classical Hamiltonian $H:\mathcal{M}\to \mathbb{C}$ is
$H(x,p)=\frac{1}{2m}|p|^2+V(x),$\index{self-adjoint operator}
where $V$ is a potential function. An example of a quantum Hamiltonian is a time-independent Schr\"{o}dinger operator 
$H=-\frac{\hbar^2}{2}\Delta+V,$
where $V:M\to \mathbb{C}$ is some potential.
\item Classical dynamics are given by the Hamiltonian flow of the vector field $X_H$, where $H:\mathcal{M}\to \mathbb{R}$ (or $\mathbb{C}$). If we take the canonical symplectic structure $\omega=\sum_{i} dx_i\wedge dp_i$, then the flow is defined by Hamilton's equations and preserves $\omega$. Quantum dynamics are given by the Schrodinger equation and the unitary flow $U^t$ (a quantized geodesic flow) coming from the Laplacian $\Delta$ acting on $\mathcal{H}$; see \S2.3.3 and \cite{Zel1} for a more rigorous treatment of quantum evolution.\index{unitary flow}
\end{itemize}

\item \textbf{Physical intuition in the semiclassical limit.} Although we can numerically take the semiclassical limit $\hbar \to 0$, in reality we need the energies to be bounded. Our expectation should be that, in the semiclassical regime, the asymptotic behavior of quantum objects is governed by classical mechanics. The semiclassical limit therefore serves as a physical passage from quantum to classical mechanics.

\item \textbf{Quantization as a bridge between the categories of Hilbert spaces and symplectic manifolds.} To actually relate quantum and classical mechanics, we must associate the Hilbert space $\mathcal{H}=L^2(M)$ to the symplectic manifold $\mathcal{M}=T^*M$ and assign operators on $\mathcal{H}$ to functions on $\mathcal{M}$. It is well-understood that a functorial procedure of doing so does not exist \cite{Hov1}, but there are certain ``nice" ways in which we can ``quantize" operators. The most convenient of these is \emph{Weyl quantization}.
In particular, the Weyl quantization formula uses the Fourier transform to associate the \emph{symbol} $a=a(x,p):\mathcal{M}\to \mathbb{C}$ to a quantum observable (pseudodifferential operator) $A(x, hD)$, where $x$ denotes position, $D$ a differential, and $h$ a semiclassical parameter. 
How do the analytic properties of the symbol $a$ dictate the functional-analytic properties of its quantization $A$? It turns out that the \emph{symbol calculus} of \S2.2 will give us a framework for manipulating pseudodifferential operators. \index{semiclassical analysis}
\index{quantization!Weyl quantization}\index{quantization}

\item \textbf{The technical framework for semiclassical analysis.} 
We use symplectic geometry to formalize the behavior of classical dynamical systems and the {Fourier transform} to relate their position and momentum variables. 
Since semiclassical quantization relies on a rescaled, semiclassical Fourier transform, analytic methods of calculating integrals and Fourier transforms will prove useful. 
Working the semiclassical calculus out on $\mathbb{R}^n$ will 
allow an extension of its tools to coordinate patches on arbitrary manifolds, ultimately leading to a proof of the quantum ergodicity theorem. 

\item \textbf{Visualizing the simple cases and understanding the interaction between structure and randomness.} As a general technique, we note that geometry is oftentimes based on visualization. It will therefore be instrumental to remember the billiard flow---one of the simplest, most well-studied models in quantum chaos---as we work out the classical-quantum correspondence in subsequent chapters. Our study of semiclassical analysis illustrates the thematic \emph{dichotomy between classical structure and quantum randomness}, about which the mathematician Terrence Tao writes in \cite{Tao2}:
\bb
\emph{The ``dichotomy between structure and randomness" seems to apply in circumstances in which one is considering a Òhigh-dimensionalÓ class of objects... one needs tools such as algebra and geometry to understand the structured component, one needs tools such as analysis and probability to understand the pseudorandom component, and one needs tools such as decompositions, algorithms, and evolution equations to separate the structure from the pseudorandomness.}
\end{enumerate}

\pagebreak

\setcounter{section}{1}
\section{An Introduction to Semiclassical Analysis}
\setcounter{itemcounter}{1}
This chapter provides a primer in one of the most basic notions of semiclassical analysis: that of symbol quantization. We start by defining the Fourier transform on $\mathbb{R}^n$ and the quantization formulas we will use, and end by proving several results that are crucial to quantum chaos and quantum ergodicity. These results include Weyl's law for the asymptotic distribution of Laplacian eigenvalues and Egorov's theorem for the correspondence between classical and quantum mechanics.  
In developing the {symbol calculus}, we will also 
describe how certain properties of symbols relate to the properties of their quantum counterparts and derive estimates for the asymptotic behavior of quantized operators. 

\subsection{Semiclassical Quantization}

From elementary physics, we know that the Fourier transform allows us to convert functions of the position variable $x$ to functions of the momentum variables $p$ in the phase space $T^*M$. Quantization is the tool that allows us to deal with both sets of variables simultaneously in the semiclassical limit. Functions of both $x$ and $p$ variables are called \emph{symbols}, and are quantized using a modified, semiclassical Fourier transform. Moreover, the pseudodifferential operators ($\psi$DOs) produced by quantization have a precise meaning as \emph{quantum observables}, the self-adjoint operators corresponding to the classical observables represented by the symbols. We therefore start with a review of the Fourier transform before proceeding to write down quantization formulas.

\subsubsection{Distributions and the Fourier Transform}

We define the Fourier transform on $\mathbb{R}^n$; the following constructions are applicable to any open subset of a smooth manifold using a partition of unity. Recall that the Fourier transform is an automorphism of the \emph{Schwartz space}, the function space of all smooth, \emph{rapidly decaying} functions $f$ in the sense that the derivatives $f^{(n)}$ decay faster than any inverse power of $|x|$.\index{Fourier transform}\index{Fourier transform!Schwartz space}
\bb
\textbf{Definition \cc} (\emph{Schwartz space}) 
Define the \emph{seminorm} as 
$||f||_{\alpha,\beta}:=\sup_{x\in \mathbb{R}^n}|x^\alpha\partial^\beta f|$
 for multiindices $\alpha=(\alpha_1,...,\alpha_n),\beta=(\beta_1,...,\beta_n)\in \mathbb{N}^n$ and functions $f\in C^\infty(\mathbb{R}^n)$, where
$$x^\alpha=\prod_{i=1}^nx_i^{\alpha_i}, \ \ \ \ \ \ \ \partial^\beta=\prod_{i=1}^n \frac{\partial^{\beta_i}}{\partial x_i^{\beta_i}}.$$
The \emph{Schwartz space in $\mathbb{R}^n$} is the set 
$\mathcal{S}=\mathcal{S}(\mathbb{R}^n)=\{f\in C^\infty(\mathbb{R}^n):||f||_{\alpha,\beta}<\infty\ \forall \alpha,\beta \in \mathbb{N}^n\}.$
With the seminorm $||\cdot||_{\alpha,\beta}$ as above, we note that $\mathcal{S}$ is a Fr\'{e}chet space over $\mathbb{C}$, and say that $f_j\to f$ in $\mathcal{S}$ if $||f_j-f||_{\alpha,\beta}\to 0$ for all multiindices $\alpha$ and $\beta$. 
\bb
\textbf{Definition \cc} (\emph{Fourier transform}) The \emph{Fourier transform} is an isomorphism of topological vector spaces (but not of Fr\'{e}chet spaces) $\mathcal{F}:\mathcal{S}\ni f \mapsto  \mathcal{F}(f)\in \mathcal{S}$, which for a function $f\in \mathcal{S}$ is denoted by either $\mathcal{F}(f)$ or $\hat{f}$ and given by 
$$\mathcal{F}(f)(p)=\int_{\mathbb{R}^n}e^{-i\langle p,x\rangle}f(x)dx$$
for $x,p\in \mathbb{R}^n$ and $f\in \mathcal{S}$, with inverse 
$$\mathcal{F}^{-1}(f)(x)=(2\pi)^{-n}\int_{\mathbb{R}^n}e^{i\langle x,p\rangle}f(p)dp.$$
Note that we will always denote the variable conjugate to $x$ as $p$. The latter equation is called the \emph{Fourier inversion formula}, and combining $\mathcal{F}$ and $\mathcal{F}^{-1}$ leads to the identity
$$f(x)=(2\pi)^{-n}\iint_{\mathbb{R}^n\times \mathbb{R}^n} e^{i\langle x-y,p\rangle}f(y)dydp.$$\vspace{5pt}\\
\textbf{Proposition \cc} (\emph{Fourier transform of an exponential of a real quadratic form}, \cite{Zwo1}) Let $Q$ be a real, symmetric, and positive-definite $n\times n$ matrix. Then 
$$\mathcal{F}(e^{-\frac{1}{2}\langle Qx,x\rangle})=\frac{(2\pi)^{n/2}}{(\det Q)^{1/2}}e^{-\frac{1}{2}\langle Q^{-1}p,p\rangle}.$$
This example is useful as a higher-dimensional generalization of the fact that in the one-dimensional case the Fourier transform of a Gaussian (a function of the form $f(x)=C\exp(-ax^2)$, where $C, a\in \mathbb{R}$) remains a Gaussian. It is also important in subsequent proofs relating quantization to the Fourier transform.

\emph{Proof.} We have from a straightforward computation that 
\begin{eqnarray*}
\mathcal{F}(e^{-\frac{1}{2}\langle Qx,x\rangle})&=&\int_{\mathbb{R}^n}e^{-\frac{1}{2}\langle Qx,x\rangle-i\langle x,p\rangle}dx=\int_{\mathbb{R}^n}e^{-\frac{1}{2}\langle Q(x+iQ^{-1}p),x+iQ^{-1}p\rangle}e^{-\frac{1}{2}\langle Qp,p\rangle}dx
\\
&=&e^{-\frac{1}{2}\langle Q^{-1}p,p\rangle}\int_{\mathbb{R}^n}e^{-\frac{1}{2}\langle Qy,y\rangle}dy
=e^{-\frac{1}{2}\langle Q^{-1}p,p\rangle}\int_{\mathbb{R}^n}e^{-\frac{1}{2}\sum_{k=1}^n \lambda_kw_k^2}dw,
\end{eqnarray*}
where the last equality follows by changing into an orthogonal set of variables $\{w_k\}$ so that $Q$ is diagonalized with entries $\lambda_1,...,\lambda_n$. The second factor is then given by
$$e^{-\frac{1}{2}\sum_{k=1}^n \lambda_kw_k^2}dw=\prod_{k=1}^n\int_{-\infty}^\infty e^{-\frac{\lambda_k}{2}w^2}dw=\prod_{k=1}^n \frac{2^{1/2}}{\lambda_k^{1/2}}\int_{-\infty}^\infty e^{-y^2}dy=\frac{(2\pi)^{n/2}}{(\lambda_1...\lambda_n)^{1/2}}=\frac{(2\pi)^{n/2}}{(\det Q)^{1/2}},
$$
and the desired result follows. $\hfill \blacksquare$
\bb
Having defined the Fourier transform $\mathcal{F}$, we deduce its following properties, which are proven rigorously in standard analysis textbooks; see, for example, \cite{Ste2} and \cite{Hor1}.
\bb
\textbf{Theorem \cc} (\emph{properties of $\mathcal{F}$}) The Fourier transform $\mathcal{F}:\mathcal{S}\to \mathcal{S}$ is indeed an isomorphism of topological vector spaces  with inverse $\mathcal{F}^{-1}$ given above. Furthermore, it possesses the following differentiation and convolution relations for all $f,g\in \mathcal{S}$:

(i) $D_p^\alpha(\mathcal{F}(f))=\mathcal{F}((-x)^\alpha f)$ and $\mathcal{F}(D_x^\alpha (f))=p^\alpha \mathcal{F}(f)$, where 
$D^\alpha_x:=\frac{1}{i^{|\alpha|}}\partial^\alpha=(-i\partial_{x_1})^{\alpha_1}...\\(-i\partial_{x_n})^{\alpha_n}$.

(ii) $\mathcal{F}(fg)=(2\pi)^{-n}\mathcal{F}(f)*\mathcal{F}(g)$.

(iii) $\langle \mathcal{F}({f}),g\rangle=\langle f,\mathcal{F}({g})\rangle.$

(iv) 
$\langle f,g\rangle=(2\pi)^{-n}\langle \mathcal{F}({f}),\mathcal{F}({g})\rangle,$ and in particular $||f||^2=(2\pi)^{-n}||\mathcal{F}(f)||^2$, where $\langle \cdot,\cdot\rangle$ and $||\cdot||$ denote by default the $L^2$-inner product and norm. This is \emph{Plancherel's theorem}.
\bb
These properties result in the following useful estimates, which are stated without proof below: 
\bb
\textbf{Proposition \cc} (\emph{estimates of $\mathcal{F}$}) Let $||f||_p$ denote the $L^p$-norm of $f$. We have

(i) $||\mathcal{F}(f)||_{\infty}\leq ||f||_1$ and $||f||_\infty\leq (2\pi)^{-n}||\mathcal{F}(f)||_1$.

(ii) $\exists C>0: \forall \alpha\in \mathbb{N}^n, ||\mathcal{F}(f)||_1\leq C\max_{|\alpha|\leq n+1}||\partial^\alpha f||_1$.\index{distribution}\bb
Extending the Fourier transform to distributions now allows us to define $\mathcal{F}$ for a broader range of generalized functions. Recall that a \emph{distribution on $\mathbb{R}^n$} is a linear functional $\varphi:C_c^\infty(\mathbb{R}^n)\to \mathbb{R}$ (or $\mathbb{C}$) such that $\lim_{n\to \infty}\varphi(f_n)=\varphi(\lim_{n\to \infty}f_n)$ in $C_c^\infty(\mathbb{R}^n)$, where the seminorm is the same as before and we remember that $C_c^\infty(\mathbb{R}^n)$ denotes the space of smooth, compactly supported functions on $\mathbb{R}^n$. The set of all distributions generalizes and forms a vector space dual to $C_c^\infty(\mathbb{R}^n)$. For example, the Dirac delta distribution $\delta$, which has the property that
$$\int_{-\infty}^\infty \delta(x)f(x)dx=f(0),$$
is given by $\delta: C_c^\infty(\mathbb{R})\ni f\mapsto \langle \delta,f\rangle=f(0)\in \mathbb{R}$, where we abuse notation by writing $\delta=\langle \delta,\cdot\rangle$ and continue to take the $L^2$-inner product $\langle f,g\rangle=\int_{\mathbb{R}^n} f(x)\overline{g}(x)dx$. Analogously, the vector space of \emph{tempered distributions}\index{distribution!tempered distribution} $\mathcal{S}'$ is defined by duality from the Schwartz space $\mathcal{S}$. Introducing tempered distributions gives, among other things, the correct vector space for a rigorous formulation of the Fourier transforms of nonsmooth functions.  
\bb
\textbf{Definition \cc} (\emph{tempered distributions}) Let the space of \emph{tempered distributions} $\mathcal{S}'$ be the set of all continuous linear functionals $\varphi:\mathcal{S}\ni f\mapsto \varphi(f):=\langle \varphi,f\rangle\in \mathbb{C}$ in the sense that $\lim_{n\to \infty}\varphi(f_n)=\varphi(\lim_{n\to \infty}f_n)$. We say that $\varphi_j\to \varphi\in \mathcal{S}'$ if $\varphi_j(f)\to \varphi(f)$ for all $f\in \mathcal{S}$, and define for any multiindex $\alpha \in \mathbb{N}^n$:

(1) $D^\alpha \varphi(f):=(-1)^{|\alpha|}\varphi(D^\alpha f).$

(2) $(x^\alpha \varphi)(f):=\varphi(x^\alpha f).$
\\
Since $\mathcal{F}: \mathcal{S}\to \mathcal{S}$ is an automorphism, $\mathcal{F}$ also extends to $\mathcal{S'}$ by
$\mathcal{F}(\varphi)(f):=\varphi(\mathcal{F}(f))$,
where $\varphi\in \mathcal{S}'$ and $f\in \mathcal{S}$. 
\bb
Thus the vector space $\mathcal{S}'$ generalizes the set of bounded, slow-growing, locally integrable functions: in particular, all $L^2$ functions and distributions with compact support are in $\mathcal{S}'$.
\bb
\textbf{Example \cc} (\emph{Heavyside step function}) Let $H:\mathbb{R}\to \mathbb{R}$ be given by $H(x)=1$ if $x\geq 0$, and $0$ otherwise. This definition of $H$ gives the tempered distribution $\langle H,\cdot\rangle$, whose derivative is the Dirac delta. Indeed, we have 
$\langle H',f\rangle = -\langle H,f'\rangle=-\int_0^\infty f'(x)dx=-f(x)|_0^\infty=f(0)=\langle \delta,f\rangle.$
\bb
\textbf{Example \cc} (\emph{Fourier transform of Dirac delta}) Viewed as a tempered distribution, $\delta$ has a Fourier transform:
$$\langle \mathcal{F}(\delta),f\rangle=\langle \delta,\mathcal{F}(f)\rangle=\mathcal{F}(f)(0)=\int_\mathbb{R} f(x)dx=\langle 1,f\rangle\Longrightarrow \mathcal{F}(\delta)=1.$$
On the other hand, for the Fourier transform of the constant function 1, we see that
$$\langle \mathcal{F}(1),f\rangle=\langle 1,\mathcal{F}(f)\rangle=\int_\mathbb{R}\hat{f}(x)dx=2\pi f(0)\Longrightarrow \mathcal{F}(1)=2\pi\delta.$$
The following proposition will be used in \S2.2 to show a result pertaining to the decomposition of a Weyl-quantized operator. 
\bb
\textbf{Proposition \cc} (\emph{Fourier transform of an imaginary quadratic exponential}, \cite{Zwo1}) Let $Q$ be a real, symmetric, and invertible $n\times n$ matrix. Then
$$\mathcal{F}(e^{\frac{i}{2}\langle Qx,x\rangle})=\frac{(2\pi)^{n/2}e^{\frac{i\pi}{4}\text{sgn}(Q)}}{|\det Q|^{1/2}}e^{-\frac{i}{2}\langle Q^{-1}p,p\rangle},$$
where $\text{sgn}(Q)=\#\{\text{positive eigenvalues of }Q\}-\#\{\text{negative eigenvalues of }Q\}$ is called the \emph{sign of $Q$}. In particular, we have an extension of Proposition 2.1.3, where the phase shift $\exp({\frac{i\pi}{4}\text{sgn}(Q)})$ comes from the complex exponential. 

\emph{Proof.} First we note that the Fourier transform $\mathcal{F}(e^{\frac{i}{2}\langle Q x,x\rangle})$ is not absolutely convergent since
$$\int_{\mathbb{R}^n}|e^{\frac{i}{2}\langle Qx,x\rangle-i\langle x,p\rangle}|dx=\int_{\mathbb{R}^n} e^{\text{Im}(\langle x,p\rangle)-\frac{1}{2}\text{Im}(\langle Qx,x\rangle)}dx=\int_{\mathbb{R}^n} dx= \infty,$$
as $Q$ is a real matrix. To ensure absolute convergence, we perturb $Q$ slightly so that $Q_\epsilon:=Q+\epsilon i I$ for some $\epsilon>0$. This gives us
$$
\int_{\mathbb{R}^n}|e^{\frac{i}{2}\langle Q_\epsilon x,x\rangle-i\langle x,p\rangle}|dx=
\int_{\mathbb{R}^n} e^{-\frac{1}{2}\text{Im}(\langle Q_\epsilon x,x\rangle)}dx=\int_{\mathbb{R}^n} e^{-\frac{\epsilon}{2}\langle x,x\rangle}dx
<\infty,$$
where the convergence follows from the argument used to prove Proposition 2.1.3. Rewriting the Fourier transform of the modified exponential $\mathcal{F}(e^{\frac{i}{2}\langle Q_\epsilon x,x\rangle})$ now gives us
$$\mathcal{F}(e^{\frac{i}{2}\langle Q x,x\rangle})=
\lim_{\epsilon \to 0}\mathcal{F}(e^{\frac{i}{2}\langle Q_\epsilon x,x\rangle})=\lim_{\epsilon \to 0}
\int_{\mathbb{R}^n}
e^{\frac{i}{2}\langle Q_\epsilon x,x\rangle-i\langle x,p\rangle}dx$$
$$=\lim_{\epsilon \to 0}
\int
_{\mathbb{R}^n}
e^{\frac{i}{2}\langle Q_\epsilon(x-Q_\epsilon^{-1}p),x-Q_\epsilon^{-1}p\rangle}e^{-\frac{i}{2}\langle Q_\epsilon^{-1}p,p\rangle}dx=
e^{-\frac{i}{2}\langle Q^{-1}p,p\rangle} \lim_{\epsilon \to 0}\int
_{\mathbb{R}^n}
e^{\frac{i}{2}\langle Q_\epsilon y,y\rangle}dy,$$
where 
$y:=x-Q^{-1}_\epsilon p$. As before, we diagonalize $Q$ so that $Q=(\lambda_{ab})$, where $\lambda_{ab}=\lambda_a$ for $a=b$ and $\lambda_{ab}=0$ otherwise. Moreover, we arrange the eigenvalues so that $\lambda_1,..,\lambda_m$ are positive and $\lambda_{m+1},...,\lambda_n$ are negative. 
Then, since
$$\int_{\mathbb{R}^n}e^{-\frac{1}{2}\langle Qy,y\rangle}dy=\int_{\mathbb{R}^n}e^{-\frac{1}{2}\sum_{k=1}^n \lambda_kw_k^2}dw$$
and $Q_\epsilon=Q+\epsilon i I$ can be diagonalized so that 
$$Q_\epsilon=
\left(\begin{array}{ccc}
\lambda_{1}+\epsilon i & & 0 \\
& \ddots & \\
0 & & \lambda_{n}+\epsilon i \\
\end{array}\right),$$
we have 
$$\lim_{\epsilon \to 0} \int_{\mathbb{R}^n}e^{\frac{i}{2}\langle Q_\epsilon y,y\rangle}dy=\lim_{\epsilon \to 0}\int_{\mathbb{R}^n}e^{\sum_{k=1}^n\frac{1}{2}(i\lambda_k-\epsilon)w^2_k}dw=\lim_{\epsilon \to 0}\prod_{k=1}^n\int_{-\infty}^\infty e^{\frac{1}{2}(i\lambda_k-\epsilon)w^2}dw.$$
\begin{center}
\begin{figure}\hspace{20pt}
  \begin{minipage}[b]{0.2\textwidth}
        \captionsetup{position=top,labelsep=none,font=footnotesize}
    \renewcommand{\thefigure}{\arabic{section}.\arabic{figure}\ $|$\vspace{2pt}\newline }
          \caption{The contours $C_k$ for $\lambda_k>0$ and $\lambda_k<0$ used in the proof of Proposition 2.1.9.}
          \label{figure}
      \vspace{130pt}
    \end{minipage}\hspace{-30pt}
      \begin{subfigure}[b]{0.8\textwidth}\centering
    \includegraphics[height=7cm]{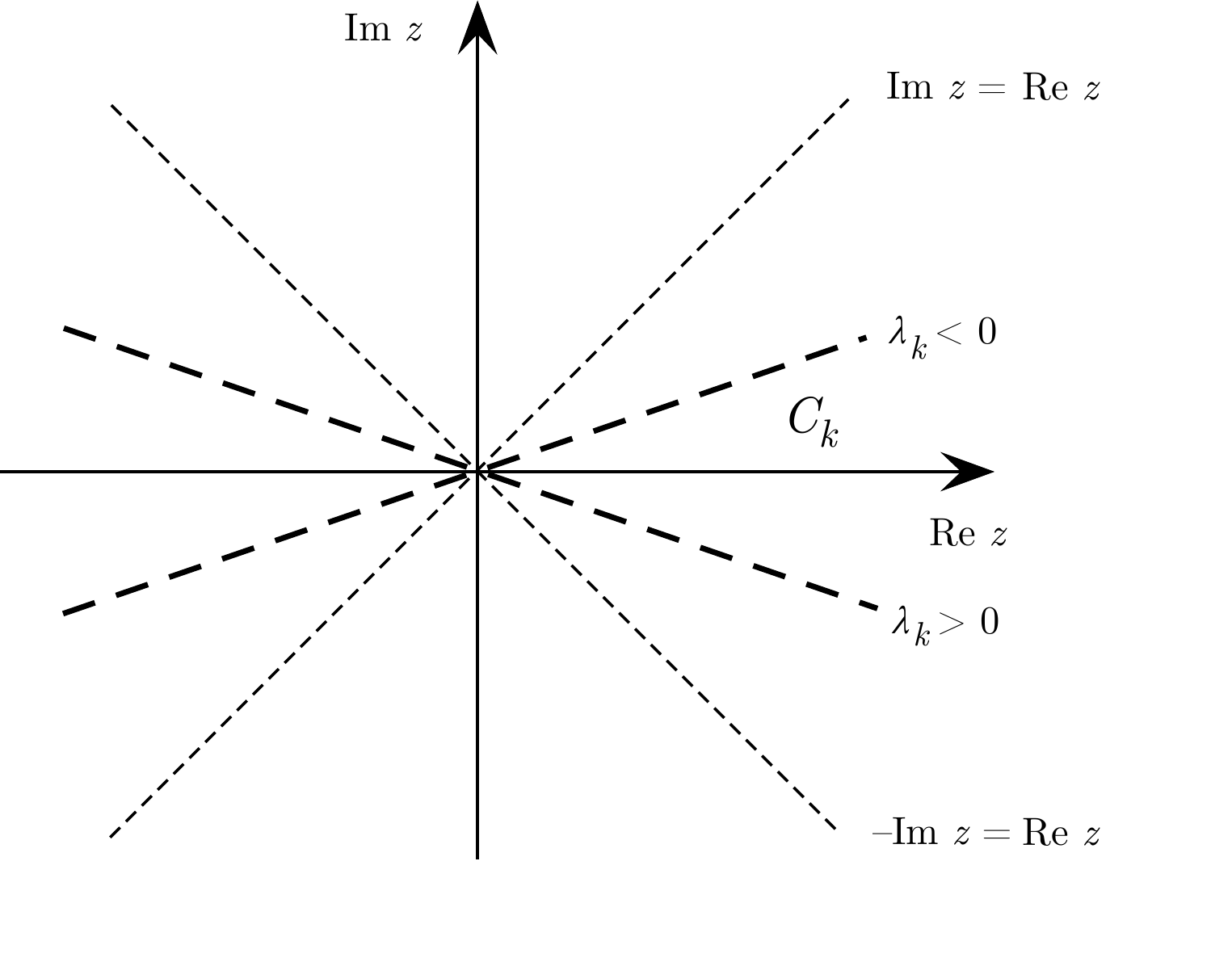}
          \label{subfig-1}
      \end{subfigure}\hspace{7pt}
  
\end{figure}
\end{center}
$\null$\vspace{-50pt}
\\
If $1\leq k\leq m$, then $\lambda_k>0$ and we make the change of variables $z=(\epsilon-i\lambda_k)^{{1}/{2}}w$, where we choose the branch of the square root for which $\text{Im}((\epsilon-i\lambda_k)^{1/2})<0$. This gives us 
$$\int_{-\infty}^\infty e^{\frac{1}{2}(i\lambda_k-\epsilon)w^2}dw=(\epsilon-i\lambda_k)^{-1/2}\int_{C_k}e^{-\frac{1}{2}z^2}dz
=(\epsilon-i\lambda_k)^{-1/2}\int_{C_k}e^{-\frac{1}{2}z^2}dz,$$
where $C_k$ is the contour in $\mathbb{C}$ shown in Figure 2.1. With $z=x+iy$, we also see that $e^{-\frac{1}{2}z^2}=e^{\frac{1}{2}(y^2-x^2)-ixy}$. The fact that $f(z)=e^{-\frac{1}{2}z^2}$ is entire and $x^2>y^2$ on $C_k$ then allows us to deform $C_k$ into the real line, so that
$$\int_{C_k}e^{-\frac{1}{2}z^2}dz=\int_{-\infty}^\infty e^{-\frac{1}{2}x^2}dx=\sqrt{2\pi}.$$
Thus, for $\lambda_k>0$, 
$$ 
\lim_{\epsilon\to 0}
\prod_{k=1}^m\int_{-\infty}^\infty e^{\frac{1}{2}(i\lambda_k-\epsilon)w^2}dw =
(2\pi)^{m/2}\lim_{\epsilon\to 0}\prod_{k=1}^m(\epsilon-i\lambda_k)^{-1/2}=
(2\pi)^{m/2}\prod_{k=1}^me^{\frac{i\pi}{4}}\lambda_k^{-1/2}.$$
Repeating the argument above for the negative eigenvalues $\lambda_k<0$ (${m}^*=m+1\leq k\leq n$) and the branch of the square root where $\text{Im}((\epsilon-i\lambda_k)^{1/2})>0$ gives us 
$$
\lim_{\epsilon\to 0}\prod_{k= m^*}^n\int_{-\infty}^\infty e^{\frac{1}{2}(i\lambda_k-\epsilon)w^2}dw =(2\pi)^{\frac{n-m}{2}}\lim_{\epsilon\to 0}\prod_{k=m^*}^n(\epsilon-i\lambda_k)^{-\frac{1}{2}}\\
=
(2\pi)^{\frac{n-m}{2}}\lim_{\epsilon\to 0}\prod_{k=m^*}^ne^{-\frac{i\pi}{4}}|\lambda_k|^{-\frac{1}{2}}.
$$
We therefore conclude that
\begin{eqnarray*}
\mathcal{F}(e^{\frac{i}{2}\langle Q x,x\rangle})&=&
\lim_{\epsilon\to 0}\mathcal{F}(e^{\frac{i}{2}\langle Q_\epsilon x,x\rangle})
=
e^{-\frac{i}{2}\langle Q^{-1}p,p\rangle}
\lim_{\epsilon\to 0}
 \int_{\mathbb{R}^n}
e^{\frac{i}{2}\langle Q_\epsilon y,y\rangle}dy\\
&=&
e^{-\frac{i}{2}\langle Q^{-1}p,p\rangle}\frac{(2\pi)^{n/2}e^{\frac{i\pi}{4}(m-(n-m))}}{|\lambda_1\cdots \lambda_n|^{1/2}}=\frac{(2\pi)^{n/2}e^{\frac{i\pi}{4}\text{sgn}(Q)}}{|\det Q|^{1/2}}e^{-\frac{i}{2}\langle Q^{-1}p,p\rangle},
\end{eqnarray*}
as desired. $\hfill \blacksquare$
\bb
The Fourier transform of tempered distributions (and in particular, $L^2$ functions) is important in quantum mechanics. For instance, it provides the mathematical basis for the Heisenberg {uncertainty principle}.\index{Heisenberg uncertainty principle}
\bb
\textbf{Example \cc} (\emph{uncertainty principle in $\mathbb{R}$}, \cite{Du1}) Consider some $\psi\in L^2(\mathbb{R})$ where $x\psi$ and $p\mathcal{F}(\psi)\in L^2(\mathbb{R})$. With the \emph{dispersion} of $\psi$ defined as
$$\mathcal{D}\psi:=\frac{\int_\mathbb{R}x^2|\psi(x)|^2dx}{\int_\mathbb{R}|\psi(x)|^2dx},$$
we see by a straightforward calculation that
$(\mathcal{D}\psi )(\mathcal{D}\mathcal{F}({\psi}))\geq 1/4$. 
In particular, using integration by parts we have
\begin{eqnarray*}
\int_{\mathbb{R}} |\psi(x)|^2dx&=&x|\psi(x)|^2|_{-\infty}^\infty-\int_{\mathbb{R}}x\psi(x)\overline{\psi'(x)}dx-\int_{\mathbb{R}}x\overline{\psi(x)}\psi'(x)dx\\
&=&x|\psi(x)|^2|_{-\infty}^\infty-2\text{Re}\left(\int_{\mathbb{R}}x\overline{\psi(x)}\psi'(x)dx\right)\\
&=&-2\text{Re}\left(\int_{\mathbb{R}}x\overline{\psi(x)}\psi'(x)dx\right),
\end{eqnarray*}
where the last equality follows from the decay properties of functions in $\mathcal{S}$. Squaring both sides and using the H\"{o}lder inequality $||\psi \phi||_1\leq ||\psi||_p||\phi||_q$ for $1/p+1/q=1$ gives
$$\left(\int_{\mathbb{R}} |\psi(x)|^2dx\right)^2\leq 4\left(\int_{\mathbb{R}}|x\overline{\psi(x)}\psi'(x)|dx\right)^2\leq 4\left(\int_\mathbb{R}x^2|\psi(x)|^2dx\right)\left(\int_\mathbb{R}|\psi'(x)|^2dx\right).$$
Theorem 2.1.4 (i) and (iv) give $\mathcal{F}(\psi'(x))=ip\mathcal{F}(\psi)(p)$ and $||\psi||^2=(2\pi)^{-1}||\mathcal{F}(\psi)||^2$, so 
$$\int_\mathbb{R}|\psi'(x)|^2dx=\frac{1}{2\pi}\int_\mathbb{R}p^2|\hat{\psi}(p)|^2 dp.$$
Thus
$$\left(\int_{\mathbb{R}} |\psi(x)|^2dx\right)\left(\frac{1}{2\pi} \int_{\mathbb{R}} |\hat{\psi}(x)|^2dx\right)
\leq
4\left(\int_\mathbb{R}x^2|\psi(x)|^2dx\right)\left(\frac{1}{2\pi}\int_\mathbb{R}p^2|\hat{\psi}(p)|^2 dp\right),
$$
and the claim follows. What the foregoing exposition tells us is that a function $\psi\in L^2(\mathbb{R}$) cannot be simultaneously highly localized in both its position and momentum variables; see the discussion following Theorem 2.1.13. 
\bb
This brief example provides motivation for the semiclassical Fourier transform. 
Since we would like to control the degree of localization and uncertainty of $\mathcal{F}$ in the semiclassical limit, we can reparameterize $\mathcal{F}$ using the semiclassical parameter $h>0$. Theorem 2.1.13 justifies the following definition.\index{Fourier transform!semiclassical Fourier transform}
\bb
\textbf{Definition \cc} (\emph{semiclassical Fourier transform}) For $h>0$, the \emph{semiclassical Fourier transform} $\mathcal{F}_h:\mathcal{S}'\to \mathcal{S}'$ is given by
$$\mathcal{F}_h(f)(p):=\mathcal{F}(f)\left(\frac{p}{h}\right)=\int_{\mathbb{R}^n}e^{-\frac{i}{h}\langle x,p\rangle}f(x)dx,$$
with inverse
$$\mathcal{F}_h^{-1}(f)(x)=h^{-n}\mathcal{F}(f)\left(\frac{x}{h}\right)=(2\pi h)^{-n}\int_{\mathbb{R}^n} e^{\frac{i}{h}\langle x,p\rangle}f(p)dp.$$
We can scale appropriately to derive properties similar to those of the usual Fourier transform $\mathcal{F}$ above. In particular, we have the following:
\bb
\textbf{Theorem \cc} (\emph{useful properties of $\mathcal{F}_h$}) Like Theorem 2.1.4, we have

(i) $(hD_p)^\alpha\mathcal{F}_h(\varphi)=\mathcal{F}_h((-x)^\alpha \varphi)$ and $\mathcal{F}_h((hD_x)^\alpha \varphi)=p^\alpha \mathcal{F}_h(\varphi)$.

(iii) $||\varphi||=(2\pi h)^{-n/2}||\mathcal{F}_h(\varphi)||$.

\emph{Proof.} We have $(hD_p)^\alpha\mathcal{F}_h(\varphi)(f)=(hD_p)^\alpha\varphi(\mathcal{F}_h(f))=(-1)^{|\alpha|}\varphi((hD_p)^\alpha\mathcal{F}_h(f))=\varphi(\mathcal{F}_h((-x)^\alpha f))=\mathcal{F}_h((-x)^\alpha \varphi(f))$.
The other statements follow similarly. $\hfill \blacksquare$ 
\bb
\textbf{Theorem \cc} (\emph{generalized uncertainty principle}, \cite{Mar1,Zwo1}) For $j=1,...,n$ and $f\in \mathcal{S}'$,\index{Heisenberg uncertainty principle}
$$\frac{h}{2}||f||\cdot ||\mathcal{F}_h(f)||\leq ||x_jf||\cdot ||p_j\mathcal{F}_h(f)||.$$

\emph{Proof.} From Theorem 2.1.12 (i), we have $p_j\mathcal{F}_h(f)(p)=\mathcal{F}_h(hD_{x_j}f)$. We also have the following commutation relation: 
$$[x_j,hD_{x_j}]f=\frac{h}{i}[\langle x_j,\partial_{j}f\rangle-\partial_{j}(x_jf)]=ihf.$$ 
Rewriting the right hand side of the equality we wish to prove, we observe that
$||x_j f||\cdot ||p_j \mathcal{F}_h (f)||=||x_j f||\cdot ||\mathcal{F}_h(hD_{x_j}f)||=(2\pi h)^{n/2}||x_jf||\cdot ||hD_{x_j}f||$. But
$$(2\pi h)^{n/2}||x_jf||\cdot ||hD_{x_j}f||\geq (2\pi h)^{n/2}|\langle hD_{x_j}f,x_jf\rangle|\geq (2\pi h)^{n/2}|\text{Im}\langle hD_{x_j}f,x_jf\rangle|,$$
and we can rewrite this with the commutation relation as $(2\pi h)^{n/2}|\text{Im}\langle hD_{x_j}f,x_jf\rangle|=\frac{1}{2}{(2\pi h)^{n/2}}|\langle [x_j,hD_{x_j}]f,f\rangle|=\frac{1}{2}{(2\pi h)^{n/2}}h||f||^2=\frac{h}{2}||f||\cdot ||\mathcal{F}_h(f)||.$ $\hfill \blacksquare$
\bb
The foregoing theorem generalizes Example 2.1.10 to the $n$-dimensional semiclassical case: we can retrieve the former by taking $n=1$ and $h=1/2$. Suppose that, in general, we have a function $\psi\in L^2(\mathbb{R}^n)$ where $1=||\psi||={(2\pi h)^{-n/2}}||\mathcal{F}_h (\psi)||$. 
As above, the localization of $\psi$ relative to $x=0$ can be gauged by $||x_j \psi||$ for $j=1,...,n$. If for example we have 
$\psi(x)=h^{-|\rho|/2}\phi(x_1/h^{\rho_1},...,x_n/h^{\rho_n})$
for some $n$-tuple $\rho$ where $0\leq \rho_j\leq 1$, $|\rho|=\rho_1+...+\rho_n$,  $\phi \in \mathcal{S}$, and $||\phi||=1$, then $\psi$ is ``localized" in $x$ to the region $N_h(\epsilon):=[-h^{\rho_1-\epsilon},h^{\rho_1-\epsilon}]\times...\times [-h^{\rho_n-\epsilon},h^{\rho_n-\epsilon}]$. Namely, for any $\epsilon>0$,
$$\int_{\mathbb{R}^n- N_h(\epsilon)}|\psi(x)|^2dx=O(h^\infty)$$
and
$||x_j \psi||\simeq h^{\rho_j}$ for all $j$. On the other hand, the semiclassical Fourier transform gives us 
$\mathcal{F}_h(\psi)(p)=h^{|\rho|/2}\mathcal{F}(\psi)(p_1/h^{1-\rho_1},...,p_n/h^{1-\rho_n})$, which implies that ${(2\pi h)^{-n/2}}||p_j\mathcal{F}_h(\psi)||\simeq h^{1-\rho_j}$. We see again that localization in $x$ is matched by delocalization in $p$, and vice-versa. What is different about this semiclassical formulation is that we can also vary the parameter $h$ to attain any desired degree of localization.

\subsubsection{Quantization Procedures}

We are now ready to write down quantization formulas, which are equations involving modified semiclassical Fourier transforms that assign symbols (classical observables) to $h$-dependent linear operators (quantum observables) acting on functions $\varphi(x)\in \mathcal{S}(\mathbb{R}^n)$. \index{quantization}\index{quantization!Weyl quantization}\index{quantization!$t$-quantization}\index{quantization!left quantization}\index{quantization!right quantization}\index{symbol}
\bb
\textbf{Definition \cc} (\emph{symbols and quantization}) Let any function $a=a(x,p)\in \mathcal{S}(\mathbb{R}^{2n})$ be called a \emph{symbol}. The \emph{Weyl quantization} $Op^w:\mathcal{S}(\mathbb{R}^{2n})\to \text{Hom}(\mathcal{S}(\mathbb{R}^n))$ of $a$ is given by
$$Op^w(a)(\varphi)(x)=(2\pi h)^{-n}\iint_{\mathbb{R}^{n}\times \mathbb{R}^n} e^{\frac{i}{h}\langle x-y,p\rangle}a\left(\frac{x+y}{2},p\right)\varphi(y)dydp,$$
where $\varphi\in \mathcal{S}(\mathbb{R}^n)$. 
In general, if $0\leq t\leq 1$, then the \emph{$t$-quantization} $Op_t$ is given by
$$Op_t(a)(\varphi)(x)=(2\pi h)^{-n}\iint_{\mathbb{R}^{n}\times \mathbb{R}^n}e^{\frac{i}{h}\langle x-y,p\rangle}a(tx+(1-t)y,p)\varphi(y)dydp.$$
Note that $Op^w(a)=Op_{{1}/{2}}(a)$, and the \emph{left and right quantizations} are given by $Op^l(a)=Op_1(a)$ and $Op^r(a)=Op_0(a)$, respectively. The left quantization $Op^l(a)$ is oftentimes referred to as the \emph{standard quantization}. Any operator of the form $Op_t(a)$ is called a \emph{semiclassical pseudodifferential operator}, and to show its dependence on both $x$ and $hD$ we will oftentimes write $Op_t(a)(x,hD)$. \index{pseudodifferential operator}
\bb
We see from the definition above that the Weyl quantization ``splits the difference" between the right and left quantizations by virtue of being defined as $Op_{1/2}(a)$. Although the left (standard) quantization is simpler to calculate since it can be rewritten with the semiclassical Fourier transform as $Op^l(a)(\varphi)(y)=\mathcal{F}_h^{-1}(a(x,y)\mathcal{F}_h\varphi(y))$, we will work predominantly with the Weyl quantization $Op^w$ since it has many useful properties. For example, $Op^w$ sends real-valued functions to symmetric operators. If $a$ is real-valued, then $\langle Op^w(a)(\psi_1),\psi_2\rangle=\langle \psi_1,Op^w(a)(\psi_2)\rangle$ because
$$\int Op^w(a)(\psi_1)(x)\overline{\psi_2(x)}dx=\iiint e^{\frac{i}{h}\langle x-y,p\rangle}a\left(\frac{x+y}{2},p\right)\psi_1(y)\overline{\psi_2(x)}dxdydp$$
$$\iiint e^{\frac{i}{h}\langle y-x,p\rangle}\overline{a\left(\frac{x+y}{2},p\right)}\psi_1(y)\overline{\psi_2(x)}dxdydp=\int \psi_1(y)\overline{Op^w(a)(\psi_2)(y)}dy.$$
\vspace{5pt}
\\
We now exhibit several examples of symbol quantization. 
\bb
\textbf{Example \cc} (\emph{quantizing a $p$-dependent symbol}) If $a(x,p)=p^\alpha$ for a multiindex $\alpha\in \mathbb{N}^n$, then 
$Op_t(a)(\varphi)(x)=a(x,hD)\varphi(x)=(hD)^\alpha \varphi(x)$, where again $hD$ is a semiclassical scaling of the usual differential operator $D^\alpha:=i^{-|\alpha|}\partial^\alpha$. 
Furthermore, if $a(x,p)=\sum_{|\alpha|\leq N}a_\alpha(x)p^\alpha$, then clearly $a(x,hD)\varphi(x)=\sum_{|\alpha|\leq N}a_\alpha(x)(hD)^\alpha \varphi(x)$. Thus we see why the operators created by quantization maps are called ``pseudodifferential": if the symbol $a$ is a polynomial in $p$, then we obtain a ``normal" differential operator. 
\bb
\textbf{Example \cc} (\emph{quantizing an inner product}) If $a(x,p)=\langle x,p\rangle$, then by definition we have $Op_t(a)(\varphi)=(1-t)\langle hD,x\varphi\rangle+t\langle x,hD\varphi\rangle$. In particular, $Op^w(a)(\varphi)=\frac{h}{2}(\langle D,x\varphi\rangle+\langle x,D\varphi\rangle)$. 
\bb
\textbf{Example \cc} (\emph{quantizing an $x$-dependent symbol}, \cite{Zwo1}) If $a(x,p)=a(x)$, then $Op_t(a)(\varphi)=a\varphi$. To see this, we take the derivative with respect to $t$ of $Op_t(a)(\varphi)$:
\begin{eqnarray*}
\partial_tOp_t(a)(\varphi)&=&(2\pi h)^{-n}\iint_{\mathbb{R}^n\times \mathbb{R}^n}e^{\frac{i}{h}\langle x-y,p\rangle}\langle \partial_t a(tx+(1-t)y),x-y\rangle\varphi(y)dydp\\
&=&\frac{h}{i}(2\pi h)^{-n}\int_{\mathbb{R}^n}\div_p\left(\int_{\mathbb{R}^n}e^{\frac{i}{h}\langle x-y,p\rangle}\partial_t a(tx+(1-t)y)\varphi(y)dy\right)dp\\
&=&\frac{h}{i}(2\pi h)^{-n}\int_{\mathbb{R}^n}\div_p\left(e^{\frac{i}{h}\langle x,p\rangle}\mathcal{F}(\psi(p))\right)dp,
\end{eqnarray*}
where $\psi(y):=\partial_t a(tx+(1-t)y)\varphi(y)$. The last expression vanishes by rapid decay ($\mathcal{F}(\psi)(p)\to 0$ as $|p|\to \infty$), so indeed $Op_t(a)\varphi$ does not depend on $t$ and $Op_t(a)\varphi=Op_1(a)\varphi=a\varphi$ for all $0\leq t\leq 1$. 
\bb
\textbf{Example \cc} (\emph{quantizing a linear symbol}, \cite{Mar1}) Let $a(x,p)=\langle x,x^*\rangle+\langle p,p^*\rangle$, where $(x^*,p^*)\in \mathbb{R}^{2n}$. Then, from the derivations above, $Op_t(a)=\langle x,x^*\rangle+\langle hD,p^*\rangle$ for all $0\leq t\leq 1$. We call $a$ a \emph{linear symbol}, and identify it with the point $(x^*,p^*)\in \mathbb{R}^{2n}$. 
\bb
We conclude this brief section by stating several theorems that will be helpful in \S2.2. The first  theorem, whose proof is omitted, tells us how quantization transforms the Schwartz space and the space of tempered distributions. 
\bb
\textbf{Theorem \cc} (\emph{properties of quantization})

(i) If $a\in \mathcal{S}(\mathbb{R}^{2n})$, then $Op_t(a)$ is a continuous map from $\mathcal{S}'(\mathbb{R}^{n})\to \mathcal{S}(\mathbb{R}^n)\ \forall t\in [0,1]$. 

(ii) If $a\in \mathcal{S}'(\mathbb{R}^{2n})$, then $Op_t(a)$ is a continuous map from $\mathcal{S}(\mathbb{R}^n)\to \mathcal{S}'(\mathbb{R}^n)\ \forall t\in [0,1]$. 

(iii) If $a\in \mathcal{S}(\mathbb{R}^{2n})$, then the adjoint of $Op_t(a)$ is $Op_{1-t}(\overline{a})$, and in particular the Weyl quantization of a real symbol is self-adjoint. 
\bb
\textbf{Theorem \cc} (\emph{relation of quantization to commutators}, \cite{Gui1,Zwo1}) We have 

(i) $Op^w(D_{x_j}a)=[D_{x_j},Op^w(a)]$

(ii) $hOp^w(D_{p_j}a)=-[x_j,Op^w(a)]$ 

\emph{Proof.} Let $\varphi \in \mathcal{S}$. Then \vspace{-10pt}
\begin{eqnarray*}Op^w(D_{x_j}a)(\varphi)&=&(2\pi h)^{-n}\iint_{\mathbb{R}^n\times \mathbb{R}^n} D_{x_j}a\left(\frac{x+y}{2},p\right)e^{\frac{i}{h}\langle x-y,p\rangle}\varphi(y)dpdy\\
&=&(2\pi h)^{-n}\iint_{\mathbb{R}^n\times \mathbb{R}^n}(D_{x_j}+D_{y_j})\ a\left(\frac{x+y}{2},p\right)e^{\frac{i}{h}\langle x-y,p\rangle}\varphi(y)dpdy\\
&=& (2\pi h)^{-1}\left( \iint_{\mathbb{R}^n\times \mathbb{R}^n}D_{x_j}a\left(\frac{x+y}{2},p\right)e^{\frac{i}{h}\langle x-y,p\rangle}\varphi(y)dpdy\right.\\
&& \left.+\iint_{\mathbb{R}^n\times \mathbb{R}^n}a\left(\frac{x+y}{2},p\right)e^{\frac{i}{h}\langle x-y,p\rangle}\left(\frac{p_j}{h}-D_{y_j}\right)\varphi(y)dpdy\right)\\
& = & D_{x_j}Op^w(a)(\varphi)-Op^w(a)(D_{x_j}\varphi)=[D_{x_j},Op^w(a)]\varphi.
\end{eqnarray*}
Assertion (ii) follows similarly. $\hfill \blacksquare$
\bb
\textbf{Theorem \cc} (\emph{conjugation by the semiclassical Fourier transform}) We have 
$$\mathcal{F}_h^{-1}Op^w(a)(x,hD)\mathcal{F}_h=Op^w(a)(hD,-x).$$
Since the proof of this last theorem follows by the definitions of $Op^w$ and $\mathcal{F}_h$, we will not write it out explicitly.

\subsection{Pseudodifferential Operators and Symbols}
\setcounter{itemcounter}{1}

Having seen the motivation and definition of quantization in \S2.1, we proceed to understand the analytic and algebraic characteristics of the resulting semiclassical $\psi$DOs. 

\subsubsection{Semiclassical Pseudodifferential Operators and their Algebra}

For simplicity, we shall deal only with the Weyl quantization in this section. Let us consider the equation 
$$Op^w(a)Op^w(b)=Op^w(c),$$
where $a,b,$ and $c$ are symbols. We want to know under which conditions this holds, and how to compute the symbol $c:=a\# b$ for the \emph{Weyl product operator} $\#$. The general procedure for answering this question involves writing the Weyl quantization of an arbitrary symbol as an expression in the quantizations of complex exponentials of linear symbols. Recall that linear symbols take on the form \index{Weyl product operator}
$l(x,p):=\langle x,x^*\rangle + \langle p,p^*\rangle$, where $(x^*,p^*)\in \mathbb{R}^{2n}$, and that we can identify the symbol $l$ with the point $(x^*,p^*)\in \mathbb{R}^{2n}$. 

We first require two lemmas, one of which deals with quantizating the complex exponentials of linear symbols and the other of which relates the Weyl quantization to the Fourier transform. 
\bb
\textbf{Lemma \cc} (\emph{quantizing an exponential of a linear symbol}, \cite{Zwo1}) Let $l(x,p)=\langle x,x^*\rangle + \langle p,p^*\rangle$ be a linear symbol. If $a(x,p)=e^{\frac{i}{h}l(x,p)}$, then 
$$Op^w(a)(x,hD)=e^{\frac{i}{h}l(x,hD)},$$
where
$l(x,hD)=Op^w(l)(x,hD)=\langle x,x^*\rangle+\langle hD,p^*\rangle$
and
$e^{\frac{i}{h}l(x,hD)}\varphi(x):=e^{\frac{i}{h}\langle x,x^*\rangle}+e^{\frac{i}{2h}\langle x^*,p^*\rangle}\varphi(x+p^*).$
Furthermore, if $l$ and $m$ are both linear symbols (identified as points on $\mathbb{R}^{2n}$), then
$$e^{\frac{i}{h}l(x,hD)}e^{\frac{i}{h}m(x,hD)}=e^{\frac{i}{2h}\sigma(l,m)}e^{\frac{i}{h}(l+m)(x,hD)},$$
for $\sigma((x,p),(y,q))=\langle p,y\rangle-\langle x,q\rangle$ being the symplectic form on $\mathbb{R}^n\times \mathbb{R}^n$.

\emph{Proof.} Let us consider the PDE with boundary condition
$$\begin{cases}
ih\partial_tv+l(x,hD)v(x,t)=0\\
v(x,t=0)=u(x),
\end{cases}$$
for $u\in \mathcal{S}$ and $t\in \mathbb{R}$. Its unique solution is given by $v(x,t)=e^{\frac{it}{h}l(x,hD)}u$ for $t\in \mathbb{R}$, while the equation above defines the operator $e^{\frac{it}{h}l(x,hD)}$ (whose action on $u$ is given by the time-evolution of the PDE above). Now if $l(x,hD)=Op^w(l)(x,hD)=\langle x,x^*\rangle+\langle hD,p^*\rangle$, then it follows that
$$v(x,t)=e^{\frac{it}{h}\langle x,x^*\rangle+\frac{it^2}{2h}\langle x^*,p^*\rangle}u(x+tp^*),$$
which gives $e^{\frac{i}{h}l(x,hD)}\varphi(x)=e^{\frac{i}{h}\langle x,x^*\rangle}+e^{\frac{i}{2h}\langle x^*,p^*\rangle}\varphi(x+p^*).$ We can then compute
\begin{eqnarray*}
Op^w(e^{\frac{i}{h}l})(u) & =& (2\pi h)^{-n}\iint_{\mathbb{R}^n\times\mathbb{R}^n} e^{\frac{i}{h}\langle x-y,p\rangle}e^{\frac{i}{h}(\langle p,p^*\rangle+\langle \frac{x+y}{2},x^*\rangle)}u(y)dydp\\
&=&(2\pi h)^{-n}{e^{\frac{i}{2h}\langle x,x^*\rangle}}\iint_{\mathbb{R}^n\times \mathbb{R}^n}e^{\frac{i}{h}\langle x-y+p^*,p\rangle}e^{\frac{i}{2h}\langle x^*,y\rangle}u(y)dydp\\
&=&(2\pi h)^{-n}e^{\frac{i}{2h}\langle x,x^*\rangle}\iint_{\mathbb{R}^n\times \mathbb{R}^n}e^{\frac{i}{h}\langle x-y,p\rangle}e^{\frac{i}{2h}\langle x^*,y+p^*\rangle}u(y+p^*)dydp\\
&=&e^{\frac{i}{h}\langle x,x^*\rangle+\frac{i}{2h}\langle x^*,p^*\rangle}u(x+p^*),
\end{eqnarray*}
since rescaling the Fourier inversion formula applied to Example 2.1.8 gives $\delta_{xy}=(2\pi h)^{-n}\\\int_{\mathbb{R}^n}e^{\frac{i}{h}\langle x-y,p\rangle}dp$ in $\mathcal{S}'$, the space of tempered distributions. Thus we have our first identity
$$Op^w(e^{\frac{i}{h}l})(x,hD)=e^{\frac{i}{h}l(x,hD)}.$$
Now suppose $l(x,p)=\langle x,x^*\rangle+\langle p,p^*\rangle$ and $m(y,q)=\langle y,y^*\rangle+\langle q,q^*\rangle$. From the equation above, we have
$e^{\frac{i}{h}m(x,hD)}u(x)=e^{\frac{i}{h}\langle x,y^*\rangle+\frac{i}{2h}\langle y^*,q^*\rangle}u(x+q^*),$ 
which implies that
$$e^{\frac{i}{h}l(x,hD)}e^{\frac{i}{h}m(x,hD)}u(x)=e^{\frac{i}{h}\langle x,x^*\rangle+\frac{i}{2h}\langle x^*,p^*\rangle}e^{\frac{i}{h}\langle y^*,x+p^*\rangle+\frac{i}{2h}\langle y^*,q^*\rangle}u(x+p^*+q^*).$$
Since
$e^{{\frac{i}{h}(l+m)(x,hD)}}u(x)=e^{\frac{i}{h}\langle x^*+y^*,x\rangle+\frac{i}{2h}\langle x^*+y^*,p^*+q^*\rangle}u(x+p^*+q^*),$
we have
$$e^{\frac{i}{h}(l+m)(x,hD)}u(x)=e^{\frac{i}{2h}(\langle x^*,q^*\rangle-\langle y^*,p^*\rangle)}e^{\frac{i}{h}l(x,hD)}e^{\frac{i}{h}m(x,hD)}u(x).$$
This gives us the desired equation $e^{\frac{i}{h}l(x,hD)}e^{\frac{i}{h}m(x,hD)}=e^{\frac{i}{2h}\sigma(l,m)}e^{\frac{i}{h}(l+m)(x,hD)}.$ $\hfill \blacksquare$
\bb
Thus, the lemma above tells us that the Weyl quantization of an exponential of a linear symbol is itself an exponential of the same linear symbol, with the difference that $p$ is converted into the differential operator $hD$ in the exponential. We also need the following:
\bb
\textbf{Lemma \cc} (\emph{Fourier decomposition of $Op^w(a)$}) Let us write
$$\mathcal{F}(a)(l):=\iint_{\mathbb{R}^n\times \mathbb{R}^n} e^{-\frac{i}{h}l(x,p)}\ a(x,p)dxdp,$$
where $a\in \mathcal{S}$ and $l(x,p)=\langle x,x^*\rangle + \langle p,p^*\rangle$ is a linear symbol, identified as a point $(x^*,p^*)\in \mathbb{R}^{2n}$. The following decomposition formula for $Op^w(a)$ holds:
$$Op^w(a)(x,hD)=(2\pi h)^{-2n}\iint_{\mathbb{R}^n\times\mathbb{R}^n}\mathcal{F}(a)(l)e^{\frac{i}{h}l(x,hD)}dl.$$

\emph{Proof.} From the Fourier inversion formula, we have 
$$a(x,p)=(2\pi h)^{-2n}\iint_{\mathbb{R}^n\times \mathbb{R}^n}e^{\frac{i}{h}l(x,p)} \mathcal{F}(a)(l)dl,$$
and applying the previous lemma gives the result. 
$\hfill \blacksquare$
\bb
We are now ready to address the problem at the beginning of this section. The following theorem, proved using the lemmas above, shows that the product of two pseudodifferential operators is a pseudodifferential operator. This fact implies that these operators form a commutative algebra, similar to how the algebra of classical observables is also commutative. 
\bb
\textbf{Theorem \cc} (\emph{quantization composition theorem}, \cite{Dim1}) If $a,b\in \mathcal{S}$, then
$Op^w(a)Op^w(b)=Op(a\# b)^w,$\index{quantization!composition theorem}
where
$$(a\# b)(x,p):=e^{ihA(D)}(a(x,p)b(y,q))|_{y=x,q=p}$$
for $A(D):=\frac{1}{2}\sigma(D_x,D_p,D_y,D_q)$, and $\sigma((x,p),(y,q))=\langle p,y\rangle-\langle x,q\rangle$ being the symplectic form on $\mathbb{R}^n\times \mathbb{R}^n$.

\emph{Proof.} Let $m$ and $l$ be linear symbols. Using the Fourier decomposition formula, we have:
\vspace{-10pt}
\begin{eqnarray*}
Op^w(a)(x,hD)&=&(2\pi h)^{-2n}\iint_{\mathbb{R}^n\times\mathbb{R}^n}\hat{a}(l)e^{\frac{i}{h}l(x,hD)}dl,\\
Op^w(b)(x,hD)&=&(2\pi h)^{-2n}\iint_{\mathbb{R}^n\times\mathbb{R}^n}\hat{b}(m)e^{\frac{i}{h}m(x,hD)}dm.
\end{eqnarray*}
Then, according to Lemma 2.2.1, 
$$
Op^w(a)(x,hD)Op^w(b)(x,hD)=(2\pi h)^{-4n}\iint_{\mathbb{R}^{2n}\times \mathbb{R}^{2n}} \hat{a}(l)\hat{b}(m)e^{\frac{i}{h}l(x,hD)}e^{\frac{i}{h}m(x,hD)}dmdl$$
$$=\frac{1}{(2\pi h)^{4n}}\iint_{\mathbb{R}^{2n}\times \mathbb{R}^{2n}} \hat{a}(l)\hat{b}(m)e^{\frac{i}{2h}\sigma(l,m)}e^{\frac{i}{h}(l+m)(x,hD)}dmdl=\frac{1}{(2\pi h)^{2n}}\int_{\mathbb{R}^{2n}}\hat{\varphi}_1(r)e^{\frac{i}{h}r(x,hD)}dr,
$$
where
$\hat{\varphi}_1(r):=(2\pi h)^{-2n}\int_{l+m=r}\hat{a}(l)\hat{b}(m)e^{\frac{i}{2h}\sigma(l,m)}dl$
is obtained from a change of variables setting $r=m+l$. Let us now show that $\varphi_1=\varphi=a\#b$ as defined above. Rewriting with $z=(x,p), w=(y,q)$, we have
$\varphi(z)=e^{\frac{ih}{2}\sigma(D_z,D_w)}a(z)b(w)|_{w=z}=e^{\frac{i}{2h}\sigma(hD_z,hD_w)}a(z)b(w)|_{w=z},$
and
$$a(z)=(2\pi h)^{-2n}\int_{\mathbb{R}^{2n}}e^{\frac{i}{h}l(z)}\hat{a}(l)dl,\ \ \  b(w)=(2\pi h)^{-2n}\int_{\mathbb{R}^{2n}}e^{\frac{i}{h}m(w)}\hat{b}(m)dm.$$
Since $l(z)=\langle l,z\rangle$ and $m(w)=\langle m,w\rangle$, we have 
$\exp({\frac{i}{2h}\sigma(hD_z,hD_w)})\exp({\frac{i}{h}(l(z)+m(w))})=\exp({\frac{i}{h}(l(z)+m(w))+\frac{i}{2h}\sigma(l,m)}),$
which implies that 
\begin{eqnarray*}
\varphi(z)&=&(2\pi h)^{-4n}\iint_{\mathbb{R}^{2n}\times\mathbb{R}^{2n}} e^{\frac{i}{2h}\sigma(hD_z,hD_w)}\left.e^{\frac{i}{h}(l(z)+m(w))}\right|_{z=w}\hat{a}(l)\hat{b}(m)dldm\\
&=&(2\pi h)^{-4n}\iint_{\mathbb{R}^{2n}\times\mathbb{R}^{2n}} e^{\frac{i}{h}(l(z)+m(z))+\frac{i}{2h}\sigma(l,m)}\hat{a}(l)\hat{b}(m)dldm.
\end{eqnarray*}
Taking the semiclassical Fourier transform of $\varphi$ yields
\begin{eqnarray*}
\mathcal{F}_h(\varphi)&=&(2\pi h)^{-2n}\iint_{\mathbb{R}^{2n}\times \mathbb{R}^{2n}}(2\pi h)^{-2n}\left(\int_{\mathbb{R}^n}e^{\frac{i}{h}(l+m-r)(z)}dz\right) e^{\frac{i}{2h}\sigma(l,m)}\hat{a}(l)\hat{b}(m)dldm\\
&=&(2\pi h)^{2n}\int_{l+m=r}e^{\frac{i}{2h}\sigma(l,m)}\hat{a}(l)\hat{b}(m)dl=\hat{\varphi}_1(r),
\end{eqnarray*}
where the penultimate equality follows since $\delta({l+m}=r)\in \mathcal{S}'$ is the term inside the parentheses. Thus we have $\varphi_1=\varphi=a\# b$ as given. $\hfill \blacksquare$
\bb
The Weyl product $a\# b$ of two symbols also admits an integral representation, as given in Theorem 2.2.5. Before discussing this, however, we require the following lemma:
\bb
\textbf{Lemma \cc} (\emph{quantizing exponentials of quadratic forms}, \cite{Zwo1}) Let $Q$ be an invertible and symmetric $n\times n$ matrix. 
\\
(i) If $a=a(x)\in \mathcal{S}(\mathbb{R}^n)$, then
$$e^{\frac{ih}{2}\langle QD,D\rangle}a(x)=\frac{|\det Q|^{-1/2}}{(2\pi h)^{n/2}}e^{\frac{i\pi}{4} \text{sgn}(Q)}\int_{\mathbb{R}^n}e^{-\frac{i}{2h}\langle Q^{-1}y,y\rangle}a(x+y)dy.$$
(ii) If $a=a(x,y)\in \mathcal{S}(\mathbb{R}^{2n})$, then 
$$e^{ih \langle D_x,D_y\rangle}a(x,y)=(2\pi h)^{-n}\iint_{\mathbb{R}^n\times\mathbb{R}^n} e^{-\frac{i}{h}\langle x_1,y_1\rangle}a(x+x_1,y+y_1)dx_1dy_1.$$
(iii) If $a=a(z,w)\in \mathcal{S}(\mathbb{R}^{4n})$, then 
$$e^{ih\sigma(D_z,D_w)}a(z,w)=(2\pi h)^{-2n}\iint_{\mathbb{R}^{2n}\times \mathbb{R}^{2n}}e^{-\frac{i}{h}\sigma(z_1,w_1)}a(z+z_1,w+w_1)dz_1dw_1.$$

\emph{Proof.} From Proposition 2.1.9, we have 
$$\mathcal{F}(e^{\frac{i}{2}\langle Qx,x\rangle})={(2\pi)^{n/2}e^{\frac{i\pi}{4}\text{sgn}(Q)}}{|\det Q|^{-1/2}}e^{-\frac{i}{2}\langle Q^{-1}p,p\rangle}.$$
In the semiclassical case, 
$$(2\pi h)^{-n}\int_{\mathbb{R}^n} e^{\frac{i}{h}\langle w,p\rangle} e^{\frac{i}{2h}\langle Qp,p\rangle}dp=\mathcal{F}^{-1}_h(e^{\frac{i}{2h}\langle Qp,p\rangle})(w)=\frac{|\det Q|^{-1/2}}{(2\pi h)^{n/2}}e^{\frac{i\pi}{4}\text{sgn}(Q)}e^{-\frac{i}{2h}\langle Q^{-1}w,w\rangle}.$$
We see from a brief computation that
$$
e^{\frac{ih}{2}\langle QD,D\rangle}f(x)=e^{\frac{i}{2h}\langle QhD,hD\rangle}f(x)=(2\pi h)^{-n}\iint_{\mathbb{R}^n\times \mathbb{R}^n}e^{\frac{i}{h}\langle x-y,p\rangle}e^{\frac{i}{2h}\langle Qp,p\rangle}f(y)dydp$$$$
=Ce^{\frac{i\pi}{4} \text{sgn}(Q)}\int_{\mathbb{R}^n}e^{-\frac{i}{2h}\langle Q^{-1}(x-y),x-y\rangle}f(y)dy=Ce^{\frac{i\pi}{4} \text{sgn}(Q)}\int_{\mathbb{R}^n}e^{-\frac{i}{2h}\langle Q^{-1}y,y\rangle}f(x+y)dy,
$$
with $C={|\det Q|^{-1/2}}{(2\pi h)^{-n/2}}$ as desired. This gives (i). For (ii) and (iii), let us write
$$A=\left(\begin{array}{cc} & I \\ I & \end{array}\right)\ \ \  \text{ and }\ \ \  B=\left(\begin{array}{cc} & -J \\ J & \end{array}\right),$$ 
where $I$ denotes the $n\times n$ identity matrix and $J$ is the $2n\times 2n$ complex structure. 
The argument for (ii) is similar to that of (i), with $2n$ instead of $n$ and $Q=A$. Here $Q$ is symmetric, $Q^{-1}=Q$, $|\det Q|=1$, $\text{sgn}(Q)=0$, and $Q(x,y)=(y,x)$, so that $\frac{1}{2}\langle Q(x,y),(x,y)\rangle=\langle x,y\rangle$. Since $D=(D_x,D_y)$, $\frac{1}{2}\langle Q^{-1}D,D\rangle=\langle D_x,D_y\rangle$ and (ii) follows. The argument for (iii) is the same, but with $4n$ instead of $n$ and 
$Q=B$. 
In particular, if $(z,w)\in \mathbb{R}^{4n}$ for $z=(x,p), w=(y,q)$, then $Q(z,w)=(-Jw,Jz)$. $Q$ is also symmetric, with the properties $Q^{-1}=Q,|\det Q|=1$, and $\text{sgn}(Q)=0$. Since $\frac{1}{2}\langle Q(z,w),(z,w)\rangle=\langle Jz,w\rangle=\sigma(z,w)$ and $D=(D_z,D_w)=(D_x,D_p,D_y,D_q)$, $\frac{1}{2}\langle Q^{-1}D,D\rangle=\sigma(D_x,D_p,D_y,D_q)$ and (iii) follows. $\hfill \blacksquare$
\bb
\textbf{Theorem \cc} (\emph{integral representation formula for composed symbols}) If $a,b\in \mathcal{S}$, then 
$$(a\# b)(x,p)=(\pi h)^{-2n} \iint_{\mathbb{R}^{2n}\times\mathbb{R}^{2n}} e^{-\frac{2i}{h}\sigma(w_1,w_2)}a(z+w_1)b(z+w_2)dw_1dw_2,$$
where $z=(x,p)$. 

\emph{Proof.} We apply (iii) in the lemma above, with $h/2$ instead of $h$. $\hfill \blacksquare$
\bb
From both Theorems 2.2.3 and 2.2.5, we see that 
\begin{eqnarray*}
(a\# b)(x,p)&:=&e^{ihA(D)}(a(x,p)b(y,q))|_{y=x,q=p}\\
&=&(\pi h)^{-2n} \iint_{\mathbb{R}^{2n}\times\mathbb{R}^{2n}} e^{-\frac{2i}{h}\sigma(w_1,w_2)}a(z+w_1)b(z+w_2)dw_1dw_2,
\end{eqnarray*}
for $z=(x,p)$, $A(D):=\frac{1}{2}\sigma(D_x,D_p,D_y,D_q)$, and $\sigma((x,p),(y,q))=\langle p,y\rangle-\langle x,q\rangle$. Since these expressions for $a\# b$ may be difficult to evaluate explicitly, it is fruitful to ask whether we can write down an approximation that is valid to any order. This is the content of the \emph{semiclassical expansion theorem}, whose statement below should come as no surprise. It is proved in Appendix II. 
\bb
\textbf{Theorem \cc} (\emph{semiclassical expansion} \cite{Mar1, Uri1,Zwo1}) Let $a,b\in \mathcal{S}$. Then for all $N$, 
$$a\#b(x,p)=\sum_{k=0}^N \frac{(ih)^k}{k!}A(D)^k(a(x,p)b(y,q))|_{y=x,q=p}+O_{\mathcal{S}}(h^{N+1}),$$
where $h\to 0$, $A(D)=\frac{1}{2}\sigma(D_x,D_p,D_y,D_q)$, and the notation $\varphi=O_\mathcal{S}(h^k)$ means that for all multiindices $\alpha$ and $\beta$, 
$|\varphi|_{\alpha,\beta}:=\sup_{\mathbb{R}^n}|x^\alpha \partial^\beta \varphi|\leq C(\alpha,\beta)h^k$ 
in the limit $h\to 0$. In particular, a first-order approximation of $a\# b$ is given by\index{semiclassical analysis!semiclassical expansion}
$$a\# b=ab+\frac{h}{2i}\{a,b\}+O_{\mathcal{S}}(h^2)$$
and 
$[Op^w(a)(x,hD),Op^w(b)(x,hD)]=\frac{h}{i}Op^w(\{a,b\})(x,hD)+O_\mathcal{S}(h^3).$
If $\text{supp}(a)\cap \text{supp}(b)=\emptyset$, then $a\# b=O_{\mathcal{S}}(h^\infty)$. 
\bb
Here we begin to recognize the significance of the classical-quantum correspondence: in the equation 
$[Op^w(a)(x,hD),Op^w(b)(x,hD)]=\frac{h}{i}Op^w(\{a,b\})(x,hD)+O_\mathcal{S}(h^3)$
above, we note that the commutator $[Op^w(a),Op^w(b)]$ of $a$ and $b$ relates to the Poisson bracket $\{a,b\}$. While the former is a quantum-mechanical construct, the latter is a classical one. We point out again that, by the middle equation, the Poisson bracket $\{a,b\}$ also factors into the first-order approximation of $a\# b$.

The tools that we have developed thus far will be generalized in the following section for \emph{symbol classes}, and used in \S2.4 to prove two essential prerequisites to the quantum ergodicity theorem. To summarize, in the current section we have defined quantization procedures, shown that the resulting quantized, pseudodifferential operators form a commutative algebra, and seen that any Weyl product can be arbitrarily approximated.

\subsubsection{Generalization to Symbol Classes}

It will oftentimes be helpful to categorize a symbol $a=a(x,p)$ into \emph{symbol classes}, which allows us to extend the symbol calculus to symbols that can depend on $h$ and have varying behavior as $(x,p)\to \infty$. The notion of symbol classes was first defined by H\"{o}rmander in analyzing PDEs and $\psi$DOs. The following presentation is adapted from \cite{Dim1} and \cite{Zwo1}, which present a simpler case of H\"{o}rmander's Weyl calculus \cite{Hor3}. \index{symbol!symbol class}

We only describe the basic definition of symbol classes for the purposes of this section, since what we have already proved for Schwartz functions will be enough to motivate the following proofs. 
We refer the reader to a more detailed treatment of symbol classes in \cite{Mar1} and \cite{Hor3}.
\bb
\textbf{Definition \cc} (\emph{order function}) A measurable function $m:\mathbb{R}^{2n}\to \mathbb{R}_{>0}$ is called an \emph{order function} if there are constants $C$ and $N$ such that
$m(w)\leq C\langle z-w\rangle^N m(z)$ for all $w,z\in \mathbb{R}^{2n}$, where $\langle z\rangle:=(1+|z|^2)^{1/2}$. \index{symbol!order function}
\bb
Trivial examples of order functions are $m(z)=1$ and $m(z)=\langle z\rangle$. It is also clear that $m(z)=\langle x\rangle^a\langle p\rangle^b$ are order functions for any $a,b\in \mathbb{R}$ and $z=(x,p)$. Finally, if $m_1,m_2$ are order functions, then by definition $m_1m_2$ is an order function as well. 
\bb
\textbf{Definition \cc} (\emph{symbol class}) Let $m(z)$ be an order function. The \emph{symbol class of $m(z)$} is given by
$S(m):=\{a\in C^\infty(\mathbb{R}^{2n}):\forall \alpha\in \mathbb{N}^{2n}, \exists C=C(\alpha)\in \mathbb{R}: |\partial^\alpha a|\leq C m\}.$
Likewise, for $0\leq \delta \leq 1/2$, we have the $(h,\delta)$-dependent symbol class
$$S_\delta(m):=\{a\in C^\infty(\mathbb{R}^{2n}):\forall \alpha \in \mathbb{N}^{2n}, \exists C=C(\alpha)\in \mathbb{R}: |\partial^\alpha a|\leq C h^{-\delta|\alpha|}m\}.$$
Note that $S_0(m)=S(m)$. These symbol classes provide the natural space in which an asymptotic symbol decomposition exists.
\bb
\textbf{Definition \cc} (\emph{asymptotic symbol decomposition}) Let $a_j\in S_\delta(m)$ for all $j\in \mathbb{N}$. $a\in S_\delta(m)$ is \emph{asymptotic} to $\sum_{j=0}^\infty h^ja_j$ if for any $N$, $a-\sum^{N-1}_{j=0}h^ja_j=O_{S_\delta(m)}(h^N)$, i.e.
$$\left|\partial^\alpha\left(a-\sum^{N-1}_{j=0}h^ja_j\right)\right|\leq C h^{N-\delta|\alpha|}m$$
for all multiindices $\alpha\in \mathbb{N}^{2n}$ and $C=C({\alpha,N})$. Note that, by itself, the formal series can diverge; the conditions above only stipulate that the expression $a-\sum_{j=0}^{N-1}h^ja_j$ and its derivatives vanish fast enough in the limit $h\to 0$. If the above holds, then we write $a\sim \sum_{j=0}^\infty h^ja_j$ and call $a_0$ the \emph{principal symbol} of the \emph{complete symbol} $a$. The notion of principal symbol will be useful in later sections. 
\bb
Borel's theorem, which we will not prove, assures us that we can always construct an asymptotic decomposition of symbols.
\bb
\textbf{Theorem \cc} (\emph{Borel}, \cite{Zwo1}) If $a_j\in S_\delta(m)$ for all $j\in \mathbb{N}$, there $\exists a\in S_\delta(m): a\sim\sum_{j=0}^\infty h^ja_j$
in $S_\delta(m)$. Furthermore, if $\mathcal{F}(a)\sim \sum_{j=0}^\infty h^ja_j$, then
$a-\mathcal{F}(a)=O_{S(m)}(h^\infty).$
\bb
One of the main benefits of extending our formulation to symbol classes is that all of our previous results are preserved. Noting that $\mathcal{S}(\mathbb{R}^{2n})\subset S(m)$ for any order function $m$, it can be shown that the Weyl quantization of symbols in $S_\delta(m)$ is also a continuous linear map $\mathcal{S}(\mathbb{R}^n)\to \mathcal{S}(\mathbb{R}^n)$ in the spirit of Theorem 2.1.19. 
\bb
\textbf{Theorem \cc} (\emph{properties of quantization for symbol classes}) If $a\in S_\delta(m)$ with $0\leq \delta \leq 1/2$, then $Op^w(a)(x,hD):\mathcal{S}(\mathbb{R}^n)\to \mathcal{S}(\mathbb{R}^n)$ and $Op^w(a)(x,hD):\mathcal{S}'(\mathbb{R}^n)\to\mathcal{S}'(\mathbb{R}^n)$ are continuous linear transformations. 
\bb
We also retain both the semiclassical expansion and the quantization composition theorems, which we will not prove in this generalized context:
\bb
\textbf{Theorem \cc} (\emph{semiclassical expansion for symbol classes}, \cite{Mar1,Zwo1}) Let $Q$ be an invertible, symmetric $n\times n$ matrix, and set $B(D)=\frac{1}{2}\langle QD,D\rangle$. If $0\leq \delta < 1/2$, then the operator on Schwartz spaces $\exp(ihB(D)):\mathcal{S}(\mathbb{R}^n)\to \mathcal{S}(\mathbb{R}^n)$ extends uniquely to an operator on symbol classes $\exp(ihB(D)):S_\delta(m)\to S_\delta(m)$, and\index{semiclassical analysis!semiclassical expansion}
$$e^{ihB(D)}a\sim \sum_{k=0}^\infty \frac{(ih)^k}{k!}(B(D))^ka$$\
for all $a\in S_\delta(m)$. 
\bb
\textbf{Theorem \cc} (\emph{quantization composition for symbol classes}, \cite{Dim1,Zwo1}) If $a\in S_\delta(m_1)$ and $b\in S_\delta(m_2)$ for $0\leq \delta < 1/2$, then $a\# b\in S_\delta(m_1m_2)$ and $Op^w(a)Op^w(b)=Op^w(a\# b)$. An approximation for $a\# b$ is given by
$$a\# b=ab+\frac{h}{2i}\{a,b\}+O_{S_\delta(m_1m_2)}(h^{1-2\delta}),$$
and furthermore we have another equation relating the commutator $[\cdot,\cdot]$ to the Poisson bracket $\{\cdot,\cdot\}$:
$[Op^w(a)(x,hD),Op^w(b)(x,hD)]=\frac{h}{i}Op^w(\{a,b\})(x,hD)+O_{S_\delta(m_1m_2)}(h^{3(1-2\delta)}).$\index{quantization!composition theorem}
\bb
Let us remind ourselves that in quantum mechanics, we are mainly concerned with the $L^2$ space of functions. It turns out that we can say even more about the Weyl quantization in this setting:  $Op^w(a)$ becomes a bounded operator, and is compact assuming certain decay conditions on the order function $m$. Recall that an operator $A:L^2\to L^2$ is said to be \emph{bounded} if there exists a $c\geq 0$ such that $||A(f)||\leq c||f||$ for all $f\in L^2$, where we set $||A||_{L^2\to L^2}:=\sup_{f\neq 0}\frac{||A(f)||}{||f||}$ (c.f. Appendix I). We end this section by stating the following:
\bb
\textbf{Theorem \cc} (\emph{$L^2$ boundedness and compactness for symbol classes}) If $a\in \mathcal{S}$, then $Op^w(a)(x,hD):L^2(\mathbb{R}^n)\to L^2(\mathbb{R}^n)$
is bounded independently of $h$. Moreover, if $a\in S_\delta(1)$ for $0\leq \delta \leq 1/2$, then $Op^w(a)(x,hD)$ is bounded with estimate 
$$||Op^w(a)(x,hD)||_{L^2\to L^2}\leq C\sum_{|\alpha|\leq Mn}h^{|\alpha|/2}\sup_{\mathbb{R}^n}|\partial^\alpha a|,$$
where $M$ and $C$ are constants. 
Finally, if $a\in S(m)$ and $\lim_{(x,p)\to \infty}m=0$, then $Op^w(a)(x,hD):L^2(\mathbb{R}^n)\to L^2(\mathbb{R}^n)$ is a compact operator.

\subsubsection{Inverses, Estimates, and G{\aa}rding's Inequality}

We now revisit our original goal of understanding the analytic and algebraic characteristics of semiclassically quantized $\psi$DOs given relevant information about the symbols. Our answer to this question comprises both a theorem about the inverses of quantized $\psi$DOs and a form of {G{\aa}rding's inequality}, which gives a lower bound for the bilinear form induced by the Weyl quantization of any symbol. This section will complete our exposition of the basic symbol calculus, after which we shall prove Weyl's law on $\mathbb{R}^n$ and extend the symbol calculus to any smooth manifold. 

To help us consider the inverses of Weyl-quantized operators, we first define the notion of {ellipticity} for symbols. As the proof of Theorem 2.2.16 shows, this condition is important for showing that an inverse to any given $\psi$DO exists.\index{symbol!ellipticity}
\bb
\textbf{Definition \cc} (\emph{elliptic symbols}) A symbol $a$ is \emph{elliptic}  in the symbol class $S(m)$ if there is a real number $c=c(a)>0$ such that $|a|\geq c m$. 
\bb
\textbf{Theorem \cc} (\emph{inverses for elliptic symbols}, \cite{Mar1,Zwo1}) Let $a\in S_\delta(m)$ for $0\leq \delta< 1/2$ be elliptic in $S(m)$. If $m\geq 1$, then there exist $h_0>0$ and $C>0$ such that 
$$||Op^w(a)(x,hD)(\varphi)||\geq C||\varphi||$$ 
for all $\varphi\in L^2(\mathbb{R}^n)$ 
and $0<h<h_0$. Furthermore, if $m=1$, then there exists some $h_0>0$ where $Op^w(a)(x,hD)^{-1}$ is a well-defined bounded linear operator on $L^2(\mathbb{R}^n)$ for $0<h\leq h_0$. 

\emph{Sketch of Proof.} We set $b:=1/a\in S_\delta(1/m)$, so that by Theorem 2.2.13 we have 
$$
\begin{cases}
a\# b=1+r_1 \\
b\# a=1+r_2,
\end{cases}$$ 
where $r_1,r_2\in h^{1-2\delta}S_\delta$ are remainder terms coming from the Weyl product. Quantizing each of these symbols gives $A=Op^w(a)(x,hD),B=Op^w(b)(x,hD),R_1=Op^w(r_1)(x,hD),R_2=Op^w(r_2)(x,hD)$, and 
$$\begin{cases}
AB=I+R_1\\
BA=I+R_2,
\end{cases}$$
with the condition that the operator norms of $R_1$ and $R_2$ decay with $h$. In particular, 
$$||R_1||_{L^2\to L^2}=O(h^{1-2\delta})\leq {1}/{2},\ \ \ \  ||R_2||_{L^2 \to L^2}=O(h^{1-2\delta})\leq {1}/{2}$$
for $0<h\leq h_0$ and some suitable $h_0$.  Now if $m\geq 1$, then for all $\varphi \in L^2(\mathbb{R}^n)$ there exists a constant $C$ such that
$$||\varphi||=||(I+R_2)^{-1}BA\varphi||\leq C||A\varphi||$$
by a combination of Theorem 2.2.14 and the fact that $b\in S(1/m)\subset S(1)$ is bounded on $L^2$. Finally, if $m=1$, then $B$ serves as both an approximate left and right inverse to $A$, and applying Theorem I.5 we conclude that the inverse $A^{-1}=Op^w(a)(x,hD)^{-1}$ exists for small enough $h$. $\hfill \Box$
\bb
We now obtain an analytic bound on the operator $Op^w(a)$ with the assumption that the corresponding symbol $a$ is nonnegative. This bound is given by both the weak and sharp versions of G{\aa}rding's inequality, for which we prove only the former:\index{G{\aa}rding's inequality}
\bb
\textbf{Theorem \cc} (\emph{weak G{\aa}rding's inequality}, \cite{Dim1}) Let $a\in S$, and suppose that there is a constant $c$ with $a\geq c>0$ on $\mathbb{R}^{2n}$. Then for all $\epsilon>0$, there exists a constant $h_0=h_0(\epsilon)>0$ such that
$$\langle Op^w(a)(x,hD)(\varphi),\varphi\rangle\geq (c-\epsilon)||\varphi||^2$$
for all $0<h\leq h_0$ and $\varphi\in L^2(\mathbb{R}^n)$.

\emph{Proof.} We first apply the previous theorem to see that the inverse $(a-\lambda)^{-1}$ is in $S$ for any $\lambda< c-\eps$. If $b:=(a-\lambda)^{-1}$, then by Theorem 2.2.13 we have 
$$(a-\lambda)\# b=1+\frac{h}{2i}\{a-\lambda,b\}+O_S(h^2)=1+O_S(h^2).$$
Quantizing this equation gives us 
$$(Op^w(a)(x,hD)-\lambda)(Op^w(b)(x,hD))=I+O_{L^2\to L^2}(h^2),$$
and we see that $Op^w(b)(x,hD)$ is an approximate right inverse of $Op^w(a)(x,hD)-\lambda$. Repeating this argument tells us that $Op^w(b)(x,hD)$ is an approximate left inverse as well, and so $Op^w(a)(x,hD)-\lambda$ is invertible for all $\lambda<c-\eps$. By the spectral theorem (Appendix I), $$\text{Spec}(Op^w(a)(x,hD)-\lambda)\subset [c-\eps,\infty).$$ It follows from Theorem I.2 that 
$\langle Op^w(a)(x,hD)(\varphi),\varphi\rangle\geq (c-\epsilon)||\varphi||^2,$
as desired. $\hfill \blacksquare$
\bb 
\textbf{Theorem \cc} (\emph{sharp G{\aa}rding's inequality}, \cite{Dim1}) If $a\in S$ and $a\geq 0$ on 
$\mathbb{R}^{2n}$, then there are constants $c\geq 0$ and $h_0>0$ such that
$$\langle Op^w(a)(x,hD)(\varphi),\varphi\rangle \geq -ch||\varphi||^2$$
for all $0<h\leq h_0$ and $\varphi\in L^2(\mathbb{R}^n)$. 
\bb
These variants of G{\aa}rding's inequality tell us that the bilinear form induced by the Weyl quantization of a nonnegative symbol as applied to a function is greater than the $L^2$-norm of that function. 
In particular, given that the symbol $a$ is nonnegative, the quantized operator $Op^w(a)$ is ``essentially positive." 
A stronger flavor of G{\aa}rding's inequality is the \emph{Fefferman-Phong inequality}, which was proved in 1979 \cite{Mar1}; nonetheless, we will use only the weak and sharp G{\aa}rding's inequality for remainder of this thesis.

Over the last two sections, we have extended our symbol calculus to symbol classes and derived inverses and analytic statements about Weyl-quantized $\psi$DOs. Our goal now is to prove \emph{Weyl's law} and \emph{Egorov's theorem} (c.f. \S1.1 and \S2.3), two assertions that will be useful for our proof of the \emph{quantum ergodicity theorem} in \S3.2. For ease of exposition, we will only prove Weyl's law in $\mathbb{R}^n$ before 
extending our symbol calculus to manifolds and proving Egorov's theorem in a more general context.

\subsection{Weyl's Law and Egorov's Theorem}
\setcounter{itemcounter}{1}

This section focuses singularly on the proofs of Weyl's law and Egorov's theorem, two insightful results that relate the semiclassical tools we have developed in \S2.2 back to the eigenvalue spacing statistics of the Laplacian and the correspondence between classical and quantum mechanics. These statements will also be useful for the proof of the quantum ergodicity theorem in \S3.

\subsubsection{Weyl's Law in $\mathbb{R}^n$}
\newcommand{\HH}{\Xi}

We begin with a real-valued \emph{potential function} $V\in C^\infty(\mathbb{R}^n)$ and define the \emph{Hamiltonian symbol}\index{Weyl's law}\index{Hamiltonian symbol}\index{Hamiltonian operator}
$\xi(x,p):=|p|^2+V(x)$
along with the corresponding \emph{Schr\"{o}dinger operator} in $n$ dimensions
$$\HH(h):=\HH(x,hD)=-h^2\Delta+V(x),$$
where $\Delta$ is the Laplacian on $\mathbb{R}^n$ as defined in \S1.2. Note that  $Op^w(\xi)(x,hD)=\HH(h)$. Our goal is to understand how properties of the symbol $\xi$ influence the asymptotic distribution of the eigenvalues of its quantization $\HH(h)$ as $h\to 0$. 

Let us first consider the potential $V(x)=x^2$ of a one-dimensional simple harmonic oscillator (SHO), so that we have $h=1$ and $\HH:=-\partial_x^2+x^2$. From elementary quantum mechanics, we know that the \emph{creation and annihilation operators} $A:=D_x+ix$ and $A^\dagger := D_x-ix$ have the property that\index{simple harmonic oscillator}
$A^*=A^\dagger, (A^\dagger)^*=A$, and
$\HH=AA^\dagger+1=A^\dagger A-1$ \cite{Gri1}.
Recall as well that we can solve for the eigenfunctions of the SHO with the Hermite polynomials $H_n(x):=(-1)^n \exp(x^2)\frac{d^n}{dx^n}\exp(-x^2)$. In particular, we have
\begin{enumerate}
\item $\langle \HH \psi,\psi\rangle\geq ||\psi||^2\Longrightarrow \HH\geq 1$ for all $\psi \in C_c^\infty(\mathbb{R}^n)$; \vspace{-5pt}
\item the function $\varphi_0(x):=\exp(-x^2/2)$ is an eigenfunction of $\HH$ corresponding to the smallest eigenvalue of $1$; \vspace{-5pt}
\item if $\varphi_n:= A^n\varphi_0$ for all $n\in \mathbb{N}$, then $\HH\varphi_n=(2n+1)\varphi_n$, and setting
$\psi_n:=\varphi_n/||\varphi_n||\in L^2(\mathbb{R}^n)$, 
we see that
$\psi_n(x)=H_n(x)\exp(-x^2/2)$; \vspace{-5pt}
\item $\langle \psi_n,\psi_m\rangle=\delta_n^m$, and the collection of eigenfunctions $\{\psi_n\}_{n=0}^\infty$ is complete in $L^2(\mathbb{R}^n)$. 
\end{enumerate}$\null$\vspace{-10pt}
\\
Generalizing this result to the case of an $n$-dimensional harmonic oscillator scaled by the semiclassical parameter $h$, we see that for 
$\HH(h):=-h^2\Delta+|x|^2$ and $\alpha \in \mathbb{N}^n$,
$$\psi_\alpha(x,h)=h^{-n/4}\prod_{i=1}^n H_{\alpha_i}(x_ih^{-1/2})\exp\left(-\frac{|x|^2}{2h}\right)$$
with corresponding eigenvalue $E_\alpha(n)=(2|\alpha|+n)h$. Thus the eigenvalue equation can be written in this context as 
$\HH(h)\psi_i(x,h)=E_i(h)\psi_i(x,h)$
after reindexing. 
\bb
\textbf{Theorem \cc} (\emph{Weyl's law for the SHO}, \cite{Zwo1}) For $0\leq a<b<\infty$, we have
$$\# \{E(h):a\leq E(h)\leq b\}=(2\pi h)^{-n} (|\{a\leq |x|^2+|p|^2\leq b\}|+o(1)),$$
where $E(h)$ is any eigenvalue of $\HH(h)$ and $|\{a\leq |x|^2+|p|^2\leq b\}|$ denotes the volume in $\mathbb{R}^{2n}$ of the set of $(x,p)$ where $a\leq |x|^2+|p|^2\leq b$. \index{Weyl's law!for the simple harmonic oscillator}

\emph{Proof.} We follow the exposition in \cite{Zwo1}. Without loss of generality, let $a=0$. Since $E(h)=(2|\alpha|+n)h$ for some multiindex $\alpha\in \mathbb{N}^n$,
$$\#\{E(h):0\leq E(h)\leq b\}=\# \{\alpha\in \mathbb{N}^n : 0\leq 2|\alpha|+n\leq b/h\}=\# \{ \alpha\in \mathbb{N}^n:|\alpha|\leq c\},$$
where $c:=(b-nh)/2h$. This implies that 
\begin{eqnarray*}
\# \{E(h):0\leq E(h)\leq b\}&=& |\{x\in \mathbb{R}^n:x_i\geq 0,\ 1\leq i\leq n\text{ and } x_1+...+x_n\leq c\}|+o(c^n)\\&=&(n!)^{-1}c^n+o(c^n),
\end{eqnarray*}
where the last equality holds as $c\to \infty$ since the volume of $\{x\in \mathbb{R}^n:x_i\geq 0,\  1\leq i\leq n\text{ and } x_1+...+x_n\leq 1\}$ is $(n!)^{-1}$. Thus $\# \{E(h):0\leq E(h)\leq b\}= \frac{b^n}{n!(2h)^n}+o(h^{-n})$
in the limit $h\to 0$. 

Now observe that $|\{|x|^2+|p|^2\leq b\}|=b^nV(2n)$, where $V(k):=\pi^{k/2}(\Gamma(k/2+1))^{-1}$ is the volume of the unit ball in $\mathbb{R}^k$. Since $V(2n)=\pi^n(n!)^{-1}$, we have 
$$\# \{E(h):0\leq E(h)\leq b\} = \frac{b^n}{n!(2h)^n}+o(h^{-n})=(2\pi h)^{-n}|\{|x|^2+|p|^2\leq b\}|+o(h^{-n}),\vspace{-5pt}$$
as desired. $\hfill \blacksquare$
\bb
Let us now use the previous result to prove Weyl's law in greater generality. Suppose that $V\in C^\infty(\mathbb{R}^n)$ satisfies
$$\begin{cases}
|\partial^\alpha V(x)|\leq C\langle x\rangle^k & \forall \alpha\in \mathbb{N}^n, C=C(\alpha)\in \mathbb{R}\\
V(x)\geq c\langle x\rangle^k & |x|\geq R,
\end{cases}$$
for positive constants $k,c,R\in \mathbb{R}$. If these properties hold, then we say that $V$ is an \emph{admissible potential function}. The only black box we will require in the proof of Weyl's law is the following proposition:
\bb
\textbf{Proposition \cc} (\emph{products of projection and quantized operators}) Let $a=a(x,p)$ be a symbol in $S$, and suppose that $\text{supp}(a)\subset \{|x|^2+|p|^2<R\}$ for some suitable $R\in \mathbb{R}$. With $\HH(h):L^2(\mathbb{R}^n)\to L^2(\mathbb{R}^n)$ given by $\HH(h):=-h^2\Delta+|x|^2$, let $\Pi$ denote the projection in $L^2$ onto $\text{span}\{\psi(x,h):\HH(h)\psi(x,h)=E(h)\psi(x,h), E(h)\leq R\}$. Then
$$Op^w(a)(x,hD)(I-\Pi)=O_{L^2\to L^2}(h^\infty)\text{ and }(I-\Pi)Op^w(a)(x,hD)=O_{L^2\to L^2}(h^\infty).$$
This tells us that there is no essential difference between an arbitrary function in $L^2(\mathbb{R}^n)$ and its projection onto the span of $L^2$ eigenfunctions of $\HH(h)$ for small $h$, at least in terms of their corresponding Weyl quantizations. For a proof of this proposition, we refer the reader to \cite{Dim1} and \cite{Zwo1}. 
\bb
\textbf{Theorem \cc} (\emph{Weyl's law for $\mathbb{R}^n$}) Suppose that $V$ is an admissible potential function, and that $E(h)$ denotes an arbitrary eigenvalue of the operator $\HH(h)=-h^2\Delta+V(x)$. Then
$$\# \{E(h):a\leq E(h)\leq b\}=(2\pi h)^{-n} (|\{a\leq |p|^2+V(x)\leq b\}|+o(1))$$
for all $a<b$ in the limit $h\to 0$. \index{Weyl's law!for $\mathbb{R}^n$}

\emph{Proof.} We again follow the proof in \cite{Zwo1}. Let $N(\lambda):=\# \{E(h):E(h)\leq \lambda\}$, and choose a $\chi \in C^\infty_c(\mathbb{R}^{2n})$ such that
$$\chi(x,p):=\begin{cases}
1 & \text{if } \xi(x,p)\leq \lambda+\eps \\
0 & \text{if } \xi(x,p)\geq \lambda+2\eps,
\end{cases}
$$ 
so that in the following $a$ will be elliptic. Indeed, for large enough $M$, we have the bound
$$a(x,p):=\xi(x,p)+M\chi(x,p)-\lambda\geq \gamma m,$$
where $m:=\langle x\rangle^m+\langle p\rangle^2$ and $\gamma=\gamma(\eps)>0$. This bound implies that $a$ is elliptic, and from Theorem 2.2.16 we know that $Op^w(a)(x,hD)$ is invertible for small enough $h$. 
\bb
\emph{Claim 1:} 
We first wish to show that
$$\langle (\HH(h)+MOp^w(\chi)-\lambda)\varphi,\varphi\rangle \geq \gamma ||\varphi||^2$$
for some $\gamma >0$ and all $\varphi \in \mathcal{H}(\mathbb{R}^n)$ where
$\mathcal{H}(\mathbb{R}^n):=\{\varphi \in \mathcal{S}':(I-h^2\Delta+\langle x\rangle^k)(\varphi)\in L^2(\mathbb{R}^n)\}.$
A straightforward application of Theorem I.6 tells us that $\mathcal{H}(\mathbb{R}^n)$ is in the domain of $\HH(h)$. To prove Claim 1, we start by taking some $b\in S(m^{1/2})$ where $b^2=a$. Then by Theorem 2.2.13, $b^2=b\# b+r_0$ where $r_0\in hS(m)$. By Theorem 2.2.16, the right inverse of $Op^w(b)(x,hD)$ exists and 
$$Op^w(b)^{-1}Op^w(r_0)Op^w(b)^{-1}=O_{L^2\to L^2}(h),$$
which implies that 
\begin{eqnarray*}
Op^w(a)&=&Op^w(b)Op^w(b)+Op^w(r_0)=Op^w(b)(1+Op^w(b)^{-1}Op^w(r_0)Op^w(b)^{-1})Op^w(b)\\
&=&Op^w(b)(1+O_{L^2\to L^2}(h))Op^w(b).
\end{eqnarray*}
Thus, for small enough $h>0$ we have 
\begin{eqnarray*}
\langle (\HH(h)+MOp^w(\chi)-\lambda)\varphi,\varphi\rangle &=& \langle Op^w(a)(x,hD)\varphi,\varphi\rangle \\ 
&\geq &||Op^w(b)(x,hD)(\varphi)||^2(1-O(h))\\
&\geq& \gamma ||\varphi||^2
\end{eqnarray*}
for some $\gamma>0$, again by Theorem 2.2.16. 
\bb
\emph{Claim 2:} We next claim that there exists a bounded linear operator $Q$ where
$$Op^w(\chi)(x,hD)=Q+O_{L^2\to L^2}(h^\infty)\ \text{ and }\ 
\text{rank}(Q)\leq (2\pi h)^{-n} (|\{\xi(x,p)\leq \lambda+2\eps \}|+\delta)$$
for all $\delta>0$. 
To prove this, we first cover the set $\{\xi(x,p)\leq \lambda+2\eps\}$ with a set of $N$ balls $\{B_j\}_{j=1}^N$ of radius $r_j^2$ centered at $(x_i,p_j)$, denoted by $B_j:=B((x_j,p_j),r_j^2)$. Note that this set of balls has the property that 
$\sum_{j=1}^N |B_j|\leq |\{ \xi(x,p)\leq \lambda+2\eps\}|+\frac{\delta}{2}$.

Next, we set a SHO shifted by $(x_j,p_j)$ as $\HH_j(h):=|hD_x-p_j|^2+|x-x_j|^2$, and
$\Pi$ as the orthogonal projection in $L^2$ onto 
$$V=\text{span}\{\varphi :\HH_j(h)\varphi=E_j(h)\varphi, E_j(h)\leq r_j\text{ for }1\leq j\leq N\}.$$ 
Observe that 
$$(I-\Pi)Op^w(\chi)(x,hD)=O_{L^2\to L^2}(h^\infty),$$
which is justified as follows: by setting 
$\chi = \sum_{j=1}^N \chi_j$ for $\text{supp}(\chi_j)\subset B((x_j,p_j),r_j^2)$ and $\Pi_j$ to be the orthogonal projection in $L^2$ onto the span of $\{u:\HH_j(h)u=E_j(h)u, E_j(h)\leq r_j\}$, we can apply Proposition 2.3.2 to see that
$(I-\Pi)Op^w(\chi)(x,hD)=O(h^\infty)$. Noting as well that $\Pi( \Pi_j)=\Pi_j$, we have
\begin{eqnarray*}
(I-\Pi)Op^w(\chi)=\sum_{j=1}^N (I-\Pi)Op^w(\chi_j)=\sum_{j=1}^N (I-\Pi)(I-\Pi_j)Op^w(\chi_j)=O_{L^2\to L^2}(h^\infty).
\end{eqnarray*}
It follows that
$$Op^w(\chi)(x,hD)=\Pi Op^w(\chi)(x,hD)+(I-\Pi)Op^w(\chi)(x,hD)=Q+O_{L^2\to L^2}(h^\infty)$$
where the bounded linear operator we desire is defined as $Q:=\Pi Op^w(\chi)(x,hD)$. To show that $Q$ is indeed bounded, we note that its rank is bounded. In particular,  
\begin{eqnarray*}
\dim (\text{im}(Q))&\leq& \dim( \text{im}(\Pi))\leq \sum_{j=1}^N \#\{E_j(h):E_j(h)\leq r_j\}=(2\pi h)^{-n}\left(\sum_{j=1}^N |B_j|+o(1)\right)\\
&\leq& (2\pi h)^{-n}(|\{p\leq \lambda+2\eps\}|+{\delta}/{2}+o(1)),
\end{eqnarray*}
where the penultimate inequality comes from Theorem 2.3.1 and the last inequality comes from the fact that $\sum_{j=1}^N |B_j|\leq |\{ \xi(x,p)\leq \lambda+2\eps\}|+\frac{\delta}{2}$. Thus we have proven Claim 2. 
\bb
Combining the two claims above, we see that
$$\langle \HH(h)\varphi,\varphi\rangle \geq (\lambda+C)||\varphi||^2-M\langle Q\varphi,\varphi \rangle-O(h^\infty)||\varphi||^2\geq \lambda||\varphi||^2-M\langle Q\varphi,\varphi\rangle,$$
where $Q$ is bounded as above. Applying Theorem I.7, we get 
$$N(\lambda)\leq (2\pi h)^{-n} (|\{\xi(x,p)\leq \lambda+2\eps \}|+\delta+o(1)),$$
and since this holds for all $\eps$ and $\delta>0$, as $h\to 0$ we have the bound 
$$N(\lambda)\leq (2\pi h)^{-n} (|\{\xi(x,p)\leq \lambda\}|+o(1)).$$
\emph{Claim 3:} 
We prove the opposite inequality. If $B_j\subset \{\xi(x,p)<\lambda\}$ and $V_j:=\text{span}\{\varphi:\HH_j(h)\varphi=E_j(h)\varphi, E_j(h)\leq r_j\}$, then for each $\varphi \in V_j$ we wish to show that
$$\langle \HH(h) \varphi,\varphi\rangle \leq (\lambda+\eps+O(h^\infty))||\varphi||^2.$$
To do this, we choose a symbol $a\in C_c^\infty(\mathbb{R}^{2n})$ with $a=1$ on $\{\xi(x,p)\leq \lambda\}$ and $\text{supp}(a)\subset \{\xi(x,p)\leq \lambda+\eps/2\}$. Setting $c:=1-a$, we see that   $$\varphi-Op^w(a)(x,hD)(\varphi)=Op^w(c)(x,hD)(\varphi)=O_{L^2}(h^\infty)$$ 
by Proposition 2.3.2, since $\text{supp}(c)\cap B_j=\emptyset$. 
Now let 
$$Op^w(b):=\HH(h)Op^w(a)(x,hD),$$ so that $\xi(x,p)\in S(m)$ and $a\in S(m^{-1})$. This implies that $b=\xi(x,p)a+O_{L^2}(h)\in S$, and it is also clear that $Op^w(b)$ is a bounded operator. Then $b\leq \lambda+\eps/2\Longrightarrow Op^w(b)(x,hD)\leq \lambda+3\eps/4\Longrightarrow$
$$\langle \HH(h)Op^w(a)(x,hD)(\varphi),\varphi\rangle = \langle Op^w(b)(x,hD)(\varphi),\varphi\rangle \leq (\lambda + {3\eps}/{4})||\varphi||^2,$$
and since $Op^w(a)(x,hD)(\varphi)=\varphi+O(h^\infty)$, we have 
$$\langle \HH(h)\varphi,\varphi\rangle \leq (\lambda+\eps+O(h^\infty))||\varphi||^2,$$
which proves Claim 3. 
\bb
To finish the proof of Theorem 2.3.3, we 
pick a set of disjoint balls $B_j\subset \{\xi(x,p)<\lambda\}$ such that
$|\{ \xi(x,p)<\lambda \}|\leq \sum_{j=1}^N |B_j|+\delta$, and 
set 
$V=V_1+...+V_N$, where as before $V=\text{span}\{\varphi :\HH_j(h)\varphi=E_j(h)\varphi, E_j(h)\leq r_j\text{ for }1\leq j\leq N\}$. 
Although $V_i$ and $V_j$ are not orthogonal for $i\neq j$, the disjointness of $B_i,B_j$ implies with Proposition 2.3.2 that 
$$\langle \varphi,\psi\rangle=O(h^\infty)||\varphi||||\psi||$$
for all $\varphi \in V_i$ and $\psi\in V_j$ with $i\neq j$. Furthermore, since each $V_j$ has an orthonormal basis of eigenvectors, Claim 3 holds for $\varphi \in V_j$, and the approximate orthogonality estimate above gives
$$\langle \HH\varphi,\varphi\rangle\leq (\lambda+\delta)||\varphi||^2$$
for all $\varphi \in V$. Thus, for small enough $h$, an application of Theorem 2.3.1 yields
\begin{eqnarray*} 
\dim V&=&\sum_{j=1}^N \dim V_j = \sum_{j=1}^N \# \{E_j(h)\leq r_j\}=(2\pi h)^{-n}\left(\sum_{j=1}^N |B_j|+o(1)\right)\\
&\geq &(2\pi h)^{-n}(|\{\xi(x,p)<\lambda\}|-\delta+o(1))
\end{eqnarray*} 
Applying Theorem I.7, we have 
$$N(\lambda)\geq (2\pi h)^{-n}(|\{ \xi(x,p)<\lambda \}|-\delta+o(1)),$$
which by the previous bound $N(\lambda)\leq (2\pi h)^{-n} (|\{\xi(x,p)\leq \lambda\}|+o(1))$ and the definition of $N(\lambda)$ implies that
$$\# \{E(h):a\leq E(h)\leq b\}=(2\pi h)^{-n} (|\{a\leq |p|^2+V(x)\leq b\}|+o(1)),$$
as desired. $\hfill \blacksquare$
\bb
Weyl's law may be physically interpreted as an approximation of the number of energy states less than a fixed energy $E_0$, i.e. $\#\{E(h):0\leq E(h)<E_0\}$, by the number of ``Planck cells" $h^n$ which fit into an accessible phase-space volume of the corresponding classical system. The generalization presented in the next section will allow us to estimate the number of energy states on any smooth manifold.

\subsubsection{Extension of Symbols and $\psi$DOs to Manifolds}

In this section, we briefly detail how our formulation of symbol calculus applies to arbitrary smooth manifolds. Although the intuition from \S2.1 and \S2.2 remains the same, we require this extension to state a manifold version of Weyl's law and prove a generalized Egorov's theorem. We will nonetheless rely on our prior intuition to justify omitting some rather uninsightful proofs. 

Let $M$ be a smooth Riemannian manifold of dimension $n$, and assume that all manifolds in the remainder of this chapter are compact. Let $\gamma: M\supset U_\gamma \to V_\gamma\subset \mathbb{R}^n$ be a $C^\infty$ diffeomorphism of open sets, and denote the set of all $C^\infty$ diffeomorphisms of $M$ as $\text{Diff}(M)$. To formulate the symbol calculus and quantization operators on $M$, we start by defining distributions on $M$.
\bb
\textbf{Definition \cc} (\emph{distributions on $M$}) Let $\varphi: C^\infty(M)\to \mathbb{C}$ be a linear map, and set 
$$\Sigma(f):= \varphi(\gamma^*(\chi f)),$$ 
where $\gamma \in \text{Diff}(M), \chi\in C^\infty_c (V_\gamma)$, and $f\in \mathcal{S}(\mathbb{R}^n)$. If $\Sigma \in \mathcal{S}'(\mathbb{R}^n)$, then $\varphi$ is a \emph{distribution on $M$}, and we write $\varphi \in \mathcal{D}'(M)$. \index{distribution!on a manifold}
\bb
\textbf{Definition \cc} (\emph{differential operators on $M$}) If 
$P=\sum X_{i_1}... X_{i_k}$, where $X_{i_j}:M\to TM$ is a smooth vector field for all $j$ and $1\leq k\leq m$, then 
$P$ is a \emph{differential operator on $M$ of order at most $m$}.
\bb
Due to the mapping properties of vector fields, we see that any differential operator $P$ maps $C^\infty(M)\to C^\infty(M)$, $C^\infty_c(M)\to C^\infty_c(M)$, and $\mathcal{D}'(M)\to \mathcal{D}'(M)$. 
\bb
Let us now turn to defining $\psi$DOs and quantization on manifolds. For any smooth manifold $M$, we may think that it is straightforward to use the standard pseudodifferential calculus on $\mathbb{R}^n$ and a partition of unity argument to define quantization operators on $M$. This construction, however, depends on the choice of local coordinates and the partition of unity.  

Thus, a natural question to begin with is determining which symbols are invariant under some diffeomorphism $\kappa:\mathbb{R}^n\to \mathbb{R}^n$: namely, if the symbol class $S(m)$ is defined by the condition that $\forall \alpha\in \mathbb{N}^n, \exists C=C(\alpha)\in \mathbb{R}: |\partial^\alpha a|\leq Cm$, then it may not be true that the pullback of $a$ by the lift of $\kappa^{-1}$ to the contangent bundle $T^*\mathbb{R}^n$  satisfies the same inequality. It turns out that the appropriate invariant class of symbols is obtained by choosing the order function $\langle p\rangle^m$ for $m\in \mathbb{Z}$, which yields the so-called \emph{Kohn-Nirenberg symbols}  (\cite{Zwo1} is the first and only source to give the symbol class this name, but we also follow this nomenclature). 
We refer the reader to \cite{Dim1,Gui1} for a rigorous treatment of invariance issues and a proof of Propositions 2.3.7 and 2.3.8.
\bb
\textbf{Definition \cc} (\emph{Kohn-Nirenberg symbols}) The \emph{Kohn-Nirenberg symbol class of order $m\in \mathbb{Z}$} is defined as\index{symbol!Kohn-Nirenberg class}
$$S^m(\mathbb{R}^{2n}):=\{a\in C^\infty(\mathbb{R}^{2n}): \forall \alpha,\beta\in \mathbb{N}^n, \exists C=C(\alpha,\beta)\in \mathbb{R}:|\partial^\alpha_x\partial^\beta_p a|\leq C\langle p\rangle^{m-|\beta|}\}.$$\vspace{-15pt}\bb\bb
\textbf{Proposition \cc} (\emph{invariance of Kohn-Nirenberg symbols under diffeomorphisms}) Let $\kappa :\mathbb{R}^n\to \mathbb{R}^n$ be a $C^\infty$ diffeomorphism satisfying the inequalities $|\partial^\alpha \kappa|\leq C$ and $|\partial^\alpha \kappa^{-1}|\leq C$ for $C=C(\alpha)$ and any $\alpha \in \mathbb{N}^n$. Then for each symbol $a\in S^m(\mathbb{R}^{2n})$, the pullback $b(x,p):=a(\kappa^{-1}(x),\partial \kappa (\kappa^{-1}(x))\cdot  p)$ under the lift of $\kappa^{-1}$ is in $S^m$.
\bb
\textbf{Proposition \cc} (\emph{Kohn-Nirenberg symbol composition}) Let $a\in S^{m_1}, b\in S^{m_2}$. Then $Op^w(a)Op^w(b)=Op^w(c)$ for $c\in S^{m_1+m_2}$ as given in Theorem 2.2.3. Furthermore, as in Theorem 2.2.6, $c$ admits the decomposition
$$c(x,p)=\sum_{k=0}^N \frac{(ih)^k}{k!}A(D)^k(a(x,p)b(y,q))|_{y=x,q=p}+O_{S^{m_1+m_2-N-1}}(h^{N+1}),$$
where $A(D)=\frac{1}{2}\sigma(D_x,D_p,D_y,D_q)$. 
\bb
Our definition of the Kohn-Nirenberg symbols allows us to define the relevant class of $\psi$DOs on any smooth manifold.
\bb
\textbf{Definition \cc} (\emph{$\psi$DOs on $M$}) A linear map $A:C^\infty(M)\to C^\infty(M)$ is called a \emph{pseudodifferential operator on $M$ of order $m$} if it can be written on each coordinate patch $U_\gamma\subset M$ as
$$\varphi A(\psi f)=\varphi \gamma^* Op^w(a_\gamma)(x,hD)(\gamma^{-1})^*(\psi f),$$
where $\gamma \in \text{Diff}(M)$, $\varphi, \psi\in C^\infty_c(U_\gamma ), f\in C^\infty(M)$, and the symbol $a_\gamma$ is in the Kohn-Nirenberg class $S^m(\mathbb{R}^{2n})$ for some order $m$. 
If $A$ is a $\psi$DO of order $m$ on $M$, then we write $A\in \Psi^m(M)$. \index{pseudodifferential operator!on a manifold}
\bb
We similarly define symbols on $T^*M$ by pullback. 
\bb
\textbf{Definition \cc} (\emph{symbols on $T^*M$}) Let $a\in C^\infty(T^*M)$, $\gamma \in \text{Diff}(M)$, and $\phi: V_\gamma \times \mathbb{R}^n \to T^*U_\gamma$ be the natural identification between the open set $V_\gamma \subset M$ and $\mathbb{R}^n$. If $\phi^*a\in S^m(V_\gamma\times \mathbb{R}^n)$, then $a$ is a \emph{symbol of order m} on $T^*M$, and we write $a\in S^m(T^*M)$. \index{symbol!on a manifold}
\bb
The definitions above provide the natural language in which to formulate corresponding versions of previous theorems for smooth manifolds. Building on the intuition developed in \S2.2, we state the following theorems without proof: 
\bb
\textbf{Theorem \cc} (\emph{quantization on $M$}, \cite{Dim1,Mar1,Zwo1}) 
If $\Psi^m(M)$ denotes the image of $S^m(T^*M)$ under $Op^w$, then there exist linear maps 
$\sigma:\Psi^m(M)\to S^m(T^*M)/ hS^{m-1}(T^*M)$ and $Op^w:S^m(T^*M)\to \Psi^m(M)$
defined by
$$\sigma(A_1A_2)=\sigma(A_1)\sigma(A_2)\ \ \text{ and } \ \ \sigma(Op^w(a))=[a]\in S^m(T^*M)/hS^{m-1}(T^*M),$$
where $[a]$ denotes the equivalence class of $a$ in $S^m(T^*M)/hS^{m-1}(T^*M)$, i.e. the principal symbol. $a=\sigma(A)$ is then the \emph{(principal) symbol of the pseudodifferential operator $A$}. \index{quantization!on a manifold}\index{pseudodifferential operator} 
\bb
\textbf{Theorem \cc} (\emph{boundedness and compactness of $\psi$DOs}) If $A\in \Psi^0(M)$, then $A:L^2(M)\to L^2(M)$ is bounded. Moreover, if $A\in \Psi^m(M)$ for $m<0$, then $A:L^2(M)\to L^2(M)$ is compact.

\emph{Sketch of Proof.} Use a partition of unity argument and apply Theorem 2.2.14. $\hfill \Box$
\bb
In practice, explicit computations for symbol quantization on manifolds can be achieved by pulling back to $\mathbb{R}^n$, but working these calculations out for any given diffeomorphism $\gamma$ is oftentimes too complex. We will avoid these calculations in the remainder of this thesis, keeping at the back of our mind that what is true in the $\mathbb{R}^n$ case (\S2.2) holds also for $M$. 

Let us conclude this section by relating our manifold formulation back to Weyl's law. With a choice of local coordinates, we consider the Schr\"{o}dinger operator $\HH(h):=-h^2\Delta+V(x)$ on the compact Riemannian manifold $(M,g)$, where $\Delta$ is the metric-induced Laplacian on $M$ and $V\in C^\infty(M)$ is real-valued. Based on the examples in \S2.1.2, the symbol of $\HH(h)$ (an operator in $\Psi^2(M)$) is given by 
$$\sigma(\HH(h))=\xi(x,p):=|p|^2_{g_x}+V(x)=g^{ij}(x)p_ip_j+V(x).$$
We note that with $V=0$ and $h=1$, $\HH(h)$ just becomes  the usual (negative) Laplacian as seen in \S1.2. We state the following important theorems without proof; c.f. \cite{Jos1,Mar1,Zwo1} and Appendix I: 
\bb
\textbf{Theorem \cc} (\emph{eigenfunctions of $\HH(h)$}) The pseudodifferential operator $\HH(h):C_c^\infty(M)\\ \to C_c^\infty(M)$ as defined above is essentially self-adjoint. 
Furthermore, for each $h>0$, there exists an orthonormal basis $\{\psi_j(h)\}_{j=1}^\infty$ of $L^2(M)$ with the property that each $\psi_j(h) \in C^\infty(M)$ is an eigenfunction of $\HH(h)$, i.e. $\HH(h)\psi_j(h)=E_j(h)\psi_j(h)$ for $j\in \mathbb{N}^+$, where the eigenvalues $\mathbb{R}\ni E_j(h)\to \infty$ as $j\to \infty$. \index{Hamiltonian operator}\index{eigenvalue}\index{eigenfunction}
\bb
The previous theorem is analogous to Theorem 1.1.5, but is adapted to the generalized case of Schr\"{o}dinger operators. 
\bb
\textbf{Theorem \cc} (\emph{generalized Weyl's law}) Let $(M,g)$ be a compact Riemannian manifold, $V\in C^\infty(M)$, $\HH(h):=-h^2\Delta+V(x)$ as before, and $E(h)$ be an arbitrary eigenvalue of the operator $\HH(h)$. Then for all $a<b$, 
$$\#\{E(h):a\leq E(h)\leq b\}=(2\pi h)^{-n} (\text{Vol}_{T^*M}\{a\leq |p|^2_{g_x}+V(x)\leq b\}+o(1))$$
as $h\to 0$. \index{Weyl's law!for compact Riemannian manifolds}
\bb
The proof of Theorem 2.3.14 is similar to that of Theorem 2.3.3, but we must develop and use the functional calculus for the Schr\"{o}dinger operator $\HH(h)$ on $M$. This is done explicitly in \cite{Dim1}, \cite{Mar1}, and \cite{Zwo1}. 
\bb
\textbf{Corollary \cc} (\emph{Weyl's law for a domain $\Omega \subset \mathbb{R}^2$}) Let $\Omega \subset \mathbb{R}^2$ be a planar domain equipped with the usual metric. If $V=0$, $h=1$, and $\HH\psi_j = E_j\psi_j$ where $\HH:=\HH(h)=-\Delta$ and $\psi_j\in L^2(\Omega)$, 
then
$$N(\lambda):=\# \{j:E_j<\lambda\} \sim \frac{\text{Area}(\Omega)}{4\pi}\lambda$$
in the limit $\lambda \to \infty$. \index{Weyl's law!for Euclidean planar domains}

\emph{Proof.} Taking $n=2$, $V=0$, and $h=1$ in Theorem 2.3.14 above, we see that
$$N(\lambda)=\#\{E(h):0\leq E(h)\leq \lambda \} = (2\pi)^{-2}(\text{Vol}_{\mathbb{R}^{4}}\{0\leq p^2\leq \lambda \}+o(1)).$$
The result follows after noting that the volume expression is simply the product of the area of $\Omega$ in position space and the area of a ball of radius $\sqrt{\lambda}$ in momentum space. $\hfill \blacksquare$

\subsubsection{Egorov's Theorem}

Egorov's theorem asserts that the quantum-mechanical time evolution of a Weyl-quantized operator can be well-approximated by transporting the operator's symbol along the classical flow generated by the (principal) symbol of the Hamiltonian operator. It therefore provides an insightful correspondence between classical and quantum mechanics. \index{Egorov's theorem}

We start formulating the theorem as follows. Let $V$ be a smooth, real-valued potential on the compact Riemannian manifold $(M,g)$, and with a choice of local coordinates on $M$ we consider again the Hamiltonian $\xi(x,p):=|p|^2_{g_x}+V(x)$, where $(x,p)\in T^*M$. Let us denote the Hamiltonian flow (c.f. \S 1.2.1) of $\xi$ as
$$\Phi^t=\exp(tX_\xi),$$
where $X_\xi$ is the Hamiltonian vector field determined by $\xi$ (c.f. \S1.1). (We write $\Phi^t$ instead of $g^t$ to remind ourselves of the flow's Hamiltonian nature.) 
By Theorem 1.2.8, we see that $\Phi^t$ identifies with the geodesic flow on the tangent bundle $TM$ under the canonical isomorphism $T^*M\cong TM$ and the condition $V:=0$. 

From the functional analysis reviewed in Appendix I, by \emph{Stone's theorem} (Theorem I.4) we know that self-adjoint operators are the infinitesimal generators of unitary groups of time evolution operators. 
So let us denote the unitary group on $L^2(M)$ generated by the (essentially) self-adjoint operator $\HH(h)$ as $F(t)=\exp(-it\HH(h)/h)$. If $A\in  \cap_{m\in \mathbb{Z}}\Psi^m(M)$ is another $\psi$DO, then its quantum evolution is given by $A(t):=F^{-1}(t)AF(t)$ for all $t\in \mathbb{R}$; this agrees with the Heisenberg picture of quantum mechanics. We state and prove the following formulation of Egorov's theorem: \index{unitary flow}
\bb
\textbf{Theorem \cc} (\emph{Egorov}, \cite{Ego1,Zwo1}) Let $\Phi^t$ denote the Hamiltonian flow of $\xi(x,p):=|p|^2_{g_x}+V(x)$ and $a_t(x,p):=a(\Phi^t(x,p))$ for some $a\in S^{-\infty}(T^*M)$. If $A=Op^w(a)(x,hD)$ and $\widetilde{A}(t):=Op^w(a_t)(x,hD)$, then for any fixed $T>0$ and $0\leq t\leq T$, 
$$||A(t)-\widetilde{A}(t)||_{L^2\to L^2}=O(h)$$
uniformly. Note that we need $a\in S^{-\infty}(T^*M)$ to guarantee that $a_t$ is in the same symbol class. If $a\in S(T^*M)$, then the symbol class is not preserved by $\varphi_t^*$ due to the fact that the flow of $\xi$ is faster at higher frequencies. \index{Egorov's theorem}

\emph{Proof.} First we note that
$\partial_ta_t=\{\xi,a_t\}$, 
where $\{\xi,a_t\}=X_\xi a_t$ is the Poisson bracket on $T^*M$. If $\sigma(P)$ denotes the symbol of a $\psi$DO $P$, then \index{Poisson bracket}
$$\sigma\left(\frac{i}{h} [\HH(h),B]\right)=\{\xi,\sigma(B)\}$$
for any $B\in \cap_{m\in \mathbb{Z}}\Psi^m(M)$. This can be checked in local coordinates with Proposition 2.3.8. An application of Theorems 2.3.11 and 2.3.12 then yields
$$\partial_t \widetilde{A}(t)=\frac{i}{h} [\HH(h),\widetilde{A}(t)]+E(t)$$
for $E(t)\in h\Psi^{-\infty}(M)$, 
with 
$||E(t)||_{L^2\to L^2}=O(h)$.
Applying the time-evolution operator on $\partial_t \widetilde{A}$, we have
\begin{eqnarray*}
\partial_t(e^{-\frac{it\HH(h)}{h}}\widetilde{A}(t)e^{\frac{it\HH(h)}{h}})&=&e^{-it\HH(h)/h}(\partial_t \widetilde{A}(t)-(i/h)[\HH(h),\widetilde{A}(t)])e^{it\HH(h)/h}\\
&=&e^{-it\HH(h)/h}((i/h) [\HH(h),\widetilde{A}(t)]+E(t)-(i/h) [\HH(h),\widetilde{A}(t)])e^{it\HH(h)/h}\\
&=&e^{-it\HH(h)/h}E(t)e^{it\HH(h)/h}=O_{L^2\to L^2}(h).
\end{eqnarray*}
Integrating both sides of this equality then gives
$||e^{-it\HH(h)/h}\widetilde{A}(t)e^{-it\HH(h)/h}-A||_{L^2\to L^2}=O(h),
$
which implies that 
$$||\widetilde{A}(t)-A(t)||_{L^2\to L^2}=||\widetilde{A}(t)-e^{it\HH(h)/h}Ae^{-it\HH(h)/h}||_{L^2\to L^2}=O(h)$$
uniformly for all $t\in [0,T]$. $\hfill \blacksquare$
\bb
Egorov's theorem will be useful for proving the quantum ergodicity theorem of Schnirelman, Zelditch, and Colin de Verdi\`{e}re in the next chapter.

\pagebreak

\setcounter{itemcounter}{1}
\setcounter{section}{2}
\section{Semiclassical Analysis and Quantum Ergodicity}

In this chapter, we accomplish our goal of proving the quantum ergodicity (QE) theorem using the semiclassical, geometric, and analytic tools we have developed thus far. Our proof of the QE theorem will naturally lead us into a discussion about quantum unique ergodicity (QUE). In particular, we examine in \S3.3 both the similarities and differences of each statement, and motivate why the problem of QUE is substantially more difficult. \index{quantum ergodicity!quantum ergodicity theorem}\index{quantum ergodicity!quantum unique ergodicity}

\subsection{Quantum Ergodicity}

Let $(M,g)$ be a compact Riemannian manifold. Recall the following setup from \S1.3: if $H$ is a Hamiltonian function that specifies the integrable system $(M,\omega,H)$ with geodesic flow $g^t$, then the \emph{level sets of $M$} are denoted $\Sigma_c:=H^{-1}(c)$ for $c\in [a,b]\in \mathbb{R}$.
By Definition 1.2.7, the Liouville measure $\mu_L^c$ is an invariant measure on each fiber $\Sigma_c$. We can discuss the ergodicity of $g^t$ by viewing it as a transformation on the measure space $(\Sigma_c,\mu_L^c)$ determined by the ``energy shell" (level set) $\Sigma_c$, and checking that $g^t$ satisfies the ergodicity conditions given in Definition 1.2.10. If $g^t$ is ergodic, then by the \emph{weak ergodic theorem} we know that
$$\lim_{T\to \infty}\int_{\Sigma_c}\left( \langle f\rangle_T-\dashint_{\Sigma_c}fd\mu_L^c\right)^2 d\mu_L^c=0$$
for all $f\in L^2(\Sigma_c)$, and furthermore by \emph{Birkhoff's ergodic theorem} that 
$$\langle f\rangle_T=\lim_{T\to \infty}\frac{1}{T}\int_0^T f(g^t(z))dt=\dashint_{\Sigma_c}fd\mu_L^c.$$
Recall as well the Hamiltonian symbol $\xi:T^*M\to \mathbb{R}, \xi(x,p):=|p|^2+V(x)$ for $V\in C^\infty(M)$ and its Weyl quantization $\Xi(h)=-h^2\Delta+V(x)$ from \S2.3.1. We will take $\xi$ to be our Hamiltonian $H$ in the previous paragraph, and use the generalization $\Xi(h)$ of our usual Laplacian $\Delta$ for the statement of Theorem 3.1.3. 

We begin by stating a historical form of the QE theorem, which incidentally is one of the clearest. Translated from the original French in Colin de Verdi\`{e}re's 1985 manuscript, this statement reads as follows:
\bb
\textbf{Theorem \cc} (\emph{quantum ergodicity, Schnirelman and Colin de Verdi\`{e}re}, \cite{Col1}) Let $M$ be a compact Riemannian manifold. If the geodesic flow on $M$ is ergodic, then there exists a subsequence $\{\lambda_{k_i}\}_{i\in \mathbb{N}}$ of density 1 of the spectrum of the Laplacian $-\Delta$ such that, for any pseudodifferential operator $A$ of order zero with principal symbol $a$, we have\index{quantum ergodicity!quantum ergodicity theorem}\index{eigenvalue!spectrum}
$$\lim_{i\to \infty} \langle A \varphi_{k_i},\varphi_{k_i}\rangle=\int_{\Sigma_c}ad\mu^c_L,$$
where $\varphi_{k_i}$ is an eigenfunction of $-\Delta$ with eigenvalue $\lambda_{k_i}$. Here the \emph{density} of a set $S\subseteq \text{Spec}(-\Delta)$ is defined as
$$D(S)=\lim_{\lambda \to \infty} \frac{\# \{\lambda_k\in S:\lambda_k\leq \lambda\}}{\#\{\lambda_k\leq \lambda\}}.$$\vspace{-5pt}\bb
We remind ourselves that by a theorem of Hopf's (Theorem 1.2.12), the geodesic flow on any compact, negatively-curved Riemannian manifold is ergodic. There are variants of Theorem 3.1.1 in the literature where the hypothesis is that $M$ is negatively curved. 
\bb
\textbf{Corollary \cc} (\emph{density of eigenfunctions}) With the notation as above, we have
$$\lim_{i\to \infty} \int_S |\varphi_{k_i}|^2=\frac{\text{Vol}(S)}{\text{Vol}(M)},$$
where $S$ is an open subset of $M$.  \index{eigenfunction}\index{Laplace-Beltrami operator}
\bb
We will in fact state and prove a more general statement than Theorem 3.1.1, involving not just the eigenfunctions of $\Delta$ but also the eigenfunctions of the pseudodifferential operator $\Xi(h)$. First, several remarks are in order: 
\begin{itemize}
\item Let us keep in mind the notation $\Delta \varphi_n+\lambda_n \varphi_n=0$. In particular, we will use $\varphi$ to denote an eigenfunction of the positive Laplacian $\Delta$, and index the set of all eigenfunctions and eigenvalues by $n$. We will also set $\lambda_n=h_n^{-2}$ when appropriate.
\item The quantity $\langle A\varphi_n,\varphi_n\rangle$ denotes the expectation value of the self-adjoint operator (quantum observable) $A$. Theorem 3.1.1 most simply states that in the high-energy limit, the expectation value of a quantum observable is just the space-average of its corresponding classical symbol. 
\item We can immediately see how the intuition behind this theorem, as given in the Introduction and Chapter 1, follows. In more qualitative terms, Theorem 3.1.1 and Corollary 3.1.2 tell us that a density 1 set of Laplacian eigenfunctions induces a measure on the set of functions which converges to the uniform measure. Though we have already seen this measure in \S1.3, we will define it more rigorously as needed in \S3.3.\index{Hamiltonian symbol}\index{Hamiltonian operator}
\item The convergence of measures above is meant in the weak-$*$ sense. Again, we relegate a rigorous definition of weak-$*$ convergence to \S3.3, and use only the working definition of convergence above. \index{weak-$*$ convergence}
\item The physical and intuitive interpretation of Theorem 3.1.1 is that wavefunctions ($L^2$-eigenfunctions of $\Delta$) on a negatively-curved compact domain \emph{equidistribute} in the high-energy limit. If the classical mechanics as described by the geodesic flow of a system is ergodic, then even in the semiclassical limit $h_n\to 0$ $(\lambda_n\to \infty)$ do most eigenfunctions of the system fail to localize in phase space. 
\end{itemize}
Colin de Verdi\`{e}re proves Theorem 3.1.1 in a manner similar to our proof of Theorem 3.1.5 below, so let us proceed to stating that theorem. 
\bb
\textbf{Definition \cc} (\emph{uniform symbol}) A symbol $a$ is \emph{uniform} if for all $c\in [a,b]$, 
$$\alpha:=\dashint_{\Sigma_c}\sigma(A)d\mu_L$$ 
assumes the same value, i.e. the averages of $a$ over each level surface $\xi^{-1}(c)$ are equal to some constant $\alpha$. A $\psi$DO is \emph{uniform} if its principal symbol is. 
Note that we will use lowercase Roman letters for both real numbers and symbols, and draw the distinction by context.\index{symbol!uniformity}
\bb
Although this uniformity condition is not standard in the literature, it is useful for our particular statements of the QE theorem below. We remark that, luckily, any symbol $b$ can be made uniform by applying the correct projection to the set of uniform symbols. This projection $\mathcal{U}$ is defined as follows. Assuming that $|\partial \xi|>\gamma$ on $\xi^{-1}([a,b])$, we also have $|\partial \xi|>\gamma/2$ on $\xi^{-1}([a-\delta,b-\delta])$ for some small enough $\delta>0$. If for $\beta=\xi(x,p)$ 
$$\mathcal{U}(b(x,p)):=b(x,p)-\dashint_{\Sigma_{\beta}}bd\mu^{\beta}_L,$$
then for any $c\in [a,b]$ we have
$$\dashint_{\xi^{-1}(c)}\mathcal{U}(b)d\mu^c_L=0.$$
Setting $\chi \in C^\infty_c(\xi^{-1}((a-\delta,b+\delta)))=1$ near $\xi^{-1}([a,b])$, we define $\mathcal{U}$ by finding 
$$\mathcal{U}(\chi b)=\chi b(x,p)-\dashint_{\Sigma_\beta}\chi b d\mu_L^\beta.$$ 
Then $\mathcal{U}:C^\infty(T^*M)\to C_c^\infty(T^*M)$ is idempotent, since $\dashint_{\xi^{-1}(\beta)}\mathcal{U}(b)d\mu^\beta_L
=0$ implies 
$$\mathcal{U}^2(\chi b)=\left( \chi b(x,p)-\dashint_{\Sigma_\beta}\chi b d\mu_L^\beta\right)-\left(\dashint_{\Sigma_\beta}\mathcal{U}(\chi b) d\mu_L^\beta\right)
=
\chi b(x,p)-\dashint_{\Sigma_\beta}\chi b d\mu_L^\beta=\mathcal{U}(\chi b).$$ 
Note that the equation $\dashint_{\xi^{-1}(c)}\mathcal{U}(\beta)d\mu^c_L=0$ holds for any choice of the symbol $b$. Thus, $\mathcal{U}$ maps arbitrary symbols to uniform symbols with zero average. In terms of $\psi$DOs, we can also find a set of uniform Weyl-quantized operators $B$ with an arbitrary average $\alpha$ by taking 
$$B:=Op(\mathcal{U} (b))+\alpha Op(\chi)+Op((1-\chi)b),$$
where $b$ is any symbol and $\chi\in S(T^*M)$ is defined to be $1$ near $\xi^{-1}([a,b])$.
\bb
As mentioned before, the following two theorems deal with the more general $\Xi(h)$ operator instead of $\Delta$. We will also denote the eigenfunctions and eigenvalues of $\Xi(h)$ by $\varphi_k$ and $\lambda_k$ respectively, so that 
$\Xi(h)\varphi_k(h)=\lambda_k(h)\varphi_k(h).$
\bb
\textbf{Theorem \cc} (\emph{quantum ergodicity, Zelditch and Zworski}, \cite{Zel2}) Let $M$ be a compact Riemannian manifold with ergodic geodesic flow. If $A\in \Psi(M)$ is uniform with principal symbol $\sigma(A)$, then 
$$(2\pi h)^{n}\sum_{a\leq \lambda_k\leq b}\left| \langle A \varphi_k,\varphi_k\rangle-\dashint_{\{a\leq \xi(x,p) \leq b\}}\sigma(A)dxdp\right|^2\to  0.$$
as $h\to 0$ for all $\lambda_k\in \{ a\leq \lambda \leq b : \lambda \in \text{Spec}(\Xi(h))\}$ and eigenfunctions $\varphi_k$.
\bb
We remark that in the case $h=1$ and $\Xi(h)=-\Delta$, Theorem 3.1.4 can be modified to imply Theorem 3.1.1 and weaker QE theorems in the literature \cite{Zel3,Shn1}. In particular, we would have to drop the uniformity assumption and argue for a finer energy localization \cite{Hel5}.   
The following version of the QE theorem, which incorporates eigenfunction densities, is stated by Zworski in \cite{Zwo1}:
\bb
\textbf{Theorem \cc} (\emph{quantum ergodicity, density version}, \cite{Zwo1}) Let $M$ be a compact Riemannian manifold with ergodic geodesic flow. If $A\in \Psi(M)$ is uniform with principal symbol $\sigma(A)$, then there exists a family of subsets $\Lambda(h)\subset \{a\leq \lambda_k\leq b\}$ such that 
$$\frac{\# \Lambda(h)}{\#\{a\leq \lambda_k\leq b\}}\to 1\ \ \ \text{ and }\ \ \  \langle A \varphi_k,\varphi_k\rangle\to  \dashint_{\{a\leq \xi(x,p) \leq b\}}\sigma(A)dxdp$$
as $h\to 0$ for all $\lambda_k \in \Lambda(h)$ and eigenfunctions $\varphi_k$. 
\bb
The QE theorems given above directly imply the equidistribution of Laplacian eigenfunctions, as noted in \S1.
\bb
\textbf{Example \cc} (\emph{equidistribution of Laplacian eigenfunctions}) Let $(M,g)$ be a compact Riemannian manifold with ergodic geodesic flow, and 
consider again the eigenfunction equation
$\Delta \varphi_k+\lambda_k \varphi_k=0$
where $\Delta=\Delta_g$, $\varphi_k\in L^2(M)$, and $k\in \mathbb{N}$. Theorem 3.1.5 then tells us that there is a subsequence $k_i\to \infty$ of density 1, i.e.  
$\lim_{N\to \infty} \frac{1}{N}\# \{i:k_i\leq N\}=1,$
such that
$$\int_M |\varphi_{k_i}|^2fdx\to \int_M fdx$$
for all $f\in C_c^\infty(M)$.\index{Laplace-Beltrami operator}
\bb
One can also say that the probability measure $|\varphi_{k_i}|^2dx$ induced by the eigenfunctions $\{\varphi_{k_i}\}$ converges as $k_i\to \infty$ to the uniform measure on $M$, and again we will make this notion more precise in \S1.3. It is important to note that this remark is the starting point of quantum \emph{unique} ergodicity: that is, under the hypothesis of Theorem 3.1.5, do these eigenfunction-induced probability measures \emph{always} converge to the uniform measure? \index{Wigner measure}

Although the QE theorems may seem difficult to prove at first glance, they follow readily from a few applications of Weyl's law, Egorov's theorem, and the weak ergodic theorem proved earlier in this thesis. 

\subsection{Proof of the Quantum Ergodicity Theorem}
\setcounter{itemcounter}{1}

This section is focused singularly on the proofs of Theorems 3.1.4 and Theorem 3.1.5 above. The proofs we present are motivated by the ones in \cite{Col1} and \cite{Zwo1}, though we endeavor to elucidate certain steps of our argument more naturally. We start by stating two lemmas without proof, one in the form of an analytic bound of Weyl-quantized operators and the other in the form of a generalized Weyl's law: 
\bb
\textbf{Lemma \cc} (\emph{$L^2$ quantization bound}) Suppose $a\in S$ and $A=Op^w(a)(x,hD)$. Then
$$||A||_{L^2\to L^2}\leq C\sup_{\mathbb{R}^{2n}}|a|+O(h^{1/2})$$
as $h\to 0$, where $C$ is a constant. 
\bb
\textbf{Lemma \cc} (\emph{generalized Weyl's law}) Let $B\in \Psi(M)$. Then
$$(2\pi h)^n \sum_{a\leq \lambda_k\leq b}\langle B\varphi_k,\varphi_k \rangle\to \iint_{\{a\leq \xi (x,p)\leq b\}}\sigma(B)dxdp$$
as $h\to 0$.
\bb
Lemma 3.2.1 is directly analogous to Theorem 2.2.14, while Lemma 3.2.2 is a further generalization of Theorem 2.3.14 (we obtain the latter by taking $B=I$ in the former). For proofs of these lemmas, we refer the reader to \cite{Zwo1}. 
\bb
\textbf{\emph{Proof of Theorem 3.1.4.}} The main idea of the proof is to proceed as follows:
\begin{enumerate}
\item First, we replace the given pseudodifferential operator $A$ with another suitable pseudodifferential operator $B$ with a uniform average of zero.
\item Next, we apply Weyl's law (Lemma 3.2.2), 
time-evolve this pseudodifferential operator, and take the time-average to rewrite its expectation value $\langle B\varphi_k,\varphi_k\rangle$ in terms of the time-average of the operator $B$.
\item Then, we 
approximate the expectation value with the time-average of an analogous classically-evolved operator by applying Egorov's theorem (Theorem 2.3.21).
\item Finally, we use the weak ergodic theorem (Theorem 1.2.11) and the ergodicity of the geodesic flow to obtain the desired bound. 
\end{enumerate}
In particular, let $C^\infty_c(T^*M)\ni\chi=1$ near $\xi^{-1}([a,b])$, $\alpha= \dashint_{\Sigma_c}\sigma(A)d\mu_L$, and $B:=\chi(\Xi)(A-\alpha I)$. Since $A$ is uniform, we have
$$\int_{\Sigma_c}\sigma(B)d\mu^c_L=0$$
for all $c\in [a,b]$. For any operator $C$ and its adjoint $C^*$, we have $|\langle C\varphi_k,\varphi_k\rangle |^2\leq \langle C^*C\varphi_k,\varphi_k\rangle$ by the Cauchy-Schwarz inequality, which along with Lemma 3.2.1 tells us 
$$(2\pi h)^n \sum_{a\leq \lambda_k \leq b}|\langle (1-\chi(\Xi))A\varphi_k,\varphi_k\rangle|^2\leq (2\pi h)^n \sum_{a\leq \lambda_k\leq b}\langle A^*(1-\chi(\Xi))^2 A\varphi_k,\varphi_k\rangle \to 0
$$
as $h\to 0$. Thus we can substitute $B$ for $A$ in the statement of Theorem 3.1.4: if 
$$\eps(h)=(2\pi h)^n \sum_{a\leq \lambda_k\leq b}|\langle B\varphi_k,\varphi_k\rangle|^2,$$
then it suffices to show that $\eps(h)\to 0$ as $h\to 0$ to obtained the desired result. Because $\Xi\varphi_k=\lambda_k \varphi_k$, applying the time evolution operator (c.f. \S2.3.3) gives us 
\begin{eqnarray*}
\langle B\varphi_k,\varphi_k \rangle &=& \langle B,e^{-it\lambda_k/h}\varphi_k,e^{-it\lambda_k/h}\varphi_k\rangle = \langle Be^{-it\Xi(h)/h}\varphi_k,e^{-it\Xi(h)/h}\varphi_k\rangle\\
&=& \langle e^{it\Xi(h)/h}Be^{-it\Xi(h)/h}\varphi_k,\varphi_k\rangle = \langle B(t)\varphi_k,\varphi_k\rangle
\end{eqnarray*}
for all $t\in \mathbb{R}$. Taking the time average, we have 
$$\langle B\varphi_k,\varphi_k\rangle = \left\langle \left(\dashint_0^T B(t)dt\right) \varphi_k,\varphi_k\right\rangle = \langle \langle B\rangle_T\varphi_k,\varphi_k\rangle,$$
where we recall that
$$\langle B\rangle_T=\frac{1}{T}\int_0^T B(t)dt=\dashint_0^T B(t)dt.$$
Along with the fact that $||\varphi_k||^2=1$, this implies that
$|\langle B\varphi_k,\varphi_k\rangle|=|\langle \langle B\rangle_T \varphi_k,\varphi_k\rangle|\leq ||\langle B\rangle_T \varphi_k||^2=\langle \langle B^*\rangle_T\langle B\rangle_T \varphi_k,\varphi_k\rangle,$
and we can therefore rewrite $\eps(h)$ as 
$$\eps(h)\leq (2\pi h)^n \sum_{a\leq \lambda_k\leq b}|\langle \langle B^*\rangle_T\langle B\rangle \varphi_k,\varphi_k\rangle|.$$
Applying Egorov's theorem (Theorem 2.3.21), we see that
$$\langle B\rangle_T=\langle \widetilde{B}\rangle_T+O_{L^2\to L^2}(h,T)\ \ \ \text{ and } \ \ \  \langle \widetilde{B}\rangle_T:=\dashint_0^T \widetilde{B}(t)dt$$
where $\widetilde{B}(t)\in \Psi(M), \sigma(\widetilde{B}(t))=(g^t)^*\sigma(B)$, and 
$$\sigma(\langle \widetilde{B}\rangle_T)=\dashint_0^T \sigma(B)\circ g^t dt=\langle \sigma(B)\rangle_T.$$
Since $\langle B\rangle_T=\langle \widetilde{B}\rangle_T+O_{L^2\to L^2}(h,T)$, we can approximately replace the time-evolved operator $e^{it\Xi(h)/h}Be^{-it\Xi(h)/h}$ by the ``geodesically-evolved" $\widetilde{B}$. But then Lemma 3.3.2 and the bound for $\eps(h)$ above imply that
\begin{eqnarray*}\limsup_{h\to 0}\eps(h)&\leq& \limsup_{h\to 0}\left( (2\pi h)^n\sum_{a\leq E_j\leq b}\langle \langle \widetilde{B}^*\rangle_T\langle \langle \widetilde{B}\rangle,u_j\rangle+O_{L^2\to L^2}(h,T)\right)\\
&=&\iint_{\{a\leq \xi\leq b\}}\sigma(\langle \widetilde{B}^*\rangle_T\langle \widetilde{B}\rangle_T)dxdp\ =\ \iint_{\{a\leq \xi\leq b\}}|\sigma(\langle B\rangle_T)|^2dxdp.
\end{eqnarray*}
This is because the symbol map $\sigma$ is multiplicative (c.f. \S2) and the quantization of complex conjugate symbols are adjoint. 
Finally, applying the weak ergodic theorem (Theorem 1.2.11) with $f=\sigma(B)$ gives us 
$$\int_{\xi^{-1}[a,b]}|\langle \sigma(B)\rangle_T|^2dxdp\to 0$$
as $T\to \infty$, and taking the $T\to \infty$ limit above gives
$$\limsup_{h\to 0}\eps(h)=\lim_{T\to \infty} \limsup_{h\to 0}\eps(h)=\lim_{T\to \infty} \int_{\xi^{-1}[a,b]}|\langle \sigma(B)\rangle_T|^2dxdp= 0,$$
as desired. $\hfill \blacksquare$
\bb
We now prove Theorem 3.1.5 in two installments:
\bb
\textbf{\emph{Proof of Theorem 3.1.5, part 1.}} First, we show that there exists a family of subsets $\Lambda(h)\subset \{a\leq \lambda_k\leq b\}$ satisfying the criteria of the theorem, where $\Lambda(h)$ \emph{also depends on the given pseudodifferential operator $A$}. We will use Theorem 3.1.4 to do this.

Let $B$ be given as before, with $B:=\chi(\Xi)(A-\alpha I)$, so that again we have
$$\int_{\{a\leq \xi \leq b\}}\sigma(B)dxdp=0\ \ \ \ \text{ and }\ \ \ \ \eps(h):= (2\pi h)^n\sum_{a\leq \lambda_k \leq b}|\langle B\varphi_k,\varphi_k\rangle|^2\to 0,$$
where the latter assertion is precisely the content of Theorem 3.1.4. Then, for 
$$\Gamma(h):=\{a\leq \lambda_k \leq b:|\langle B\varphi_k,\varphi_k\rangle|^2\geq \eps(h)^{1/2}\},$$
we have $(2\pi h)^n \# \Gamma(h)\leq \eps(h)^{1/2}$. We will define our desired subset $\Lambda(h)$ by removing $\Gamma(h)$ from the bounded part of $\text{Spec}(\Xi)$, namely
$$\Lambda(h):=\{a\leq \lambda_k \leq b\}\backslash \Gamma(h).$$
This choice of $\Lambda(h)$ gives us the appropriate set: indeed, if $\lambda_k \in \Lambda(h)$, then we have
$$|\langle B\varphi_k,\varphi_k \rangle|\leq \eps(h)^{1/4}\Longrightarrow |\langle A\varphi_k,\varphi_k\rangle-\alpha|\leq \eps(h)^{1/4},$$
and
$$\frac{\# \Lambda(h)}{\# \{a\leq \lambda_k\leq b\}}=1-\frac{\#\Gamma(h)}{\# \{a\leq \lambda_k \leq b\}}.$$
Then an application of Weyl's law (Theorem 2.3.14 and Lemma 3.2.2) yields
$$\frac{\# \Gamma(h)}{\# \{a\leq \lambda_k \leq b\}}=\frac{(2\pi h)^n\# \Gamma(h)}{\text{Vol}(\{a\leq \xi(x,p)\leq b\})+o(1)}\leq C\eps(h)^{1/2}\to 0,$$
as $h\to 0$ for some suitable constant $C$. Thus $\Lambda(h)$ saturates $\{a\leq \lambda_k\leq b\}$, as desired. $\hfill \blacksquare$ 
\bb
\textbf{\emph{Proof of Theorem 3.1.5, part 2.}} Our proof proceeds as follows: 
\begin{enumerate}
\item By defining an appropriate density-1 eigenvalue subset $\Lambda_\infty(h)$ as the limit of all $\Lambda_l(h)$'s in any enumeration $\{A_l\}_{l=1}^\infty$ of $\psi$DOs, we show by a simple density argument that $\Lambda_\infty(h)$ is the density-1 subset we seek in the statement of the theorem.
\item We construct an appropriate set of $\psi$DOs $\{A_l\}_{l=1}^\infty$ which satisfy the theorem, with the additional criterion that $\{A_l\}_{l=1}^\infty$ is dense in the set of uniform $\psi$DOs. By another density argument, we can show that the theorem holds for any $\psi$DO. 
\end{enumerate}
In particular, let $\{A_l\}_{l=1}^\infty\subset \Psi(M)$ be any family of uniform pseudodifferential operators, each with average $\alpha_l$. From part 1 above, we can define some appropriate $\Lambda_l(h)$ corresponding to $A_l$ so that
$$\frac{\# \Lambda_l(h)}{\#\{a\leq \lambda_k\leq b\}}\to 1\ \ \ \text{ and }\ \ \  \langle A_l \varphi_k,\varphi_k\rangle\to  \dashint_{\{a\leq \xi(x,p) \leq b\}}\sigma(A_l)dxdp$$
as $h\to 0$, for all $\lambda_k \in \Lambda_l(h)$ and eigenfunctions $\varphi_k$. From $\{A_l\}_{l=1}^\infty$ and $\{\Lambda_l(h)\}_{l=1}^\infty$ as defined, we wish to construct a set $\Lambda(h)$ so that the conditions above are satisfied by an arbitrary pseudodifferential operator $A$; in other words, we wish to remove the dependency of $\Lambda(h)$ on $A$ from above. 

We start by observing that, since $\Lambda_l(h)$ and $\Lambda_m(h)$ both have density 1 in the bounded spectrum of $\Xi$, their intersection $\Lambda_l(h)\cap \Lambda_m(h)$ also has density 1. Thus, we may order the sets $\Lambda_l(h)$ so that 
$\Lambda_{l+1}(h)\subset \Lambda_l(h)$
for all $l$. 
For each $l$, we shall choose $h=h(l)> 0$ small enough so that
$$\frac{\# \Lambda_l(h)}{\# \{a\leq \lambda_k\leq b\}}\geq 1-\frac{1}{l}$$
for $0< h< h(l)$. Taking $h(l)> h(l+1)\to 0$ as $l\to \infty$, we define 
$\Lambda_\infty(h):=\lim_{l\to \infty} \Lambda_l(h)$, where $h(l+1)\leq h<h(l).$ (In other words, $\Lambda_\infty(h)$ is defined to be the ``smallest" set of $\{\Lambda_l(h)\}_{l=1}^\infty$, where the semiclassical parameter $h$ for $\Lambda_\infty(h)$ is bounded appropriately.) Then 
$$\frac{\# \Lambda_\infty(h)}{\# \{a\leq \lambda_k \leq b\}}\geq 1-\frac{1}{l}$$
for all $0<h<h(l)$, and taking $l\to \infty \Longrightarrow h(l)\to 0$ gives us 
$$\lim_{h\to 0}\frac{\# \Lambda_\infty(h)}{\# \{a\leq \lambda_k\leq b\}}= 1.$$
Now we observe that, for any pseudodifferential operator $A_l$ in our set above, we have 
$$\langle A_l \varphi_k,\varphi_k \rangle \to \dashint_{\{a\leq \xi(x,p) \leq b\}}\sigma(A_l)dxdp$$
as $h\to 0$, where $\lambda_k \in \Lambda_\infty(h)$. This is because the fact that $\Lambda_{l+1}(h)\subset \Lambda_l(h)$ for all $l$ and the definition of $\Lambda_\infty$
implies that 
$\Lambda_\infty(h) \subset \Lambda_l(h)$ for $h<h(l)$, and the above limit holds for $\lambda_k\in \Lambda_l(h)$. 

Let us now choose a set $\{A_l\}_{l=1}^\infty$ which is dense in $\mathcal{A}=\{A\in \Psi^{-\infty}(M)
: \dashint_{\Sigma_c}\sigma(A)d\mu^c_L=\alpha\text{ independently of }c\in [a,b]\}$. 
By \emph{dense in $\mathcal{A}$}, we mean that for any $A\in \Psi^{-\infty}(M)$ and $\eps>0$, there are choices of $l$ and $h_0$ such that the time-average of the difference of the $\psi$DOs are small:
$$\dashint_{\{a\leq \xi(x,p)\leq b\}}|\sigma(A_l-A)|dxdp<\eps\ \ \ \text{ and }\ \ \ ||A_l-A||_{L^2\to L^2}<\eps.$$
for $0<h<h_0$. As we shall see, finding such a set $\{A_l\}_{l=1}^\infty$ with this property implies the result. 

To find the desired $\{A_l\}_{l=1}^\infty$, note that by Lemma 3.2.1 and the weak ergodic theorem, we have 
$$||A-A_l||_{L^2\to L^2}\leq ||\sigma(A)-\sigma(A_l)||_{L^\infty(T^*M)}+Ch^{1/2}$$
for $C=C(a,a_l)$, 
and thus
$$\dashint_{a\leq \xi(x,p) \leq b}|\sigma(A-A_k)|dxdp\leq C||\sigma(A)-\sigma(A_l)||_{L^\infty(T^*M)}.$$
In terms of symbols, we must therefore find the set $\{a_l\}_{l=1}^\infty\subset S^{-\infty}(T^*M)$ satisfying (independently of $c\in [a,b]$) 
$$\dashint_{\Sigma_c}a_l(x,p)d\mu_L^c=\alpha,$$
so that for an arbitrary choice of $a\in S^{-\infty}(T^*M)$ and all $\eps>0$, there is a choice of $l$ such that
$$||a-a_l||_{L^\infty(T^*M)}<\eps.$$
Another way to think of this problem is that we need to find 
$$\{a_l\in C_c^\infty(T^*M)\text{ satisfying the above bound}\}$$
dense in the space 
$$C_0(T^*M):=\{\text{continuous functions vanishing at }\infty\text{ and satisfying the above bound}\}.$$
The means for this construction is provided by the uniformity projection $\mathcal{U}$ given in \S3.1: since $\mathcal{U}$ is continuous, taking a dense set of $b_l$'s in both $C_c^\infty(\mathbb{R}^{2n})$ and $C_0(\mathbb{R}^{2n})$ and setting $a_l=\mathcal{U}(b_l)$ gives a set $\{a_l\}_{l=1}^\infty$ of zero-average symbols that is also dense in $C_0(T^*M)$. 
Adding $\alpha \chi$ for $C_c^\infty(\mathbb{R}^{2n})\ni \chi=1$ near $\xi^{-1}([a,b])$ to $a_l$ then produces a symbol with average $\alpha$. 

Thus, we have found a set $\{A_l\}_{l=1}^\infty$ which is dense in $\mathcal{A}=\{A\in \Psi^{-\infty}(M)
: \dashint_{\Sigma_c}\sigma(A)d\mu^c_L=\alpha\text{ independently of }c\in [a,b]\}$. Along with the fact that 
$$\langle A_l \varphi_k,\varphi_k \rangle \to \dashint_{\{a\leq \xi(x,p) \leq b\}}\sigma(A_l)dxdp$$
as $h\to 0$ for $\lambda_k \in \Lambda_\infty(h)$, we see that for 
all $\lambda_k \in \Lambda_\infty(h)$,
$$\left| \limsup_{h\to 0}\left( \langle A\varphi_k,\varphi_k\rangle-\dashint_{\{a\leq \xi(x,p) \leq b\}}\sigma(A)dxdp\right)\right|<2\eps$$
and
$$\left| \liminf_{h\to 0}\left( \langle A\varphi_k,\varphi_k\rangle-\dashint_{\{a\leq \xi (x,p)\leq b\}}\sigma(A)dxdp\right)\right|<2\eps.$$
These bounds yield the desired result for 
arbitrary $A\in \Psi^{-\infty}(M)$. To extend this result to arbitrary $A_0\in \Psi(M)$, we replace $A\in \Psi^{-\infty}(M)$ above with $\eta(\Xi)A_0\in \Psi^{-\infty}(M)$, where $A_0\in \Psi(M)$ and $C^\infty_c(\mathbb{R})\ni\eta=1$ on $[a,b]$. 
Running the argument back with $A=\eta(\Xi)A_0$ and using the multiplicative properties of $\psi$DOs developed in \S2, we see that the desired result holds for any $A_0\in \Psi(M)$. $\hfill \blacksquare$

\subsection{Quantum Unique Ergodicity}
\setcounter{itemcounter}{1}

Let us pause the mathematical exposition for a brief dialogue. Generally speaking, the quantum ergodicity theorems answer a question pertaining to the asymptotics of Laplacian eigenfunctions when the geodesic flow $g^t$ is ergodic or chaotic: namely, how does the chaotic dynamics of the classical geodesic flow translate to quantum-mechanical eigenfunctions, which by Egorov's theorem must converge to the classical limit as the energy levels become infinite? \index{quantum ergodicity!quantum unique ergodicity}
The QE theorems particularly examine the quantities $\langle A\varphi_k,\varphi_k \rangle$, where $A$ is a semiclassical $\psi$DO or quantum observable with principal symbol $\sigma(A)=a$.  These expectation values are the most obvious asymptotic quantities to analyze: they are, mathematically and physically, the most accessible aspects of high-energy eigenfunctions, whose analytic expressions may be difficult or even impossible to write down. Furthermore, they lead us to a natural definition of an eigenfunction-induced \emph{measure}, which will motivate the question of QUE. 
In particular, the measure on $T^*M$ defined by 
$$\mu_k(a):=\langle A\varphi_k,\varphi_k\rangle$$
is called the \emph{Wigner measure} of the state $\varphi_k$. The projection of $\mu_k$ onto the configuration manifold $M$ is equal to the probability measure $\mu^M_k:=|\varphi_k(x)|^2dx$, and it is for this reason that $\mu_k$ is called the \emph{microlocal lift} of the measure $\mu_k^M$. As a measure on the phase space $T^*M$, $\mu_k$ also encodes phase information about $\varphi_k$: it describes the local momentum of a particle measured at the $h$ scale. Since this construction is fundamental to the statement of QUE, we summarize it in the following definition: \index{Wigner measure}\index{microlocal lift}
\bb
\textbf{Definition \cc} (\emph{Wigner measure and microlocal lift}) Let $A\in \Psi(M)$ have principal symbol $a=\sigma(A)$, and as before let us denote the eigenfunctions of $\Xi(h)$ as $\varphi_k$. Then the measure on $T^*M$ defined by 
$$\mu_k(a(x,p))=\langle A\varphi_k,\varphi_k\rangle$$
is called the \emph{Wigner measure} of $\varphi_k$. For a fixed quantization $a\mapsto A$, we can write $\mu_k=\mu_k(a)$. The transformation of measures defined by
$$\mu_k^M=|\varphi_k(x)|^2dx\mapsto \langle A \varphi_k,\varphi_k\rangle=\mu_k$$
is called a \emph{microlocal lifting} of the eigenfunction-induced measure $\mu_k^M$, and $\mu_k$ is called the \emph{microlocal lift} of $\mu_k^M$. 
\bb
In general, it is difficult to analyze the Wigner measures of individual eigenfunctions. As the QE theorem suggests, we should instead concentrate on the limit of a family of Wigner measures under some suitable topology. This topology is given by the \emph{weak-$*$ topology}, which we 
define for our purposes as follows:
\bb
\textbf{Defintion \cc} (\emph{weak-$*$ convergence}) A sequence of measures $\{\mu_k\}$ on $M$ \emph{converges to $\mu$ in the weak-$*$ topology }if for all $f\in C^\infty(M)$,\index{weak-$*$ convergence}
$$\lim_{k\to \infty}\int fd\mu_k=\int fd\mu.$$\vspace{-10pt}
\bb
\textbf{Definition \cc} (\emph{quantum limit measure}) A weak-$*$ limit $\mu$ of a sequence of \index{quantum ergodicity!quantum limit measure}Wigner measures $\{\mu_k\}$ is called a \emph{quantum limit measure}. 
\bb
Technically speaking, most manifolds possess no global coordinate chart, but our definition of $\psi$DOs depends on local coordinates. We must therefore ensure that microlocal lifts and their limit measures are also well-defined if the latter exists. Microlocal lifts are generally not unique depending on the lower-order terms of $A$; nonetheless, because we know that these lower-order terms ``do not matter," it should not be surprising that sequences of microlocal lifts converge to the same quantum limit regardless of lower-order terms in the fixed quantization $a\mapsto A$. For a more detailed discussion of these technical issues, we refer the reader to \cite{Gui1}, \cite{Mar1}, and \cite{Hor3}. 

We will also take it for granted that any limit measure $\mu$ possesses the following properties:
\begin{itemize}
\item $\mu$ is indeed a probability measure on any given energy shell $\Sigma_c$;
\item $\mu$ is invariant under the geodesic flow $g^t$ on $M$. 
\end{itemize}
The first point can be proved rigorously by some convergence argument, while the second point directly follows from Egorov's theorem (Theorem 2.3.16). \index{Egorov's theorem}
We now rephrase the QE theorem as follows:
\bb
\textbf{Theorem \cc} (\emph{quantum ergodicity for Wigner measures}) If $M$ is a compact Riemannian manifold with ergodic geodesic flow, then there exists a density-1 subsequence $\{k_i\}\subset \mathbb{N}$ such that $\mu_{k_i}\to \mu_L^c$ on $\Sigma_c$ as $i\to \infty$.\index{quantum ergodicity!quantum ergodicity theorem}
\bb
The main difference between QE and QUE is that QUE posits $\mu_L^c$ as the \emph{only} limit to which the $\mu_{k_i}$ converge. We can intuitively appreciate this statement as follows: if $C=Op^w(\mathbbm{1}_S)$ is the Weyl quantization of a characteristic function for some set $S\subset \Sigma_c$, then $\langle C\varphi_k,\varphi_k\rangle$ is the probability amplitude that a particle in energy state $\lambda_k$ lies in $S$. If the only quantum limit measure of $\{\mu_{k_i}\}$ is the uniform Liouville measure $\mu_L^c$, then 
$$\langle Op^w(\mathbbm{1}_S)\varphi_{k_i},\varphi_{k_i}\rangle \to \frac{\mu_L^c(S)}{\mu_L^c(\Sigma_c)}$$
as $i\to \infty$, so that the particle becomes completely diffuse on the energy shell $\Sigma_c$. As \cite{Non1} notes, this is a quantum analogue of the phenomenon that chaotic trajectories 
equidistribute. 
Thus, one way to view the fact that QUE is much stronger than QE is that the
eigenfunctions of a quantized Hamiltonian $\Xi(h)$ become diffuse on the energy surface $\Sigma_c$, and not just on the configuration manifold $M$. \index{quantum ergodicity!quantum unique ergodicity}

If QUE fails, then there may exist a sparse, exceptional sequence of Wigner measures that does not converge uniquely in the weak-$*$ topology to the Liouville measure: the corresponding eigenfunction sequence must, in particular, exhibit a form of singular concentration. 
The \emph{QUE conjecture}, first proposed by Zeev Rudnick and Peter Sarnak in 1993 \cite{Rud1}, proposes that QUE does not fail on any manifold with an ergodic geodesic flow (or negative curvature). It is stated as a direct analogue of the QE theorem as follows.
\bb
\textbf{Conjecture \cc} (\emph{quantum unique ergodicity}, \cite{Rud1}) If $M$ is a compact Riemannian manifold with negative curvature, then the sequence of Wigner measures $\mu_k$ induced by the eigenfunctions of $\Xi(h)$ converges to the Liouville measure $\mu_L^c$ on any energy shell $\Sigma_c:=\xi^{-1}(c)\subset T^*M$. \index{quantum ergodicity!quantum unique ergodicity conjecture}
\bb
The truth of this conjecture would imply that, at both the quantum level and its semiclassical limit, whether or not a Hamiltonian system is classically ergodic does not influence its dynamics: there is little manifestation of chaotic, classical behavior in both these regimes. In particular, the quantum mechanics of these chaotic systems would ``not reflect the finer features of the classical mechanics" \cite{Sar2}. 
\pagebreak

\begin{figure}
  \begin{minipage}[b]{0.24\textwidth}
        \captionsetup{position=top,labelsep=none,font=footnotesize}
    \renewcommand{\thefigure}{\arabic{section}.\arabic{figure}\ $|$\vspace{2pt}\newline }
          \caption{Time evolution of localized, ``scarring" Laplacian eigenfunctions on the Bunimovich stadium (in triplet sequences) \cite{Kin1}.}
    \vspace{15pt}
          \label{figure}
      
    \end{minipage}\hspace{15pt}
      \begin{subfigure}[b]{0.33\textwidth}\centering
    \includegraphics[height=3.5cm]{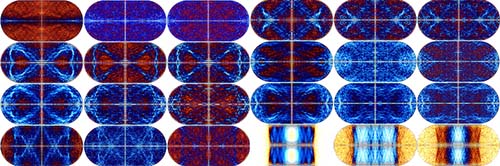}
    \captionsetup{font=footnotesize}\vspace{-20pt}

          \label{subfig-1}
      \end{subfigure}\hspace{7pt}\vspace{10pt}
  \end{figure}
The reason why QUE is substantially harder than QE is that it involves a convergence of measures and the elimination of all possible exceptional eigenfunctions, both of which avoid the scope of such standard semiclassical-analytic tools as Weyl's law. Instead of controlling the behavior of a density-1 family of eigenfunctions of $\Xi$, we must control the behavior of \emph{all} its eigenfunctions. Without such familiar techniques, 
the task of describing all the ways in which eigenfunctions can localize seems daunting. 

Another source of difficulty is the lack of explicit, easily-described examples. The only manifolds for which one can explicitly compute quantum limits are mostly limited to ones with completely integrable classical dynamics, such as the sphere, torus, or other forms of symmetric surfaces. Nonetheless, these examples are anything but classically ergodic, and any effort to find new 
examples of QE or QUE requires a considerable amount of work. 

In concluding this section, we note below several examples and non-examples of QE and QUE in the literature, and proceed to segue into \S4. 
\bb
\textbf{Example \cc} (\emph{non-QE for $S^1$}) The circle $S^1$ with Laplacian eigenfunctions $\{e^{in\theta}\}$ 
is clearly not QE or QUE; see \cite{Ana3} and \cite{Has4} for a more detailed discussion and explicit computations. 
\bb
\textbf{Example \cc} (\emph{QE for a class of billiards}) Zelditch and Zworski prove in \cite{Zel2} that compact Riemannian manifolds with piecewise smooth boundaries and ergodic billiard flows are QE. Examples of this class include the Bunimovich stadium and the Sinai billiard (c.f. Figure 1.3). 
Quantum ergodicity for the stadium billiard was first proven by G\'{e}rard and Leichtman in \cite{Ger1}. 
\bb
\textbf{Example \cc} (\emph{QE and non-QUE for stadium billiards}) Although the Bunimovich stadium is QE, it is one of the few systems proven to not be QUE in a recent work by Hassell \cite{Has1}. We present this proof in \S4.1. Unlike the geodesic flow on a negatively curved manifold, Bunimovich billards turn out to have a family of periodic orbits that correspond to \emph{bouncing ball} modes, which are eigenfunctions that describe the motion of a particle bouncing up and down the rectangular sides. The difficulty of showing QUE is proving that these bouncing ball modes exist in the high-eigenvalue limit. The Bunimovich stadium is clearly not classically uniquely ergodic due to the vertical bouncing ball modes, and so one would conjecture that it is not QUE either. \index{billiards!Bunimovich stadium}
\bb
\textbf{Example \cc} (\emph{QUE on arithmetic surfaces}) It has been shown by Lindenstrauss in \cite{Lin1,Lin2} that QUE holds in the arithmetic case, i.e. for arithmetic surfaces that arise as the quotient of the Poincar\'{e} half-disk $\mathbb{H}$ by certain congruent co-compact lattices $\Gamma$. As mentioned in \S1.1, this case is interesting because there exists a class of Laplacian-like operators, called Hecke operators, that share common eigenfunctions with the Laplacian. Nonetheless, arithmetic QUE is beyond the scope of this thesis, and we refer the interested reader to \cite{Mar2}, \cite{Sar2}, and \cite{Sou1}. \index{quantum ergodicity!arithmetic quantum ergodicity}
\bb
Further examples and non-examples involve detailed computations. To gather a sense of how extensive these calculations are, we refer the reader to \cite{Zel1}, which works out the failure of QE and QUE on the flat torus. 

\pagebreak

\setcounter{itemcounter}{1}
\setcounter{section}{3}
\section{Recent Developments in Quantum Unique Ergodicity}

In this largely expository final chapter, we discuss Hassell's disproof of QUE for the Bunimovich stadium, a topic within the scope of our thesis. 
We then conclude with a brief 
survey of recent progress in QUE research on several fronts, some of which have already been mentioned. 
We will avoid technical details in all cases, and instead aim to explain 
how the advances we describe are significant. 

\subsection{No Quantum Unique Ergodicity on the Bunimovich Stadium} 
Hassell's disproof of QUE on the Bunimovich stadium (and more generally, \emph{partially rectangular surfaces}) is the only known counterexample to QUE for billiard systems where the geodesic flow exhibits ``chaoticity" and the quantum dynamics are known to be QE \cite{Bun1}. Let us give a semi-formal definition of the Bunimovich stadium as follows:
\bb
\textbf{Definition \cc} (\emph{Bunimovich stadium})\index{billiards!Bunimovich stadium} The \emph{Bunimovich stadium} $X_t$ is a 2-dimensional Riemannian manifold formed by adjoining a rectangle with aspect ratio $t$ to two circles. With normalization, it is given explicitly as $([-t\pi /2,t\pi /2]\times [-\pi/2,\pi/2])\cup B_{\pi/2}((\pm t\pi/2,0))\subset \mathbb{R}^2$, where $B_r(c)$ denotes the ball of radius $r$ around center $c$.
\bb
As evidenced by Figure 1.3, billiards on $X_t$ are unlike the geodesic flow on a negatively curved manifold in the sense that the rectangular part of $X_t$ gives rise to seemingly integrable dynamics: that is, billiards on $X_t$ exhibit a family of 1-dimensional periodic orbits that correspond to billiards bouncing back and forth orthogonally against the rectangular walls. As mentioned before, this is termed a \emph{bouncing ball mode}, and ``scarring" involving these modes was studied by Heller as early as 1984 \cite{Hel1}. Although $X_t$ is QE because these orbits form a set of (Liouville) measure zero, $X_t$ fails to be QUE because they exist in the high-eigenvalue limit. Hassell's main result shows that some high-eigenvalue eigenstates of $X_t$ do indeed have a positive mass on the bouncing ball orbits. More specifically:
\bb
\textbf{Theorem \cc} (\emph{Hassell}, \cite{Has1}) For every $\eps>0$, there exists a subset $B_\eps\subset [1,2]$ of measure at least $1-4\eps$ and a constant $m(\eps)>0$ with the following property: for every $t\in B_\eps$, there exists a quantum limit formed from Dirichlet eigenfunctions of $\Delta$ on the stadium $X_t$ that gives probability mass at least $m(\eps)$ to the bouncing ball trajectories.
\bb
Numerical computations, as well as construction of \emph{quasi-modes}---approximate solutions for eigenfuctions given by Wentzel-Kramers-Brillouin (WKB) methods---had already indicated before Hassell's work that there exists a subsequence of modes whose Wigner measures converge to the singular measure supported on all bouncing balls. The main difficulty was showing that such bouncing ball modes exist in the limit: there may be many true eigenfunctions whose eigenvalues are close to that of a given quasi-mode's. 

\begin{figure}
\hspace{40pt}
  \begin{minipage}[b]{0.3\textwidth}
        \captionsetup{position=top,labelsep=none,font=footnotesize}
    \renewcommand{\thefigure}{\arabic{section}.\arabic{figure}\ $|$\vspace{2pt}\newline }
          \caption{A bouncing ball mode (blue) and chaotic trajectory (red) on the Bunimovich stadium $X_t$, for $t\in [1,2]$. The notation is described in the text.}
    \vspace{0pt}
          \label{figure}
      
    \end{minipage}\hspace{30pt}
      \begin{subfigure}[b]{0.33\textwidth}\centering
    \includegraphics[height=3.2cm]{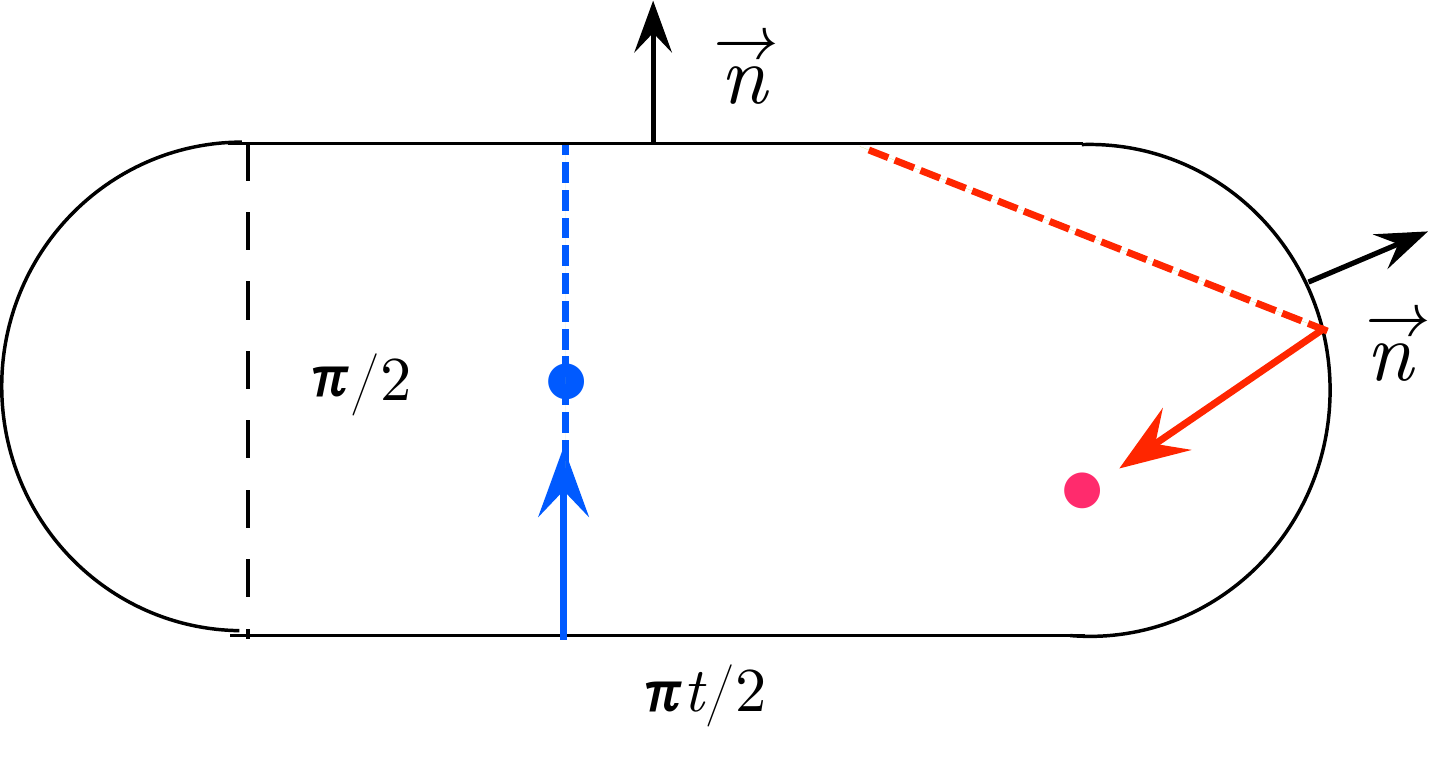}
    \captionsetup{font=footnotesize}\vspace{-20pt}

          \label{subfig-1}
      \end{subfigure}\hspace{7pt}\vspace{10pt}
\end{figure}

It is interesting to note that billiards on the Bunimovich stadium is one of the few model systems for which we have a firm understanding of QE and QUE. 
There have been no further disproofs of QUE on any billiard system since Hassell's result, 
though it is currently believed that QUE holds for billiards on the Barnett stadium (Figure 1.3). This is largely due to the work of Barnett, who verified QUE in this setting up to the 70,000th eigenfunction. 
Though it is a clear counterexample to QUE, the Bunimovich stadium illustrates the key themes of physical intuition and geometric visualization we mentioned in \S1.3, and 
is therefore an important model system to remember.

\subsection{Frontiers in Semiclassical Analysis and Quantum Chaos}

In concluding this thesis, we mention the following research areas on QUE and its related disciplines. As noted before, there are many works that lie completely outside the scope of our thesis; see \cite{Ana4, Non1, Sar2, Zel1} for more detailed surveys. Below, we aim to provide a summary of several topics that lie within the scope of our exposition, so that the reader can readily pursue further research in the following areas.
\bb
\textbf{Microlocal and semiclassical analysis.} There are several areas of microlocal and semiclassical analysis relevant to PDE analysis; examples include heat kernel methods, Fourier integral operators, the Fourier-Bros-Iagolnitzer (FBI) transform, and stationary phase approximations. We refer the reader to \cite{Dui1}, \cite{Mar1}, and \cite{Zwo1} for standard treatments of these topics. A survey of recent research in microlocal and semiclassical analysis can be found in \cite{Mel1} and \cite{Uri1}.
\bb
\textbf{Failure of QUE for cat maps.} \index{ergodicity!Arnold's cat map} It is oftentimes useful to extend the ergodicity of the geodesic flow to the ergodicity of any map on the symplectic phase space $T^*M$, where $(M,g)$ is a Riemannian manifold. Instead of a flow, ``classical" dynamics may be provided by a discrete time transformation $f:T^*M\to T^*M$ that preserves the symplectic structure. Such maps can be reconstructed from flows by considering Poincar\'{e} sections transversal to the flow, and allow us to gain insight into such topics as hyperbolic symplectomorphisms on the 2-dimensional torus. An explicit example of this subject is given by \emph{Arnold's cat map} (Figure 4.2a), defined by the action of the matrix
$$\Gamma=\left(\begin{array}{cc}
1 & 1 \\
1 & 2 \\
\end{array}\right)$$
on $\mathbb{T}^2=\mathbb{R}^2/\mathbb{Z}^2$, i.e. $f_\Gamma:\mathbb{T}^2\ni (x,p)\mapsto (x+p,x+2p)\text{ mod }1\in \mathbb{T}^2$. It has been shown that, although such maps possess the Anosov (strongly chaotic) property, the analogue of QUE for quantizations of hyperbolic toral symplectomorphisms fails to hold. 
This suggests that
there may be something unique about the geodesic flow on a negatively curved manifold, and perhaps the QUE conjecture (Conjecture 3.3.5) is already the best possible statement one can hope for. 
The reader may consult \cite{Fau1} and \cite{Fau2} for further reading.

\begin{figure}
\vspace{-10pt}
  \begin{minipage}[b]{0.13\textwidth}
        \captionsetup{position=top,labelsep=none,font=footnotesize}
    \renewcommand{\thefigure}{\arabic{section}.\arabic{figure}\ $|$\vspace{2pt}\newline }
          \caption{Illustrations for two current research areas in QUE.}
          \label{figure}
      \vspace{120pt}
    \end{minipage}\hspace{7pt}
      \begin{subfigure}[b]{0.42\textwidth}\centering
    \includegraphics[height=4.5cm]{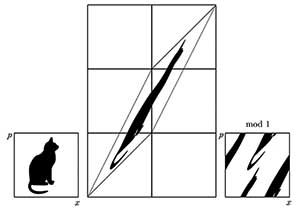}
    \captionsetup{font=footnotesize}
          \caption{A graphical representation of Arnold's cat map, the model hyperbolic toral symplectomorphism. Aside from being a test case for QUE, it is also used in dynamical systems theory and image processing.}
          \label{subfig-1}
      \end{subfigure}\hspace{12pt}
      \begin{subfigure}[b]{0.37\textwidth}\centering
    \includegraphics[height=5cm]{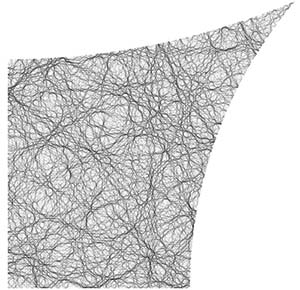}
    \captionsetup{font=footnotesize}
        \caption{A numerically-computed density plot of $|\varphi_n|^2$ on the Barnett stadium, where $n\approx 5\times 10^4$ and $\lambda_n\approx 10^6$, as calculated by Barnett's method in \cite{Bar1}.} \vspace{11.5pt}
          \label{subfig-2}
      \end{subfigure}
  \vspace{-10pt}
  \end{figure}
  $\null$
$\null$
\bb
\bb
\textbf{Numerical simulations of QUE.} \index{billiards!Barnett stadium} 
Given the technical difficulty of proving the QUE conjecture outright, many researchers have been interested in finding numerical evidence for it. The most notable project is Barnett's simulation of 70,000 eigenfunctions of the Barnett billiard (Figure 4.2b), a uniformly hyperbolic planar Euclidean billiard system, which produced evidence that QUE holds in this case. Barnett studied the rate of equidistribution of Dirichlet eigenfunctions of $\Delta$ on the Barnett stadium by examining the diagonals of the matrix elements $\langle \varphi_n,A\varphi_m\rangle$, where $A$ is some suitably defined test $\psi$DO and $\varphi_n$ is an eigenfunction of $\Delta$ with eigenvalue $\lambda_n$. In using an efficient scaling method to compute these quantities up to $n,m\approx 7\times 10^5$, Barnett found that his sample variance decayed with eigenvalue magnitude as a power of $0.48\pm 0.01$, and demonstrated a $10^2$ improvement of eigenvalue magnitude over previous studies. We refer the reader to \cite{Bar1}, \cite{Bar3}, and \cite{Non1} for a more detailed treatment of this topic. 
%
\bb
\textbf{Spectral statistics.} This tangentially-related area 
deals with the inverse problem of describing the geometric information we obtain from the eigenpairs of $\Delta$ using tools from fields such as random matrix theory. 
Although it is rarely possible to explicitly write the spectral information of $\Delta$ on any domain, one may hope to gain statistical and asymptotic information about the eigenpairs given detailed knowledge of the classical billiard or geodesic flow. Weyl's law (Corollary 2.3.15) is an example of how this can be done: it tells us that as $\lambda \to \infty$, the counting function $N(\lambda)=\#\{\lambda_n\leq \lambda\}$ satisfies the asymptotic relation
$N(\lambda)\sim \frac{\text{Area}(\Omega)}{4\pi}\lambda,$
where $\Omega$ is a planar Euclidean domain. 
In the same way that the area of $\Omega$ determines the {asymptotic mean density} of eigenvalues, 
we may introduce the \emph{consecutive level spacing distribution} \index{spectral statistics}\index{spectral statistics!consecutive level spacing distribution} as a function that counts the fraction of eigenvalues $\lambda_n$ less than $\lambda$, whose distance from the next eigenvalue $\lambda_{n+1}$ is less than $s$:
$P(s,N):=\frac{1}{N}\sum_{j=1}^N\mathbbm{1}(s>\lambda_{n+1}-\lambda_n).$
If the sequence $\{\lambda_n\}$ is sufficiently ``randomized," then there should exist a limit distribution $P(s)=\lim_{N\to \infty} P(s,N)$ so that
$$\lim_{N\to \infty}\int_0^\infty P(s,N)h(s)ds=\int_0^\infty P(s)h(s)ds.$$
for any nice function $h$. The 1977 \emph{Berry-Tabor conjecture}, \index{spectral statistics!Berry-Tabor conjecture} which has been proven in many cases, asserts that for ``generic integrable systems" 
the limit distribution $P(s)$ is equal to the waiting-time distribution for a Poisson process (i.e. $P(s)=\exp(-s)$) \cite{Ber1}.  
On the other hand, for ``generic" systems with an ergodic geodesic flow 
the 1984 \emph{Bohigas-Giannoni-Schmidt conjecture} \index{spectral statistics!Bohigas-Giannoni-Schmidt conjecture} 
proposes that the limit distribution $P(s)$ is equal to the consecutive level spacing distribution of a suitable \emph{Gaussian ensemble of Hermitian random variables} \cite{Boh1}. Numerical experiments have shown that the eigenvalue spacing statistics for the Sinai billiard correspond to those of the Gaussian orthogonal ensemble (GOE) distribution \cite{Bou4,Rud2}. One hopes that studying these statistics will lead to an increased understanding between how integrable and chaotic systems differ in the semiclassical limit. 
\bb
\textbf{Other research areas.} Aside from the topics mentioned above, there has also been progress in QUE research in the  
general case for negatively curved manifolds, 
which is exemplified in Anantharaman and Nonnenmacher's lower bound on the
 ``entropy" of any quantum limit on 
 such manifolds \cite{Ana1,Ana2,Ana3}. This implies that quantum limits cannot be too localized, although the result does not prevent the limits from 
 having
some highly localized ergodic components. 
Finally, another major 
result in QUE research is the proof of QUE in the arithmetic case; see Example 3.3.9 for a brief discussion of Lindenstrauss's work. 

\subsection{Conclusion} 

We end this thesis with a review of the semiclassical tools and results on quantum ergodicity we have developed, before proceeding to discuss the relevance of what we have hitherto studied and the future of quantum chaos at large. Having started from first principles in differential and symplectic geometry, we formalized the Laplace-Beltrami operator, motivated the study of quantum chaos---and in particular, the questions of quantum ergodicity and quantum unique ergodicity---and addressed these topics by developing the fundamental tools of semiclassical analysis. These results include the notion of Weyl quantization, which relates a function defined on phase space to a self-adjoint operator acting on functions in configuration space; the symbol calculus, which formalizes the analytic and algebraic properties of Weyl-quantized pseudodifferential operators; Weyl's law, which controls the asymptotic mean density and behavior of the eigenvalues of a generalized Hamiltonian operator or Laplacian; and Egorov's theorem, which relates the classical time evolution given by the geodesic flow on a manifold to the quantum-mechanical time evolution given by a unitary propagator. We then proved the quantum ergodicity theorem of Schnirelman, Zelditch, and Colin de Verdi\`{e}re using a combination of the symbol calculus, Weyl's law, and Egorov's theorem, after which we rigorously formulated the quantum unique ergodicity conjecture and briefly discussed current research areas in quantum chaos.

As we have mentioned before, our approach to semiclassical analysis and quantum ergodicity emphasizes intuition over straightforward proofs. The key themes introduced in \S1.3, which serve as general guidelines for thinking about problems related to semiclassical analysis and quantum chaos, illustrate this approach. For instance, we have seen that the {billiard flow} serves as a model dynamical system whose trajectories may be regular or chaotic depending on the geometry of the domain, and that the classical-quantum correspondence provides a physical heuristic as to when the QE property may hold. 
It is also important to stress that the tools we have developed in this thesis find broader applications beyond QE and QUE. In regard to semiclassical analysis, we can apply such notions as Weyl quantization and G{\aa}rding's inequality to the analysis of PDEs.
On the quantum chaos side, we may examine direct applications of Egorov's theorem to problems in quantum mechanics, as well as model the behavior of $h$-dependent and chaotic Hamiltonians. 
Lastly, in considering spectral statistics, we can endeavor to relate the distributional properties of Laplacian eigenvalues to the classical behaviors of corresponding systems. References for the foregoing applications have been mentioned in \S4.2. 

Let us now step backwards and examine QE and QUE as they relate to other fields. Our predominant emphasis on QE begs the question of how relevant QE is as a physical concept. \emph{Why, in particular, do we care that eigenfunctions of the Laplacian equidistribute? How physically useful are the notions of QE and QUE, and why should we study them if statements like the QUE conjecture are seemingly so far out of reach?} The answer is twofold. The one appealing to most mathematicians is that QE leads to, and is inspired by, many beautiful mathematical theories, some of which (like semiclassical analysis) have found broader utility which justifies their development. The second answer, which appeals to most physicists and pragmatists, is that the subject may find applications in condensed matter or quantum physics. For example, the basic idea of ergodicity is used in statistical mechanics as a measure of how chaotic and mixing an ensemble of particles is, and may also serve as a factor influencing the higher-level material properties of a system. 
Chaotic effects are manifest in atomic systems where a particle can roam freely, as in electron scattering.
Heat kernel methods, which are related to semiclassical analysis, can be computed using Feynman path integrals as expansions over geodesic terms \cite{Bar2}. \index{semiclassical analysis}
Other areas of application include wavepacket dynamics and many-body systems; the equidistribution of eigenfunctions may be important here if we specifically want particles to scar disjoint subspaces of the configuration manifold, so that they become well-described by stable periodic orbits. 
\emph{It may even be argued that chaotic---as opposed to regular---systems are the bread and butter of science: in nature, nothing is ever ideal, and the classical dynamics of a system are almost never completely integrable.}\index{ergodicity}\index{Laplace-Beltrami operator}

Although the resolution of the QUE conjecture seems to lie beyond the range of our current techniques, 
a full proof or disproof of the QUE conjecture for negatively curved manifolds would undoubtedly lead to new insights into quantum mechanics, and may introduce new techniques applicable to other areas of analysis and geometry. \index{quantum chaos}
In aiming to formulate the questions of quantum ergodicity and quantum unique ergodicity, it is our hope that the mathematics developed in this thesis will contribute toward an increased understanding of semiclassical analysis and quantum chaos. 

\lstset{numberstyle=\ttfamily}
\captionsetup[lstlisting]{format=listing,labelfont=black,textfont=black,labelsep=none,labelfont=bf}
\renewcommand\lstlistingname{Script}

\pagebreak

\appendix

\setcounter{secnumdepth}{0}

\section{Appendices}
\renewcommand{\cc}{\getcurrentref{subsection}.\theitemcounter.\stepcounter{itemcounter}}


\subsection*{Appendix I: Results from Functional Analysis}
\addcontentsline{toc}{subsection}{I \hspace{14pt} Results from Functional Analysis}

In this appendix, we provide a summary of the functional-analytic tools needed to address the Laplacian, which are also used in proofs throughout the thesis. A basic background in functional analysis is assumed; more detailed treatments than the one we present below can be found in standard analysis textbooks. See, for example, \cite{Bou1}, \cite{Dav1}, \cite{Hel4}, \cite{Hor1}, \cite{Hor2}, \cite{Red1}, and \cite{Ste3}. 

Recall that a topological space $S$ is said to be \emph{separable} if it contains a countable, dense subset, i.e. there exists a sequence $\{x_n\}_{n=1}^\infty$ of elements of $S$ such that every nonempty open subset of $S$ contains at least one element of $\{x_n\}_{n=1}^\infty$. Let $(H,\langle \cdot,\cdot\rangle)$ denote a separable complex Hilbert space. We remind ourselves that the \emph{spectrum} of a bounded linear operator $T:H\to H$ is given by the set of all $\lambda$ such that $T-\lambda I$ has a kernel. ($T$ is \emph{bounded} if its operator norm, written below, is finite.) $\lambda \in \text{Spec}(T)$ is an \emph{eigenvalue} of $T$ if $\exists x\neq 0: Tx=\lambda x$. Recall as well that $T$ is \emph{compact} if it maps a bounded set of $H$ to a relatively compact (the closure is compact) subset of $H$. The space of compact operators forms a closed ideal in the space of bounded operators in the topology induced by the operator norm $||T||=\min\{c\geq 0: ||Tx||\leq c||x||\ \forall x\in H\}$: in particular, if $T$ is compact and $S$ is bounded, then $TS$ and $ST$ are also compact; if there is a sequence $\{T_n\}$ such that $||T_n-T||\to 0$ as $n\to \infty$ and the $T_n$'s are all compact, then $T$ is also compact. \index{self-adjoint operator}\index{spectral theorem}\index{eigenvalue}\index{eigenvalue!spectrum}

The \emph{adjoint} of $T:H\to H$ is the operator $T^*$ satisfying $\langle Tx,y\rangle=\langle x,Ty\rangle$ for $x,y\in H$. $T$ is said to be \emph{self-adjoint} if $T=T^*$. The standard \emph{spectral theorem} for a bounded self-adjoint operator $T:H\to H$, which is proven in \cite{Dav1, Red1}, states that there is an orthonormal basis of $H$ consisting of eigenvectors of $T$ with real eigenvalues. The analogue for unbounded operators is more complicated, and requires some technical detail to get right.
\bb
\textbf{Theorem I.1.} (\emph{spectral theorem for bounded operators}) For each self-adjoint bounded operator $T:H\to H$, there exists a measure space $(X,\mathfrak{A},\mu)$, a real-valued function $f\in L^\infty(X,\mu)$, and a unitary operator $U:H\to L^2(X,\mu)$ such that $U^*M_fU=A$, there $M_f:x\mapsto fx$ for $x\in L^2(X,\mu)$ is the multiplication operator.
\bb
\textbf{Theorem I.2.} (\emph{spectrum of self-adjoint operators}) If $T:H\to H$ is a self-adjoint bounded operator, then $(T-\lambda I)^{-1}$ exists and is a bounded linear operator on $H$ for all $\lambda \in \mathbb{C}/\text{Spec}(T)$, and if $\text{Spec}(T)\subset [c,\infty)$, then $\langle Tx,x\rangle\geq c||x||^2$ for $x\in H$. 
\bb
Proofs of these two theorems can also be found in \cite{Red1} and \cite{Ste3}.
\bb
Now recall that an unbounded operator $S:H\to H$ is a linear operator defined on a subspace $D(S)\subset H$ called the \emph{domain} of $S$; in particular, $S$ is a linear transformation from $D(S)\to H$. $S$ is said to be \emph{densely defined} in $D(S)$ is dense as a subset in $H$. The \emph{graph} of $S$ is defined as the set $\mathcal{G}(S):=\{(x,Sx):x\in D(S)\}\subset H\times H$, and $S$ is \emph{closed} if $\mathcal{G}(S)$ is a closed subspace of $H\times H$ with norm $||(x,y)||^2=||x||^2+||y||^2$. $S$ is \emph{closable} if there is a closed unbounded operator $\overline{S}$ s.t. $D(S)\subseteq D(\overline{S})$ and $L=\overline{S}$ on $D(S)$. A standard result of functional analysis is that $\overline{S}$ is uniquely defined; call it the \emph{closure of $S$}. 

If $S$ is a unbounded densely defined operator, then it always has an unbounded adjoint operator $S^*$ such that $\langle S^*x,y\rangle=\langle x,Ly\rangle$ for all $x\in D(S^*),y\in D(S)$, where
$D(S)=\{y\in H:|\langle Sx,y\rangle|\leq C_y||x||\ \forall x\in D(S)\}$. Furthermore, $S^*$ is always closed, and if it is densely defined, then $S$ is closable with $\overline{S}=(S^*)^*, \overline{S}^*=S^*$. 
See \cite{Ste3} for a proof of this assertion. 
\bb
Finally, we recall that an unbounded, densely defined operator $S:H\to H$ is \emph{symmetric} if $\langle Sx,y\rangle=\langle x,Sy\rangle$ for all $x,y\in D(S)$; i.e. $S\subseteq S^*$, where $S\subseteq S^*$ means that $D(S)\subseteq D(S^*)$ and $S=S^*$ on all $x\in D(S)$. $S$ is \emph{self-adjoint} if $S=S^*$, and if $S$ is symmetric, then it is \emph{essentially self-adjoint} if $\overline{S}=S^*$. 

In the form of a multiplication operator, the \emph{spectral theorem} for unbounded operators can be stated as follows:
\bb
\textbf{Theorem I.3.} (\emph{spectral theorem for unbounded operators}) For each self-adjoint unbounded operator $S:H\to H$, there exists a measure space $(X,\mathfrak{A},\mu)$, a real-valued measurable function $f$, and a unitary operator $U:H\to L^2(X,\mu)$ such that $x\in D(S)$ iff $M_f(Ux)\in L^2(X,\mu)$, and $x\in D(L)\Longrightarrow U(Ax)=M_f(Ux)$, where $M_f:x\mapsto fx$ is the (unbounded) multiplication operator on $X$. 
\bb
Proofs of this result can be found in \cite{Dav1}, \cite{Red1}, and \cite{Ste2}.
\bb
Note that the Laplace-Beltrami operator on $H=L^2(M)$ is densely defined, with domain $C^\infty_c(M)$. This can be seen with an application on the \emph{closed graph theorem}, which, albeit standard functional analysis-fare, will not be presented here. 

Let us state several more theorems used in proofs of other results throughout our thesis. One useful result from functional analysis is \emph{Stone's theorem} on one-parameter unitary groups, which establishes a one-to-one correspondence between self-adjoint operators on $H$ and one-parameter families $U(t), t\in \mathbb{R}$ of unitary operators that are both strongly continuous and are homomorphisms. More precisely:
\bb
\textbf{Theorem I.4.} (\emph{Stone}) If $T:D(T)\to H$ is a bounded or unbounded self-adjoint operator, then $U(t):=\exp(-it T)$ for $t\in \mathbb{R}$ is a strongly continuous unitary group of homomorphisms, i.e. 
$$\forall t_0\in \mathbb{R}, x\in H, \lim_{t\to t_0}U(t)x=U(t_0)x$$
and 
$$\forall s,t\in \mathbb{R}, U(t+s)=U(t)U(s), U(t)^*=U(-t).$$
Furthermore, we have $\frac{d}{dt}(U(t)x)+U(t)Tx=0$ for all $t\in \mathbb{R}$ and $x\in H$. If there is a group of unitary operators satisfying the continuity and homomorphism conditions above, then we can also produce a self-adjoint operator $T$ such that all the conditions above hold. 
\bb
A proof of Stone's theorem can be found in \cite{Red1} and \cite{Ste3}. 
We also prove the following theorem for reference: 
\bb
\textbf{Theorem I.5.} (\emph{inverses of bounded linear operators}) Let $X$ and $Y$ be Hilbert spaces, $A:X\to Y$ be a bounded linear operator, and $B_1,B_2:Y\to X$ be bounded linear operators satisfying
$$
AB_1=I+C_1\ \ \  \text{ and }\ \ \ B_2A = I+C_2, $$
where $C_1:Y\to Y$, $C_2:X\to X$, $||C_1||<1$, and $||C_2||<1$. Then there exists $A^{-1}:Y\to X: AA^{-1}=A^{-1}A=I$.

\emph{Proof.} Note that, since $||C_1||<1$, $\sum_{i=0}^\infty (-1)^i C_1^i$ converges and is an inverse to the operator $I+C_1$. Thus $AA_0=I$, where $A_0:=B_1(I+C_1)^{-1}$. Similarly, $A_1A=I$ where $A_1:=(I+C_2)^{-1}B_2$. Since $A$ has both a left and right inverse, from basic functional analysis we know that it is invertible with $A^{-1}=A_0=A_1$. $\hfill \blacksquare$
\bb
Let us end the appendix by citing two useful theorems from \cite{Zwo1}. These results are needed in the proofs of Theorems 2.2.16 (inverses for elliptic symbols) and 2.3.3 (Weyl's law). The first theorem gives an explicit expression for the eigenvalues of any suitable self-adjoint operator, while the second theorem deals primarily with bounding the eigenvalue counting function $N(\lambda)$, adapted for operators beyond $\Delta$. 
\bb
\textbf{Theorem I.6.} (\emph{Courant-Fischer}) Let $T:H\to H$ be a self-adjoint operator with $\langle Tx,x\rangle\geq c\langle x,x\rangle$, and suppose that the right inverse $(T-2c)^{-1}:H\to H$ is a compact operator. Then $\text{Spec}(T)$ is discrete and countable, and $D(T)=(T-2c)^{-1}H$. If we order the eigenvalues in ascending order as $\lambda_1\leq \lambda_2\leq...$, then we also have
$$\lambda_j=\max_{\substack{V\subset H \\ \dim V<j}}
\min_{\substack{w\in D(T) \\ \dim w\perp V\neq 0}}\frac{\langle Tw,w\rangle}{||w||^2}=
\min_{\substack{V\subset D(T) \\ \dim V\geq j}}
\max_{\substack{w\in V \\ w\neq 0}}\frac{\langle Tw,w\rangle}{||w||^2},$$
where $V$ is a linear subspace of $H$. 
\bb
The proof of Theorem I.6 can be found in \cite{Red1,Cou1}. For the following statement, let $\text{rank }T=\dim T(H)$, where $T:H\to H$ is a bounded linear operator. If $S:H\to H$ is an operator that possesses a real and discrete spectrum, then we set $N_S(\lambda):=\#\{\lambda_n:\lambda_n\leq \lambda\}$. 
\bb
\textbf{Theorem I.7.} (\emph{estimates of $N(\lambda)$}) Let $T:H\to H$ be a self-adjoint operator with $\langle Tx,x\rangle\geq c\langle x,x\rangle$, and suppose that the right inverse $(T-2c)^{-1}:H\to H$ is a compact operator. 

(i) If there exists some $\delta>0$ and an operator $S:D(T)\to H$ with rank $\leq k$ such that
$$\langle Tx,x\rangle \geq (\lambda+\delta)||x||^2-\langle Sx,x\rangle$$
for $x\in D(T)$, then $N_T(\lambda)\leq k$. 

(ii) If for all $\delta>0$ there exists a subspace $V\subset D(T)$ with $\dim V\geq k$ such that
$$\langle Tx,x\rangle \leq (\lambda+\delta)||x||^2$$
for $x\in V$, then $N_T(\lambda)\geq k$. 
\bb
The proof of Theorem I.7 uses Theorem I.6, and can be found in \cite{Zwo1}.

\pagebreak

\subsection*{Appendix II: Quantization Formulas and Proof of the Expansion Theorem}
\addcontentsline{toc}{subsection}{II \hspace{9.5pt} Quantization Formulas and Proof of the Expansion Theorem}

We compile the following quantization procedures, for $a=a(x,p)\in \mathcal{S}(\mathbb{R}^{2n})$ being a real-valued symbol:\vspace{5pt}
\\\\
{\footnotesize
\begin{tabular}{p{4cm} p{10.3cm} }    
Quantization & Formula   \\\midrule
Weyl quantization    &  $Op^w(a)(\varphi)(x)=(2\pi h)^{-n}\iint_{\mathbb{R}^{n}\times \mathbb{R}^n} e^{\frac{i}{h}\langle x-y,p\rangle}a\left(\frac{x+y}{2},p\right)\varphi(y)dydp$  \\ 
 left (standard) quantization & $Op^l(a)(\varphi)(x)=(2\pi h)^{-n}\iint_{\mathbb{R}^{n}\times \mathbb{R}^n}e^{\frac{i}{h}\langle x-y,p\rangle}a(x,p)\varphi(y)dydp$ \\
 right quantization & $Op^r(a)(\varphi)(x)=(2\pi h)^{-n}\iint_{\mathbb{R}^{n}\times \mathbb{R}^n}e^{\frac{i}{h}\langle x-y,p\rangle}a(y,p)\varphi(y)dydp$ \\
 $t$-quantization, $t\in [0,1]$ & $Op_t(a)(\varphi)(x)=(2\pi h)^{-n}\iint_{\mathbb{R}^{n}\times \mathbb{R}^n}e^{\frac{i}{h}\langle x-y,p\rangle}a(tx+(1-t)y,p)\varphi(y)dydp$ 
 \\\bottomrule
\end{tabular}}
\vspace{4pt}

\hfill {\small \textbf{Table 1} $|$ A compilation of quantization procedures}
$\null$\vspace{5pt}
\bb
The properties of Weyl quantization are discussed in \S2.1.2 and \S2.2. The following table lists explicit formulas for the quantizations of certain symbols, as shown in examples throughout the thesis (c.f. Appendix VI): \vspace{5pt} 
\bb
{\footnotesize
\begin{tabular}{p{2.7cm} p{8.4cm} p{2.7cm}}    
Symbol & Quantization  & Location \\\midrule
$p^\alpha$    &  $Op_t(a)(\varphi)(x)=a(x,hD)\varphi(x)=(hD)^\alpha \varphi(x)$ & Example 2.1.15  \\ 
$\sum_{|\alpha|\leq N}a_\alpha(x)p^\alpha$ & $Op_t(a)(\varphi)(x)=a(x,hD)\varphi(x)=\sum_{|\alpha|\leq N}a_\alpha(x)(hD)^\alpha \varphi(x)$& Example 2.1.15  \\
$\langle x,p\rangle$ & $Op_t(a)(\varphi)=(1-t)\langle hD,x\varphi\rangle+t\langle x,hD\varphi\rangle$\\
&  $Op^w(a)(\varphi)=\frac{h}{2}(\langle D,x\varphi\rangle+\langle x,D\varphi\rangle)$& Example 2.1.16 \\
$a(x)$ & $Op_t(a)(\varphi)=a\varphi$& Example 2.1.17  \\
$\langle x,x^*\rangle+\langle p,p^*\rangle$ & $Op_t(a)=\langle x,x^*\rangle+\langle hD,p^*\rangle$& Example 2.1.18  \\
$e^{\frac{i}{h}\langle x,x^*\rangle + \langle p,p^*\rangle}$ & $Op^w(a)(x,hD)=e^{\frac{i}{h}l(x,hD)}$ 
& Lemma 2.2.1 \\\bottomrule 
\end{tabular}}
\vspace{4pt}

\hfill {\small \textbf{Table 2} $|$ A compilation of quantization examples}
$\null$\vspace{5pt}
\bb
We devote the remainder of this appendix to proving Theorem 2.2.6 (semiclassical expansion), following the proof in \cite{Zwo1}.
\bb
\textbf{Lemma II.1} (\emph{quadratic phase asymptotics}, \cite{Zwo1}) Let $Q$ be an invertible, symmetric real matrix. Then for each $N\in \mathbb{Z}^+$, we have
$$\int_{\mathbb{R}^n}e^{\frac{i}{2h}\langle Qx,x\rangle}a(x)dx=(2\pi h)^{{n}/{2}}\frac{e^{\frac{i \pi}{4}\text{sgn}(Q)}}{|\det Q|^{1/2}}\left(\sum_{k=0}^{N-1}\frac{h^k}{k!}\left(\frac{\langle Q^{-1}D,D\rangle}{2i}\right)^ka(0)+O(h^N)\right),$$
where $a\in C^\infty_c(\mathbb{R}^n)$ is real-valued. 

\emph{Proof.} We combine Proposition 2.1.9 and Plancherel's theorem to see that 
$$\int_{\mathbb{R}^n}e^{\frac{i}{2h}\langle Qx,x\rangle}a(x)dx=\left(\frac{h}{2\pi}\right)^{n/2} \frac{e^{\frac{i \pi}{4}\text{sgn}(Q)}}{|\det Q|^{1/2}}\int_{\mathbb{R}^n}e^{-\frac{ih}{2}\langle Q^{-1}p,p\rangle}\hat{a}(p)dp.$$
Setting $I(h,a):=\int_{\mathbb{R}^n}\exp(-\frac{ih}{2}\langle Q^{-1}p,p\rangle)\hat{a}(p)dp$, we see that
$$\frac{\partial I(h,a)}{\partial h}=\int_{\mathbb{R}^n}e^{\frac{-ih}{2}\langle Q^{-1}p,p\rangle}\left( -\frac{i}{2}\langle Q^{-1}p,p\rangle \hat{a}(p)\right)dp=I(h,Pa),$$
where $P=-\frac{i}{2}\langle Q^{-1}D,D\rangle$. Thus
$$I(h,a)=\sum_{k=0}^{N-1}\frac{h^k}{k!}I(0,P^ka)+\frac{h^N}{N!}R_N(h,a),$$
where $R_N(h,a)$ is the remainder term given by
$$R_n(h,a)=N\int_0^1 (1-t)^{N-1}I(th,P^Na)dt.$$
Applying the Fourier inversion identity gives
$$I(0,P^ka)=\int_{\mathbb{R}^n}\left(-\frac{i}{2}\langle Q^{-1}p,p\rangle\right)^k \hat{a}(p)dp=(2\pi )^nP^ka(0),$$
and using Proposition 2.1.5 gives a bound on the remainder as 
$$|R_N|\leq C_N||\widehat{P^Na}||_1\leq C_N\sup_{|\alpha|\leq 2N+n+1}|\partial^\alpha a|,$$
as desired. $\hfill \blacksquare$
\bb
\textbf{Proposition II.2} (\emph{stationary phase formula}, \cite{Zwo1}) If $a\in C^\infty_c(\mathbb{R}^{4n})$, then 
$$\iint_{\mathbb{R}^{2n}\times \mathbb{R}^{2n}}e^{\frac{i}{h}\sigma(z,w)}a(z,w)dzdw=(2\pi h)^{2n}\left( \sum_{k=0}^{N-1}\frac{h^k}{k!}\left( \frac{\sigma(D_x,D_p,D_y,D_q)}{i}\right)^k a(0,0)+O(h^N)\right)$$
for each $N\in \mathbb{Z}^+$, where $z=(x,p)\in \mathbb{R}^{2n}$ and $w=(y,q)\in \mathbb{R}^{2n}$.

\emph{Proof.} Let 
$$Q=\left(\begin{array}{cc}
& -J \\
J & \\
\end{array}\right).$$
As in Lemma 2.2.4, we see that $\frac{1}{2}\langle Q(z,w),(z,w)\rangle=\sigma(z,w)$ and 
$$\frac{1}{2}\langle Q^{-1}D,D\rangle=\sigma(D_x,D_p,D_y,D_q).$$ Applying Lemma II.1 above gives the result. $\hfill \blacksquare$
\bb
\textbf{\emph{Proof of Theorem 2.2.6.}} 
The main idea of this proof is to use Proposition II.2 above but take $h/2$ instead of $h$ and $-\sigma$ instead of $\sigma$, and substitute this into the integral representation formula (Theorem 2.2.5). Thus, if $a,b\in \mathcal{S}$, then 
$$(f\# g)(x,p)=(\pi h)^{-2n} \iint_{\mathbb{R}^{2n}\times\mathbb{R}^{2n}} e^{-\frac{2i}{h}\sigma(w_1,w_2)}a(z+w_1)b(z+w_2)dw_1dw_2,$$
where $z=(x,p)$. It remains to show that the remainders are in $O_\mathcal{S}(h^{N+1})$. For that we note from the composition theorem that
$$e^{ihA(D)}=\sum_{k=0}^N \frac{(ih)^k}{k!}A(D)^k+\frac{i^{N+1}h^{N+1}}{N!}\int_0^1 (1-t)^N e^{ithA(D)}A(D)^{N+1}dt,$$
where $e^{ithA(D)}A(D)^{N+1}$ is multiplication by $\mathcal{F}^{-1}e^{itA(\zeta)}A(\zeta)^{N+1}\mathcal{F}$ (where $\mathcal{F}$ is the Fourier transform and $\zeta\in \mathbb{R}^{4n}$). Hence $e^{ithA(D)}A(D)^{N+1}:\mathcal{S}(\mathbb{R}^{4n})\to \mathcal{S}(\mathbb{R}^{4n})$ uniformly in $h$ and $t$, which proves the estimate on the remainder. For the other equations, we simply compute
\begin{eqnarray*}
a\# b &=& ab+ihA(D)(a(x,p)b(x,p)|_{y=x,q=p}+O(h^2)\\
&=&ab+\frac{ih}{2}(\langle D_pa,D_yb\rangle-\langle D_xa,D_qb\rangle)|_{y=x,q=p}+O(h^2)\\
&=&ab+\frac{h}{2i}(\langle \partial_pa,\partial_xb\rangle-\langle \partial_xa,\partial_pb\rangle)+O(h^2)=ab+\frac{h}{2i}\{a,b\}+O(h^2).
\end{eqnarray*}
and
\begin{eqnarray*}
[Op^w(a),Op^w(b)]&=&Op^w(a)Op^w(b)-Op^w(b)Op^w(a)=Op^w(a\# b-b\# a)\\
&=&Op^w\left( ab+\frac{h}{2i}\{a,b\}+\frac{h^2}{2}A(D)^2(ab)|_{y=x,q=p}+O_\mathcal{S}(h^3)\right)\\
&&-Op^w\left(ba+\frac{h}{2i}\{b,a\}+\frac{h^2}{2}A(D)^2(ba)|_{y=x,q=p}+O_\mathcal{S}(h^3)\right)\\
&=&\frac{h}{i}Op^w(\{a,b\})+Op^w(O_\mathcal{S}(h^3)),
\end{eqnarray*}
where the final equality follows because
$$(A(D)^2(a(x,p)b(y,q))-A(D)^2(b(x,p)a(y,q)))|_{y=x,q=p}=0.$$
Finally, if $\text{supp}(a)\cap\text{supp}(b)=\emptyset$, then each term in the expansion above vanishes for arbitrary $N$. This gives the result. $\hfill \blacksquare$

\pagebreak

\subsection*{Appendix III: Source Code for Figures and Numerical Simulations}
\addcontentsline{toc}{subsection}{III \hspace{6pt} Source Code for Figures and Numerical Simulations}

This appendix provides source code for Figures 1.1 and 1.3, some of which is original work. The reader is encouraged to experiment with the scripts that generate billiard trajectories on different domains. The vector graphics language used in these simulations is Asymptote 2.24, which is available for download at \url{http://asymptote.sourceforge.net/}. \vspace{10pt}
\begin{lstlisting}[caption={Dirichlet Laplacian eigenfunctions on a square (Figure 1.1a, Mathematica 8.0+).}]
maxj = 3;
maxk = 6;
For[j = 1, j <= maxj, j++,
 For[k = 1, k <= maxk, k++,
  list[j][k] = 
    ContourPlot[Sin[(j*Pi/1)*x] Sin[(k*Pi/1)*y], {x, 0, 1}, {y, 0, 1},
      Axes -> False, Frame -> False, 
     ColorFunction -> "SunsetColors"];]]
argstring = "{";
For[j = 1, j <= maxj, j++,
 stringadd = "{";
 For[k = 1, k <= maxk, k++,
  stringadd = 
    stringadd <> "list[" <> ToString[j] <> "][" <> ToString[k] <> 
     "]" <> ",";];
 stringadd = stringadd <> "}";
 argstring = 
  argstring <> StringReplace[stringadd, {",}" -> "}"}] <> ","]
argstring = StringReplace[argstring <> "}", {"},}" -> "}}"}];
otput = GraphicsGrid[ToExpression[argstring], ImageSize -> 2000];
\end{lstlisting}

\begin{lstlisting}[caption={Dirichlet Laplacian eigenfunctions on a circle (Figure 1.1b, Mathematica 8.0+).}]
maxj = 3;
maxk = 6;
For[j = 1, j <= maxj, j++,
 For[k = 0, k <= maxk, k++,
  list[j][k] = With[{kk = k, jj = j},
     ContourPlot[(Cos[kk* phi] + Sin[kk*phi]) BesselJ[kk, 
         BesselJZero[kk, jj]* r] /. {r -> Norm[{x, y}], 
        phi -> ArcTan[x, y]}, {x, -1, 1}, {y, -1, 1}, Contours -> 50, 
      ContourLines -> False, RegionFunction -> (#1^2 + #2^2 < 1 &), 
      ColorFunction -> "SunsetColors", Axes -> False, Frame -> False]];]]
argstring = "{";
For[j = 1, j <= maxj, j++,
 stringadd = "{";
 For[k = 0, k <= maxk, k++,
  stringadd = 
    stringadd <> "list[" <> ToString[j] <> "][" <> ToString[k] <> 
     "]" <> ",";];
 stringadd = stringadd <> "}";
 argstring = 
  argstring <> StringReplace[stringadd, {",}" -> "}"}] <> ","]
argstring = StringReplace[argstring <> "}", {"},}" -> "}}"}];
otput = GraphicsGrid[ToExpression[argstring], ImageSize -> 2000];
\end{lstlisting}

\begin{lstlisting}[language=python,caption={Rectangle billiard trajectories (Figure 1.3a, Asymptote 2.24).}]
if(!settings.multipleView) settings.batchView=false; 
settings.tex="pdflatex";
defaultfilename="rectangle";
if(settings.render < 0) settings.render=4; settings.outformat=""; 
settings.inlineimage=true; settings.embed=true; settings.toolbar=false;
viewportmargin=(2,2); import graph; size(200);

path bill = (-50,-30)--(50,-30)--(50,30)--(-50,30)--cycle;
real phi = 2*pi*0.123456;
draw(bill);
pair s = (20,20), db, dt = exp(I*phi), e = s+200*dt;
path traj = s--e;
real [] c;

for(int i=0; i<50; ++i) {
  c = intersect(bill, traj);
  e = point(traj, c[1]);
  db = dir(bill, c[0]);
  draw(s--e,red);
  dot(e,blue);
  dt = -dt + 2*dot(dt,db)*db;
  s = e;
  e = s + 200*dt;
  traj = (s+dt)--e;
}
\end{lstlisting}

\begin{lstlisting}[language=python,caption={Circle billiard trajectories (Figure 1.3b, Asymptote 2.24).}]
if(!settings.multipleView) settings.batchView=false; 
settings.tex="pdflatex";
defaultfilename="circle";
if(settings.render < 0) settings.render=4; settings.outformat=""; 
settings.inlineimage=true; settings.embed=true; settings.toolbar=false;
viewportmargin=(2,2); import graph; size(200);

path bill = Circle((0,0),90.0);
real phi = 2*pi*0.23456;
draw(bill);
pair s = (20,20), db, dt = exp(I*phi), e = s+200*dt;
path traj = s--e;
real [] c;

for(int i=0; i<50; ++i) {
  c = intersect(bill, traj);
  e = point(traj, c[1]);
  db = dir(bill, c[0]);
  draw(s--e,red);
  dot(e,blue);
  dt = -dt + 2*dot(dt,db)*db;
  s = e;
  e = s + 200*dt;
  traj = (s+dt)--e;
}
\end{lstlisting}

\begin{lstlisting}[language=python,caption={Bunimovich billiard trajectories (Figure 1.3c, Asymptote 2.24).}]
if(!settings.multipleView) settings.batchView=false; 
settings.tex="pdflatex";
defaultfilename="bunimovich";
if(settings.render < 0) settings.render=4; settings.outformat=""; 
settings.inlineimage=true; settings.embed=true; settings.toolbar=false;
viewportmargin=(2,2); import graph; size(200);

path bill = (-50,-50)--(50,-50)--arc((50,0), 50, -90, 90)
  --(50,50)--(-50,50)--arc((-50,0), 50, 90, 270)--cycle;
real phi = 2*pi*0.123456;
draw(bill);
pair s = (20,20), db, dt = exp(I*phi), e = s+200*dt;
path traj = s--e;
real [] c;

for(int i=0; i<50; ++i) {
  c = intersect(bill, traj);
  e = point(traj, c[1]);
  db = dir(bill, c[0]);
  draw(s--e,red);
  dot(e,blue);
  dt = -dt + 2*dot(dt,db)*db;
  s = e;
  e = s + 200*dt;
  traj = (s+dt)--e;
}
\end{lstlisting}

\begin{lstlisting}[language=python,caption={Sinai billiard trajectories (Figure 1.3d, Asymptote 2.24).}]
if(!settings.multipleView) settings.batchView=false; 
settings.tex="pdflatex";
defaultfilename="sinai";
if(settings.render < 0) settings.render=4; settings.outformat=""; 
settings.inlineimage=true; settings.embed=true; settings.toolbar=false;
viewportmargin=(2,2); import graph; size(200);

path bill = (-90,-90)--(90,-90)--(90,90)--(-90,90)--cycle;
path inner = reverse(Circle((0,0),30.0));
real phi = 2*pi*0.05;
filldraw(bill^^inner,lightgray,black);
pair s = (30,30), db, dt = exp(I*phi), e = s+200*dt;
path traj = s--e;
real [] co;
real [][] ci;

for(int i=0; i<20; ++i) {
co = intersect(traj, bill);
ci = intersections(traj, inner);
if(ci.length > 0) {
e = point(traj, ci[0][0]);
db = dir(inner, ci[0][1]);
} else {
e = point(traj, co[0]);
db = dir(bill, co[1]);
}
draw(s--e,red);
dot(e,blue);
dt = -dt + 2*dot(dt,db)*db;
s = e;
e = s + 200*dt;
traj = (s+dt)--e;
}
size(284.52756pt,284.52756pt,keepAspect=true);
\end{lstlisting}

\begin{lstlisting}[language=python,caption={Barnett billiard trajectories (Figure 1.3e, Asymptote 2.24).}]
if(!settings.multipleView) settings.batchView=false;
settings.tex="pdflatex";
defaultfilename="barnett";
if(settings.render < 0) settings.render=4; settings.outformat=""; 
settings.inlineimage=true; settings.embed=true; settings.toolbar=false;
viewportmargin=(2,2); import graph; size(200);

pair A,B,C,D;
A = (70,-80);
D = (120,120);
B = (70,10);
C = (85,85);
guide g=A..controls B and C..D;
draw(g);

pair AA,BB,CC,DD;
AA = (-80,80);
DD = (120,120);
BB = (100,100);
CC = (-50,80);
guide gg=DD..controls BB and CC..AA;
draw(gg);

path bill = (-80,-80)--(70,-80)--g--(120,120)--gg--(-80,80)--cycle;
real phi = 2*pi*0.123456;
draw(bill);

pair s = (0,20), db, dt = exp(I*phi), e = s+300*dt;
path traj = s--e;
real [] c;

for(int i=0; i<30; ++i) {
c = intersect(bill, traj);
e = point(traj, c[1]);
db = dir(bill, c[0]);
draw(s--e,red);
dot(e,blue);
dt = -dt + 2*dot(dt,db)*db;
s = e;
e = s + 300*dt;
traj = (s+dt)--e;
}
size(284.52756pt,284.52756pt,keepAspect=true);
\end{lstlisting}

\begin{lstlisting}[language=python,caption={Triangle billiard trajectories (Figure 1.3f, Asymptote 2.24).}]
if(!settings.multipleView) settings.batchView=false; 
settings.tex="pdflatex";
defaultfilename="triangle";
if(settings.render < 0) settings.render=4; settings.outformat=""; 
settings.inlineimage=true; settings.embed=true; settings.toolbar=false;
viewportmargin=(2,2); import graph; size(200);

path bill = (-80,-80)--(120,-80)--(-80,80)--cycle;
real phi = 2*pi*0.123456;
draw(bill);

pair s = (-70,0), db, dt = exp(I*phi), e = s+300*dt;
path traj = s--e;
real [] c;

for(int i=0; i<60; ++i) {
c = intersect(bill, traj);
e = point(traj, c[1]);
db = dir(bill, c[0]);
draw(s--e,red);
dot(e,blue);
dt = -dt + 2*dot(dt,db)*db;
s = e;
e = s + 300*dt;
traj = (s+dt)--e;
}
size(284.52756pt,284.52756pt,keepAspect=true);
\end{lstlisting}
As mentioned in \S4.2, Barnett's recent numerical computation of the eigenfunctions and eigenvalues of the Dirichlet Laplacian on the Barnett stadium is a notable work. A list of the first 62,076 eigenvalues in the range $[0,1276900]$ is available at \url{http://www.math.dartmouth.edu/~ahb/qugrs_n62076.Es}. 


\pagebreak


\twocolumn
\subsection*{Appendix IV: Index of Notation}
\addcontentsline{toc}{subsection}{IV \hspace{6pt} Index of Notation}

\vspace{10pt}
{\footnotesize
\begin{supertabular}{p{1.5cm} p{4.0cm} >{\hfill}p{.4cm} }    
\multicolumn{2}{l}{List of symbols, in alphabetical order} \\\midrule
$\mathbbm{1}_S, \mathbbm{1}(S)$ & indicator function of the set or event $S$ & 60 \\
$a$ & a symbol & 35 \\
$A$ & Weyl-quantization of $a$ & 36 \\
$\mathfrak{A}$ & $\sigma$-algebra & 18\\
$A(D)$ & the differential operator $\frac{1}{2}\sigma(D_x,D_p,D_y,D_q)$ & 35 \\
$||A||_{L^2\to L^2}$ & operator norm of $A:L^2\to L^2$ & 37 \\
$\text{Area}(D)$ & area of a Euclidean domain $D\subset \mathbb{R}^n$ & 4 \\
$B^*$ & adjoint of an operator $B$ & 54 \\
$B_r(c)$ & ball of radius $r$ around $c$ & 63 \\
$\langle B\rangle_T$ & time-average of an operator $B$ & 55 \\
$\mathbb{C}$ & complex numbers & 7 \\ 
$C^1(M)$ & space of continuously differentiable functions on $M$ & 8\\
$C^\infty(M)$ & space of smooth functions on $M$ & 8 \\
$C_c^\infty(M)$ & space of smooth, compactly supported functions on $M$ & 10 \\
$\mathbb{CP}$ & complex projective space & 20 \\
$D_x$ & directional derivative with respect to $x$ & 35 \\
$D_x^\alpha$ & $\frac{1}{i^{|\alpha|}}\partial^\alpha$ & 23 \\
$\mathcal{D}'(M)$ & space of all distributions on $M$ & 45 \\
$\mathcal{D}\psi$ & dispersion of a function $\psi\in L^2(\mathbb{R})$ & 27 \\
$\partial_n,\partial_{x_n}$ & partial derivative in the $n$th local coordinate & 8 \\
$\partial^\beta$ & $\prod_{i=1}^n \frac{\partial^{\beta_i}}{\partial x_i^{\beta_i}}$ & 22 \\
${dx^i}$ & basis coordinates for the cotangent space & 9 \\ 
$\Delta,\Delta_g,\Delta_\Omega$ & Laplace-Beltrami operator, with dependencies on the Riemannian metric $g$ and domain $\Omega$  & 7 \\
$\Delta_D,\Delta_N$ & Laplace-Beltrami operator with Dirichlet and Neumann boundary conditions & 8 \\
$\delta_{xy}$ & Dirac delta of $x-y$ & 8 \\
$\delta,\delta(x)$ & Dirac delta of $x$ & 24 \\
$\text{Diff}(M)$ & space of all $C^\infty$ diffeomorphisms of $M$ & 45 \\
$\null$ \\
$\null$ \\
$\null$\vspace{5pt} \\
$\null$ \\
$\dim$ & dimension & 15 \\
$\text{div}$ & divergence &  8\\
$E_n$ & energy level (eigenvalue of $\Delta$) indexed by $n$ & 13 \\
$\exp(x)$ & exponential map $\sum_{n=1}^\infty \frac{1}{n!}x^n$ & 13 \\
$f^*$ & pullback map of $f$ & 18 \\
$||f||_p$ & norm of $f$ in $L^p$ space & 27 \\
$||f||_{\alpha,\beta}$ & seminorm of $f\in C^\infty$ with respect to multiindices $\alpha$ and $\beta$ & 22 \\
$\flat$ & flat musical isomorphism & 8\\
$\mathcal{F}(f), \hat{f}$ & Fourier transform of a function $f\in \mathcal{S}$ & 22 \\
$\mathcal{F}_h$ & semiclassical Fourier transform with parameter $h$ & 27 \\
$g_{ij}$ & covariant metric tensor & 8\\
$g^{ij}$ & contravariant metric tensor & 8\\
$|g|$ & determinant of the covariant metric tensor & 9 \\
$\Gamma^\infty$ & space of smooth sections & 8 \\
$\text{grad}, \nabla$ & gradient & 8\\
$g^t$ & geodesic flow & 18 \\
$h$ & a semiclassical parameter & 27 \\
$H$ & a Hamiltonian function & 14 \\
$\hbar$ & quantum-mechanical normalized Planck's constant & 5 \\
$H_n(x)$ & $n$th Hermite polynomial & 40 \\
$\text{Hom}(S)$ & space of homomorphisms of $S$ & 29 \\
$I$ & identity matrix & 34 \\
$\text{im}$ & image of a map & 43 \\
$\text{Im}(z)$ & imaginary part of $z$ & 25 \\
$\dashint$ & average integral & 18 \\
$\iota_{X}$ & contraction map with respect to the vector field $X$ & 14 \\
$J$ & complex structure & 34 \\
$J_k$ & $k$th Bessel function & 11 \\
$L^2$ & space of square-integrable functions & 7\\
$\lambda_n$ & eigenvalue of $\Delta$ indexed by $n$ & 4 \\
$\mathcal{L}_{X}$ & Lie derivative with respect to the vector field $X$ & 14 \\
$(M,g)$ & Riemannian manifold with metric & 8 \\
$\widetilde{M}$ & covering space of a manifold $M$ & 8 \\
$\mu$ & a measure & 18 \\
$\mu_k$ & Wigner measure of the Laplacian eigenfunction $\varphi_k$ & 59 \\
$\mu_L^c$ & Liouville measure on $\Sigma_c$ & 17 \\
$\mathbb{N}$ & natural numbers & 24 \\
$\omega$ & symplectic structure & 14 \\
$Op^l$ & left, or standard, quantization & 29 \\
$Op^r$ & right quantization & 29 \\
$Op_t$ & $t$-quantization & 29 \\
$Op^w$ & Weyl quantization & 29 \\
$\oplus$ & direct sum & 8\\
$o(f)$ & little-o of $f$ & 43 \\
$O_V(f)$ & big-O of $f$ in the vector space $V$, or w.r.t. the norm in $V$ & 35 \\
$p$ & canonical momentum coordinate & 14 \\
$\mathbb{P}$ & projective space & 20 \\
$\varphi_n,\psi_n$ & eigenfunctions of $\Delta$ indexed by $n$ & 4 \\
$\Phi^t$ & a Hamiltonian flow & 14 \\
$\#$ & Weyl product operator & 31 \\
$\# S$ & number of elements in $S$ & 4 \\
$\psi$DO & pseudodifferential operator & 22 \\
$\Psi(M)$ & $\Psi^0(M)$ & 46 \\
$\Psi^m(M)$ & space of all $\psi$DOs of order $m$ on $M$ & 46 \\
$\mathbb{R}$ & real numbers & 4 \\
$\text{Re}(z)$ & real part of $z$ & 26 \\
$\mathcal{S}$ & Schwartz space of $\mathbb{R}^n$ & 22 \\
$\mathcal{S}'$ & space of tempered distributions & 24 \\
$S(m)$ & symbol class of the order function $m$ & 36 \\
$S_\delta(m)$ & $\delta$-dependent symbol class of the order function $m$ & 36 \\
$S^m(\mathbb{R}^{2n})$ & Kohn-Nirenberg symbol class of order $m$ over $\mathbb{R}^{2n}$ & 45 \\
$S^n$ & sphere of dimension $n$ & 61 \\
$\text{sgn}(Q)$ & sign of a matrix $Q$ & 25 \\
$\sharp$ & sharp musical isomorphism & 9\\
$\sigma$ & symplectic form on $\mathbb{R}^{2n}$, given by $(x,p),(y,q)\mapsto \langle p,y\rangle-\langle x,q\rangle$ & 33 \\
$\sigma$ & map that takes $\psi$DOs to the equivalence class of symbols determined by the principal symbol & 46 \\
$\Sigma_c$ & fiber of the Hamiltonian map $H$ at value $c$ & 16 \\
$\text{Spec}(P)$ & spectrum of an operator $P$ \\
$\text{supp}(f)$ & support of a function $f$ & 35 \\
$\mathbb{T}^n$ & torus of dimension $n$ & 65 \\
$T_xM$ & tangent space of a manifold $M$ at a point $x$ & 8\\
$TM$ & tangent bundle of a manifold $M$ & 8\\
$T^*_xM$ & cotangent space of a manifold $M$ at a point $x$ & 8 \\
$T^*M$ & cotangent bundle of a manifold $M$ & 8 \\
$\mathcal{U}$ & projection operator on the set of uniform symbols & 52 \\
$V$ & a potential function & 15 \\
$\text{Vol}_MS$ & volume of the set $S$ in $M$ & 47 \\
$x$ & canonical position coordinate & 14 \\
$[x]$ & equivalence class of an element $x$ & 46 \\
$||x||$ & $\sqrt{\langle x,x\rangle}$ & 7\\
$x^\alpha$ & $\prod_{i=1}^n x_i^{\alpha_i}$ & 22 \\
$X_H$ & a Hamiltonian vector field & 14 \\
$X_t$ & Bunimovich stadium with aspect ratio $t$ & 63 \\
$\xi(x,p)$ & the Hamiltonian symbol $|p|^2+V(x)$, where $V$ is a potential & 40 \\
$\Xi(x,hD)$ & Weyl quantization of $\xi(x,p)$ & 40 \\
$\langle x,y\rangle$ & inner product of vectors $x$ and $y$ in some inner product space & 7 \\
$[X,Y]$ & Lie bracket of two vector fields $X$ and $Y$ & 15 \\
$\{X,Y\}$ & Poisson bracket of two vector fields $X$ and $Y$ & 15 \\
$\overline{z}$ & complex conjugate of $z$ & 7\\
$\langle z\rangle$ & the function $(1+|z|^2)^{1/2}$ & 36 \\
$\mathbb{Z}$ & integers & 10 \\\bottomrule
\end{supertabular}}
\newcommand{\DD}{Definition }
\newcommand{\EE}{Example }
\newcommand{\TT}{Theorem }
\newcommand{\PP}{Proposition }
\newcommand{\LL}{Lemma }
\newcommand{\FF}{Figure }
\newcommand{\CC}{Corollary }
\newcommand{\ttt}{Table }
\newcommand{\SC}{Script }
\newgeometry{top=1in,bottom=1.25in,left=1.25in,right=1.25in}
\onecolumn
\subsection*{Appendix V: Index of Mathematical Objects and Figures}
\addcontentsline{toc}{subsection}{V \hspace{10pt} Index of Mathematical Objects and Figures}
\vspace{-4pt}
{\small
\begin{longtable}{p{1.8cm} p{1.1cm} p{10.5cm} >{\hfill}p{.4cm} }    
\multicolumn{3}{l}{List of theorems and definitions, in temporal order} \\
\midrule
\DD & 1.1.1 & $L^2$ space of functions & 7 \\
\EE & 1.1.2 & Riemannian metrics & 8 \\
\DD & 1.1.3 & Laplace-Beltrami operator & 9 \\
\PP & 1.1.4 & Green's second identity & 10 \\
\TT & 1.1.5  & spectral theorem and eigenfunction basis of $\Delta$ & 10 \\
\EE &1.1.6 & Laplacian on $S^1$ & 10 \\
\EE &1.1.7 & Laplacian on a rectangular domain & 11 \\
\EE &1.1.8 & Laplacian on $\mathbb{D}$ & 11 \\
\EE &1.2.1 & circular and Sinai billiard flows & 12 \\
\EE &1.2.2 & Newton's second law & 15 \\
\DD &1.2.3 & integrability & 15 \\
\TT &1.2.4 & Liouville-Arnold, \cite{Can1} & 16 \\
\EE &1.2.5 & integrability of two-dimensional systems & 16 \\
\EE &1.2.6 & Hamiltonian nature of billiards & 16 \\
\DD &1.2.7 & Liouville measure & 17 \\
\TT &1.2.8 & geodesic and cogeodesic flow, \cite{Mil1} & 17 \\
\EE& 1.2.9 & geodesic integrability of surfaces of revolution, \cite{Kiy1} & 17 \\
\DD &1.2.10 & ergodicity and mixing & 18 \\
\TT &1.2.11 & weak ergodic theorem & 18 \\
\TT &1.2.12 & ergodicity of geodesic flow on a negatively curved manifold, \cite{Bal1} & 19 \\
\DD &2.1.1 & Schwartz space & 22 \\
\DD &2.1.2 & Fourier transform & 22 \\
\PP &2.1.3 & Fourier transform of an exponential of a real quadratic form, \cite{Zwo1} & 23 \\
\TT &2.1.4 & properties of $\mathcal{F}$ & 23 \\
\PP& 2.1.5 & estimates of $\mathcal{F}$ & 24 \\
\DD &2.1.6 & tempered distributions & 24 \\
\EE &2.1.7 & Heavyside step function & 24 \\
\EE &2.1.8 & Fourier transform of Dirac delta & 24 \\
\PP &2.1.9 & Fourier transform of an imaginary quadratic exponential, \cite{Zwo1} & 25 \\
\EE &2.1.10 & uncertainty principle in $\mathbb{R}$, \cite{Du1} & 27 \\
\DD &2.1.11 & semiclassical Fourier transform & 27 \\
\TT &2.1.12 & useful properties of $\mathcal{F}_h$ & 28 \\
\TT &2.1.13 & generalized uncertainty principle, \cite{Mar1,Zwo1} & 28 \\
\DD &2.1.14 & symbols and quantization & 29 \\
\EE &2.1.15 & quantizing a $p$-dependent symbol & 29 \\
\EE &2.1.16 & quantizing an inner product & 29 \\
\EE &2.1.17 & quantizing an $x$-dependent symbol, \cite{Zwo1} & 30 \\
\EE &2.1.18 & quantizing a linear symbol, \cite{Mar1} & 30 \\
\TT &2.1.19 & properties of quantization & 30 \\
\TT &2.1.20 & relation of quantization to commutators, \cite{Gui1,Zwo1} & 30 \\
\TT &2.1.21 & conjugation by the semiclassical Fourier transform & 31 \\
\LL &2.2.1 & quantizing an exponential of a linear symbol, \cite{Zwo1} & 31 \\
\LL &2.2.2 & Fourier decomposition of $Op^w(a)$ & 32 \\
\TT &2.2.3 & quantization composition theorem, \cite{Dim1} & 33 \\
\LL &2.2.4 & quantizing exponentials of quadratic forms, \cite{Zwo1} & 34 \\
\TT &2.2.5 & integral representation formula for composed symbols & 35 \\
\TT &2.2.6 & semiclassical expansion \cite{Mar1,Uri1,Zwo1} & 35 \\
\DD &2.2.7 & order function & 36 \\ 
\DD &2.2.8 & symbol class & 36 \\
\DD &2.2.9 & asymptotic symbol decomposition & 36 \\
\TT &2.2.10 & Borel & 36 \\
\TT &2.2.11 & properties of quantization for symbol classes & 37 \\
\TT &2.2.12 & semiclassical expansion for symbol classes, \cite{Mar1,Zwo1} & 37 \\
\TT& 2.2.13 & quantization composition for symbol classes, \cite{Dim1,Hor3} & 37 \\
\TT &2.2.14 & $L^2$ boundedness and compactness for symbol classes & 37 \\
\DD &2.2.15 & elliptic symbols & 38 \\
\TT &2.2.16 & inverses for elliptic symbols, \cite{Mar1,Zwo1} & 38 \\
\TT &2.2.17 & weak G{\aa}rding's inequality, \cite{Dim1} & 39 \\
\TT &2.2.18 & sharp G{\aa}rding's inequality, \cite{Dim1} & 39 \\
\TT &2.3.1 & Weyl's law for the SHO, \cite{Zwo1} & 40 \\
\PP &2.3.2 & products of projection and quantized operators & 41 \\
\TT &2.3.3 & Weyl's law for $\mathbb{R}^n$ & 41 \\
\DD &2.3.4 & distributions on $M$ & 45 \\
\DD &2.3.5 & differential operators on $M$ & 45 \\
\DD &2.3.6 & Kohn-Nirenberg symbols & 45 \\
\PP &2.3.7 & invariance of Kohn-Nirenberg symbols under diffeomorphisms & 46 \\
\PP &2.3.8 & Kohn-Nirenberg symbol composition & 46 \\
\DD &2.3.9 & $\psi$DOs on $M$ & 46 \\
\DD &2.3.10 & symbols on $T^*M$ & 46 \\
\TT &2.3.11 & quantization on $M$, \cite{Dim1,Mar1,Zwo1} & 46 \\
\TT &2.3.12 & boundedness and compactness of $\psi$DOs & 46 \\
\TT &2.3.13 & eigenfunctions of $\Xi(h)$ & 47 \\
\TT &2.3.14 & generalized Weyl's law & 47 \\
\CC &2.3.15 & Weyl's law for a domain $\Omega \subset \mathbb{R}^2$ & 47 \\
\TT &2.3.16 & Egorov, \cite{Ego1} & 48 \\
\TT &3.1.1 & quantum ergodicity, Schnirelman and Colin de Verdi\`{e}re, \cite{Col1} & 50 \\
\CC &3.1.2 & density of eigenfunctions & 51 \\
\DD &3.1.3 & uniform symbol & 51 \\
\TT &3.1.4 & quantum ergodicity, Zelditch and Zworski, \cite{Zel2} & 52 \\
\TT &3.1.5 & quantum ergodicity, density version, \cite{Zwo1} & 53 \\
\EE &3.1.6 & equidistribution of Laplacian eigenfunctions & 53 \\
\LL &3.2.1 & $L^2$ quantization bound & 53 \\
\LL &3.2.2 & generalized Weyl's law & 54 \\
\DD &3.3.1 & Wigner measure and microlocal lift & 59 \\
\DD &3.3.2 & weak-$*$ convergence & 59 \\
\DD &3.3.3 & quantum limit measure & 59 \\
\TT &3.3.4 & quantum ergodicity for Wigner measures & 60 \\
Conjecture &3.3.5 & quantum unique ergodicity, \cite{Rud1} & 60 \\
\EE &3.3.6 & non-QE for $S^1$ & 61 \\
\EE &3.3.7 & QE for a class of billiards & 61 \\
\EE &3.3.8 & QE and non-QUE for stadium billiards & 61 \\
\EE &3.3.9 & QUE on arithmetic surfaces & 61 \\
\DD &4.1.1 & Bunimovich stadium & 63 \\
\TT &4.1.2 & Hassell, \cite{Has1} & 63 \\
\TT & I.1 & spectral theorem for bounded operators & 68 \\
\TT & I.2 & spectrum of self-adjoint operators & 68 \\
\TT & I.3 & spectral theorem for unbounded operators & 69 \\
\TT & I.4 & Stone & 69 \\
\TT &I.5 & inverses of bounded linear operators & 69 \\
\TT &I.6 & Courant-Fischer & 70 \\
\TT &I.7 & estimates of $N(\lambda)$ & 70 \\
\LL &II.1& quadratic phase asymptotics, \cite{Zwo1} & 71 \\
\PP & II.2 & stationary phase formula, \cite{Zwo1} & 72 \\
\FF &1.1 & contour plots of Dirichlet Laplacian eigenfunctions & 11\\
\FF &1.2 & trajectories on the circular and Sinai billiards & 13 \\
\FF &1.3 & billiard flow on different domains & 13 \\
\FF &2.1 & contours $C_k$ used in the proof of \PP 2.1.9 & 26 \\
\FF &3.1 & scarring eigenfunctions on the Bunimovich stadium, \cite{Kin1} & 61 \\
\FF &4.1 & trajectories on the Bunimovich stadium & 64\\
\ttt & 1 & a compilation of quantization procedures & 71 \\
\ttt & 2 & a compilation of quantization examples & 71 \\
\SC & 1 & Dirichlet Laplacian eigenfunctions on a square & 74 \\
\SC & 2 & Dirichlet Laplacian eigenfunctions on a circle & 74 \\
\SC & 3 & rectangle billiard trajectories & 75 \\
\SC & 4 & circle billiard trajectories & 75 \\
\SC & 5 & Bunimovich billiard trajectories & 76 \\
\SC & 6 & Sinai billiard trajectories & 77 \\
\SC & 7 & Barnett billiard trajectories & 77 \\
\SC & 8 & triangle billiard trajectories & 78 
\\\bottomrule
\end{longtable}
}




{\small
\printindex}



\end{document}